\documentclass[a4paper,twocolumn,11pt9,accepted=2023-10-25]{quantumarticle}
\pdfoutput=1
\usepackage[utf8]{inputenc}
\usepackage[T1]{fontenc}
\usepackage{graphicx}
\usepackage{amssymb}
\usepackage{amsmath}
\usepackage{amsthm}
\usepackage{subcaption}
\usepackage{algorithm}
\usepackage{algpseudocode}
\usepackage{booktabs}
\usepackage{hyperref}
\usepackage{colortbl}
\usepackage[affil-it]{authblk}
\usepackage{url}
\usepackage{appendix}
\usepackage[htt]{hyphenat}
\usepackage[sort, compress]{cite}

\usepackage{array}
\newcolumntype{H}{>{\setbox0=\hbox\bgroup}c<{\egroup}@{}}

\newcommand{\gate}[1]{\texttt{#1}}
\newcommand{\eg}{e.\,g., }
\newcommand{\Eg}{E.\,g., }
\newcommand{\ie}{i.\,e.\@ }
\newcommand{\Ie}{I.\,e., }

\DeclareMathOperator{\tr}{tr}
\DeclareMathOperator{\rank}{rank}
\makeatletter
\renewcommand{\fnum@figure}{Fig.\ \thefigure}
\renewcommand{\fnum@table}{Tab.\ \thetable}
\makeatother

\title{Quantum Circuit Compiler for a Shuttling-Based Trapped-Ion Quantum Computer}

\author{Fabian Kreppel}
\affiliation{Institute of Computer Science, Johannes Gutenberg University, Staudingerweg 9, 55128 Mainz, Germany}
\email{f.kreppel@uni-mainz.de}
\orcid{0000-0002-8523-2112}
\author{Christian Melzer}
\affiliation{Institute of Physics, Johannes Gutenberg University, Staudingerweg 7, 55128 Mainz, Germany}
\email{chmelzer@uni-mainz.de}
\orcid{0000-0001-7482-7682}
\author{Diego Olvera Millán}
\affiliation{Institute of Physics, Johannes Gutenberg University, Staudingerweg 7, 55128 Mainz, Germany}
\email{dolveram@uni-mainz.de}
\orcid{0000-0003-0491-1576}
\author{Janis Wagner}
\affiliation{Institute of Physics, Johannes Gutenberg University, Staudingerweg 7, 55128 Mainz, Germany}
\email{wajanis@uni-mainz.de}
\orcid{0000-0001-8789-9079}
\author{Janine Hilder}
\affiliation{Institute of Physics, Johannes Gutenberg University, Staudingerweg 7, 55128 Mainz, Germany}
\email{hilder@uni-mainz.de}
\orcid{0000-0001-9129-1314}
\author{Ulrich Poschinger}
\affiliation{Institute of Physics, Johannes Gutenberg University, Staudingerweg 7, 55128 Mainz, Germany}
\email{poschin@uni-mainz.de}
\orcid{0000-0001-5341-7860}
\author{Ferdinand Schmidt-Kaler}
\affiliation{Institute of Physics, Johannes Gutenberg University, Staudingerweg 7, 55128 Mainz, Germany}
\email{fsk@uni-mainz.de}
\orcid{0000-0002-5697-2568}
\author{André Brinkmann}
\affiliation{Institute of Computer Science, Johannes Gutenberg University, Staudingerweg 9, 55128 Mainz, Germany}
\email{brinkman@uni-mainz.de}
\orcid{0000-0003-3083-2775}

\begin{document}

\maketitle

\begin{abstract}

The increasing capabilities of quantum computing hardware and the challenge of realizing deep quantum circuits require fully automated and efficient tools for compiling quantum circuits. To express arbitrary circuits in a sequence of native gates specific to the quantum computer architecture, it is necessary to make algorithms portable across the landscape of quantum hardware providers. In this work, we present a compiler capable of transforming and optimizing a quantum circuit targeting a shuttling-based trapped-ion quantum processor. It consists of custom algorithms set on top of the quantum circuit framework Pytket. The performance was evaluated for a wide range of quantum circuits and the results show that the gate counts can be reduced by factors up to 5.1 compared to standard Pytket and up to 2.2 compared to standard Qiskit compilation.

\end{abstract}

\section{Introduction}
\label{sec-introduction}
The current rapid maturation of quantum information processing platforms \cite{Arute2019} brings meaningful scientific and commercial applications of quantum computing within reach. While it seems unlikely that fault-tolerant devices \cite{Postler2022, Egan2021} will scale to sufficiently large numbers of logical qubits in the near future, noisy intermediate scale quantum (NISQ) devices are predicted to lead the way into the era of applied quantum computing \cite{Preskill2018quantumcomputingin}. The quantum compiler stack of such platforms will crucially determine their capabilities and performance: First, fully automated, hardware-agnostic front-ends will enable access for non-expert users from various scientific disciplines and industry. Second, optimizations performed at the compilation stage will allow overcoming limitations due to noise and limited qubit register sizes, thereby increasing the functionality of a given NISQ platform.

As quantum hardware scales to larger qubit register sizes and deeper gate sequences, the platforms become increasingly complex and require tool support and automation. It is no longer feasible to manually design quantum circuits and use fixed decomposition schemes to convert the algorithm input into a complete set of qubit operations which the specific quantum hardware can execute, referred to as its \emph{native gate set}. Dedicated quantum compilers are required for the optimized conversion of large circuits into low-level hardware instructions.

\begin{figure*}
    \centering
	\includegraphics[width=\textwidth,keepaspectratio]{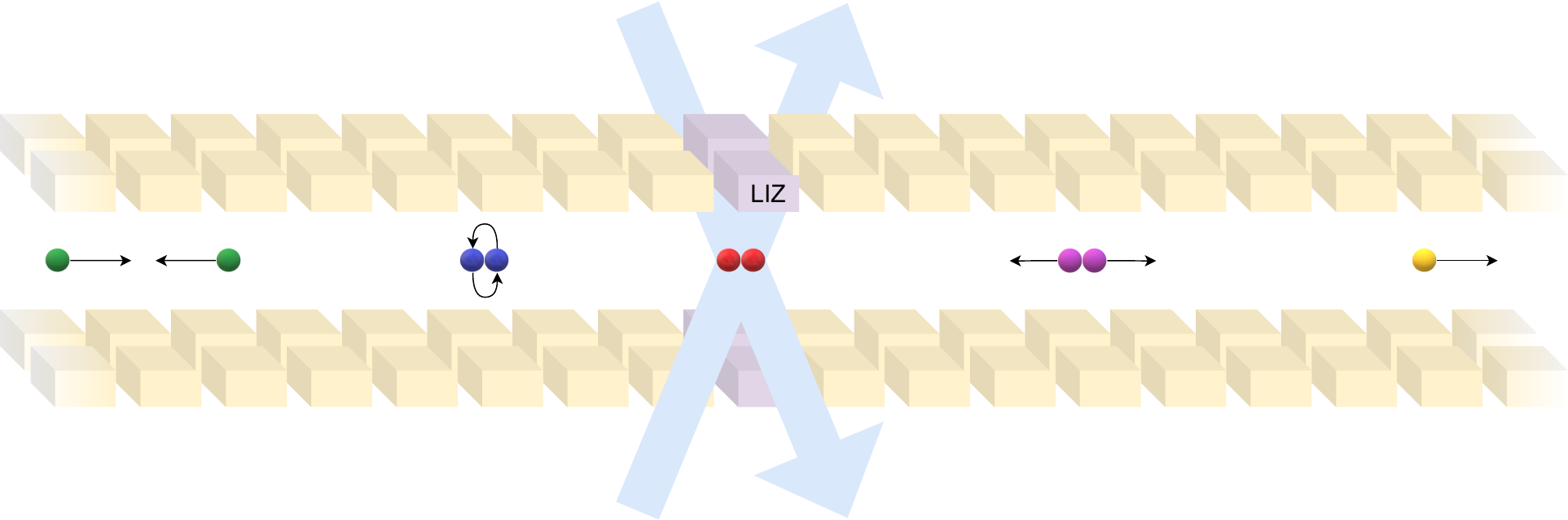}
	\caption{Linear shuttling-based segmented ion trap architecture: The ions are stored in small groups at different segments of the architecture. The lasers performing the gate operations (here on the red ions) are directed only to a specific segment, the laser interaction zone (LIZ, purple segment). The following operations reconfigure the ion positions (from left to right): merging ions into a new group (green ions), physically swapping ions (blue ions), splitting a group (purple ions), and translating ions between different segments (yellow ion).}
	\label{fig-trap-archictecture}
\end{figure*}

In this work, we present such a circuit compiler developed for a shuttling-based trapped-ion quantum computing platform \cite{KIELPINSKI2002,ShuttlingArchitecture,Pino2021,Murali20}. This platform encodes qubits in long-lived states of atomic ions stored in segmented microchip ion traps, as shown in \autoref{fig-trap-archictecture}. To increase potential scalability, subsets of the qubit register (a belonging set of qubits) are stored at different trap segments. Shuttling operations, performed by changing the voltages applied to the trap electrodes, dynamically reconfigure the register between subsequent gate operations.
Gates are executed at specific trap sites and are driven by laser or microwave radiation. While this approach provides all-to-all connectivity within the register and avoids crosstalk errors, the shuttling operations incur substantial timing overhead, resulting in rather slow operation timescales in the order of tens of microseconds per operation \cite{FaultTolerantReadoutMainz}. This can lead to increased error rates because the reduced operation speed aggravates dephasing. Furthermore, the shuttling operations can lead to reduced gate fidelities due to shuttling-induced heating of the qubit ions.
Therefore, besides compiling a given circuit into a native gate set, the main task of the gate compiler stage of a shuttling-based platform is to minimize the required amount of shuttling operations. This is achieved by minimizing the overall gate count and by arranging the execution order of the gates in a favorable way. A subsequent \emph{Shuttling Compiler} stage, which is beyond the scope of this work, handles the task of generating schedules of shuttling operations based on the compilation result \linebreak
\cite{umz-shuttling, durandau2022, muzzletheshuttle, backend-compiler, qubitrouting}.

This paper focuses on taking into account the properties of the shuttling-based quantum computing hardware when optimizing the circuit. It provides insights into how the hardware architecture can be exploited to further improve the fidelity of the compiled circuit. Since we use many state-of-the-art transformations and algorithms as the basis for our circuit compiler, much of this work is also applicable to more general quantum circuit compilers.

The structure of this paper is as follows: \autoref{sec-related-work} reviews existing circuit optimization techniques. \autoref{sec-graph-description} defines the representation of quantum circuits used in this work. This is followed in \autoref{sec-transformations} by a detailed description of all circuit transformation algorithms used. Parameterized circuits and their compilation are discussed in \autoref{sec-parameterized-circuits}. An evaluation of the methods is presented in \autoref{sec-evaluation} and shows the benefits of our circuit compiler.

\section{Background}
\label{sec-related-work}
Due to the increasing size and complexity of quantum circuits, automatic circuit compilation is required to execute quantum circuits on different platforms. For this purpose, powerful frameworks for quantum computing \cite{tket, pennylane, qiskit} have been developed. Although their features vary widely, all frameworks provide some kind of built-in optimization.

While the simplest form of circuit compilation replaces gates with predefined sequences of other gates (often referred to as decomposition), more advanced techniques minimize the number of gates. A common strategy is to reduce the overall gate count, with a particular focus on expensive two-qubit gates \cite{Estebanizer16, Nguyen2021, Maslov2017}. One such approach uses a different circuit representation called ZX-calculus \cite{ZX2011}, which allows simplifications at the functional level. Another algorithm searches for common circuit patterns, called templates, and replaces them with shorter or otherwise preferable but functionally identical gate sequences \cite{Maslov2017}.

When compiling quantum circuits, the qubit mapping is often considered as well. Ideally, each qubit can interact with any other qubit, allowing two-qubit gates to be executed between any pair of qubits. However, for existing platforms, interactions are limited to nearest neighbor topology or full connectivity within subsets of limited size. Mapping the qubits from the algorithmic circuit to physical qubits subjected to these hardware constraints is called the routing problem \cite{tket}. To make arbitrary two-qubit gates executable on the quantum hardware, \gate{SWAP} gates must be inserted into the circuit \cite{8970267, 8382253, cowtan_et_al:LIPIcs:2019:10397, DBLP:conf/asplos/LiDX19}. In the case of ion trap quantum computers, ion positions can be physically swapped to establish dynamic all-to-all connectivity. Consequently, no computational \gate{SWAP} gates need to be inserted at this stage.

The Pytket framework \cite{tket} provides a wide variety of circuit transformation algorithms and therefore we use it as the operational basis for the custom circuit compiler described in this paper. Functionality such as the removal of redundancies and the rebasing of arbitrary gates into the native gate set is mainly realized using Pytket's built-in functions. Since Pytket is designed for superconducting architectures, we have additionally developed and implemented some specific functionalities for trapped-ion quantum computers. These include concatenating multiple local rotations into global rotations, restricting gate parameters to a fixed set of values, and improving gate ordering.

Previous approaches to quantum circuit compilers have focused on different architectures such as photonic \cite{photonic_compiler} and superconducting quantum computers \cite{Shi2019}. These compilers share similarities with our approach, such as the use of the ZX-calculus \cite{zxcalculus} to optimize the circuits. There are also several Pytket extensions for different quantum devices \cite{pytket-extensions}. However, there are inherent differences in the kind of parallelism offered by the hardware and thus should be used to get the best results. The same applies to the native operations (like the physical ion swap in our case).

\section{Graph description of the quantum circuit}
\label{sec-graph-description}
\begin{figure}
    \centering
	\includegraphics[width=\columnwidth,keepaspectratio]{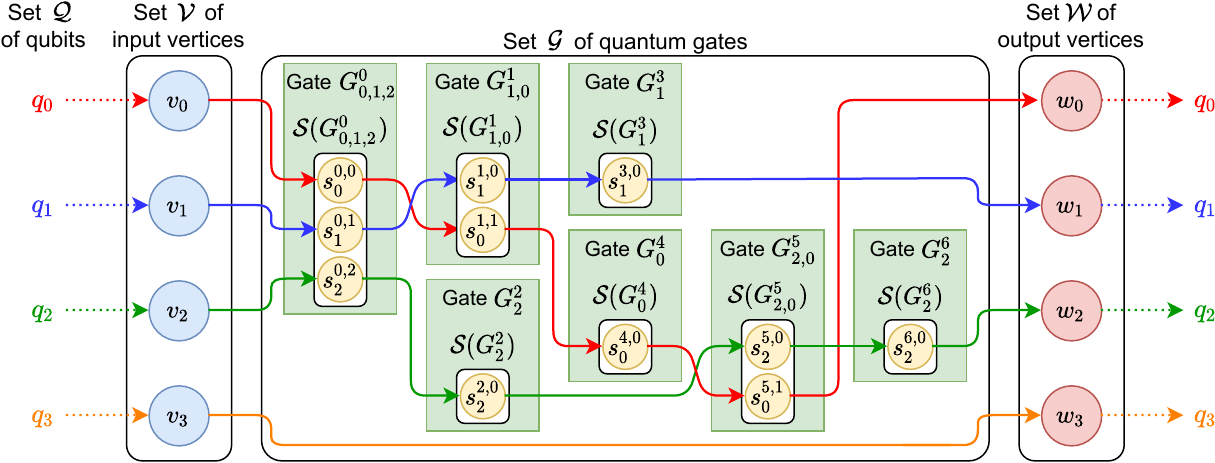}
	\caption{Example of a quantum circuit graph with four qubits and seven gates. The four qubits are depicted on both sides, and each qubit $q_i$ is assigned to its input vertex $v_i$. In the middle are the gates, which are executed on the qubits. Each gate has as many subvertices as the number of qubits it affects. To the right are the output vertices for each qubit. The directed edges show the order in which the gates are executed on each qubit, represented by the different edge colors. No gate is executed on $q_3$.}
	\label{fig-graph-construction}
\end{figure}

This section describes a quantum circuit as a directed acyclic graph (DAG), which is the data structure on which Pytket and our custom subroutines operate. The first subsection defines the DAG, and the second subsection constructs it. Such a graph is depicted in \autoref{fig-graph-construction}. At the end of this section, we describe the native gate set of our platform.

\subsection{Graph definition}
\label{subsec-graph-definition}
We consider a quantum circuit consisting of a set $\mathcal{Q} = \{ q_0, \dots, q_{n-1} \}$ of $n$ qubits. The circuit is represented as a directed acyclic graph $C=\left(\mathcal{V} \cup \mathcal{G} \cup \mathcal{W}, \mathcal{E}\right)$ with sets of vertices $\mathcal{V}$, $\mathcal{G}$ and $\mathcal{W}$ which are pairwise disjoint and defined as follows:
\begin{itemize}
    \item $\mathcal{V}$ is the set of input vertices. For each qubit $q_i \in \mathcal{Q}$ there is exactly one vertex $v_i \in \mathcal{V}$, so $|\mathcal{V}| = n$ holds. Each vertex $v_i$ has exactly one outgoing edge.
    \item $\mathcal{G}$ is the set of quantum gates of the circuit. If a quantum gate operates on $m \ge 1$ different qubits $q_{i_0}, \dots, q_{i_{m-1}} \in \mathcal{Q}$, the vertex has exactly $m$ incoming and $m$ outgoing edges. Additionally, each gate $G \in \mathcal{G}$ depends on $a \ge 0$ parameters, which are angle parameters with values of $0 \le \phi_0, \dots, \phi_{a-1} < 2\pi$. In the following, a gate $G$ acting on the qubits $q_{i_0}, \dots, q_{i_{m-1}}$ and depending on the parameters $\phi_0, \dots, \phi_{a-1}$ is denoted as
    \begin{align}
        G_{i_0, \dots, i_{m-1}}^j\left(\phi_0, \dots, \phi_{a-1}\right),
    \end{align}
    where $j$ is a unique identifier for the gate.
    \item $\mathcal{W}$ is the set of output vertices. For each qubit $q_i \in \mathcal{Q}$ there is exactly one vertex $w_i \in \mathcal{W}$, so $|\mathcal{W}| = n$ holds. Each vertex $w_i$ has exactly one incoming edge.
\end{itemize}

\subsection{Graph construction}
\label{subsec-graph-construction}
Each quantum circuit $C$ starts with the input vertices $v_0, \dots, v_{n-1} \in \mathcal{V}$ and each qubit $q_i \in \mathcal{Q}$ is assigned to its vertex $v_i$. The outgoing edge $\left(v_i, G\right) \in \mathcal{E}$ leads to the vertex $G \in \mathcal{G}$ which represents the quantum gate to be executed first on $q_i$. If the circuit does not contain a quantum gate to be executed on $q_i$, the outgoing edge goes directly to the output vertex $w_i \in \mathcal{W}$, so $\left(v_i, w_i\right) \in \mathcal{E}$ holds.

All vertices in $\mathcal{G}$ represent the inner vertices of the circuit $C$. If a gate $G^j \in \mathcal{G}$ is executed directly before a gate $G^k \in \mathcal{G}$ on the same qubit $q_i$, a directed edge $\left(G^j, G^k\right) \in \mathcal{E}$ connects $G^j$ and $G^k$.

The output vertices $w_0, \dots, w_{n-1} \in \mathcal{W}$ form the end of the circuit $C$. Each qubit $q_i \in \mathcal{Q}$ is assigned to its vertex $w_i$. The incoming edge $\left(G, w_i\right) \in \mathcal{E}$ comes from the vertex $G \in \mathcal{G}$ which represents the last quantum gate executed on $q_i$. If the circuit does not contain a quantum gate executed on $q_i$, the incoming edge comes directly from the input vertex $v_i \in \mathcal{V}$.

In contrast to the input and output vertices, which all have exactly one outgoing or one incoming edge, the inner vertices have $m \geq 1$ incoming and outgoing edges. To manage these edges, each inner vertex $G^j \in \mathcal{G}$ consists of $m$ subvertices $s^{j,0}, \dots, s^{j, m-1}$. Each of these subvertices has exactly one incoming and one outgoing edge. In the following we denote the set of subvertices of $G^j$ as $\mathcal{S}\left(G^j\right)$.

Assume that $G^j$, $G^k$ and $G^l$ with $G^j \in \mathcal{V} \cup \mathcal{G}$, $G^k\in \mathcal{G}$ and $G^l \in \mathcal{G} \cup \mathcal{W}$ are executed sequentially on a qubit $q_i \in \mathcal{Q}$, so that the edges $e_{\textnormal{in}}=\left(G^j, G^k), e_{\textnormal{out}} = (G^k, G^l\right) \in \mathcal{E}$ exist. Instead of connecting the incoming edge $e_{\textnormal{in}}$ directly to $G^k$, it is connected to a subvertex $s^{k, \alpha} \in \mathcal{S}\left(G^k\right)$. The outgoing edge $e_{\textnormal{out}}$ starts at the same subvertex $s^{k, \alpha}$. A special kind of gates acting on $m > 1$ qubits are controlled gates, where $\mu$ qubits with $1 \leq \mu < m$ act as control and $m - \mu$ qubits act as target qubits. The gate is only executed on the target qubits if and only if all $\mu$ control qubits are one, whereby the control qubits can also be in a superposition. If $G^k$ is a controlled gate, the $\mu$ control qubits are connected to the subvertices $s^{k, 0}, \dots, s^{k, \mu-1}$ and the $m - \mu$ target qubits are connected to the subvertices $s^{k, \mu}, \dots, s^{k, m-1}$. Since only $q_i$ is connected to the subvertex $s^{k, \alpha}$, the subvertex can be denoted as $s_i^{k, \alpha}$.

After constructing the complete graph, there are $n$ disjoint, well-defined paths in $C$ starting at an input vertex $v_i\in \mathcal{V}$ and ending at an output vertex $w_i \in \mathcal{W}$. Each path from $v_i$ to $w_i$ describes the order in which the gates are applied to the qubit $q_i \in \mathcal{Q}$. An example of a graph representation of a quantum circuit is shown in \autoref{fig-graph-construction}.

\subsection{Native gate set}
\label{subsec-native-gate-set}
The state of an $n$ qubit register is commonly represented by a normalized complex-valued vector with $2^n$ entries, corresponding to the probability amplitudes of the logical basis states. The gates then act as unitary transformations, represented by unitary matrices of dimension $2^n \times 2^n$, on the state.

Products of unitary operators or their corresponding matrices represent the serial execution of gates on different qubits, where the products are read from right to left. Note that we use a less strict notation throughout the paper, where operator products are written as simple products, even though the operators may act on different subsets of the qubit register, and tensor products are not always written explicitly. Since global phases of the quantum states do not affect the measurement outcomes, equality of unitaries and states means equality up to a global phase.

To execute the circuit $C$ on a given hardware platform, it must be transformed into an equivalent circuit consisting only of gates from a native gate set. Our platform \cite{FaultTolerantReadoutMainz} implements the native gate set  
\begin{align}
    \mathcal{M} = \left\{ 
    \gate{R}\left(\theta, \phi\right), 
    \gate{Rz}\left(\phi\right), 
    \gate{ZZ}\left(\theta\right) 
    \right\},
\end{align}
where each gate is parameterized by up to two rotation angles $\theta$ and $\phi$ with $0 \le \theta, \phi < 2\pi$. Due to their meaning for the actual operation, they are referred to as the pulse area and the phase, respectively. The gates from $\mathcal{M}$ are defined in terms of the Pauli operators $X$, $Y$ and $Z$ as follows:
\begin{subequations}
\label{eq-native-gates}
    \begin{align}
        \gate{R}\left(\theta, \phi\right) &= \exp\left(-i\tfrac{\theta}{2}\left(\cos \phi X + \sin \phi Y\right)\right),
        \label{eq-r-gate} \\
        \gate{Rz}\left(\phi\right) &= \exp\left(-i\tfrac{\phi}{2}Z\right),  \label{eq-rz-gate} \\
        \gate{ZZ}\left(\theta\right) &= \exp\left(-i\tfrac{\theta}{2} Z \otimes Z\right). \label{eq-zz-gate}
    \end{align}
\end{subequations}

The gates \gate{R} and \gate{Rz} are single-qubit gates, while \gate{ZZ} is a two-qubit gate. This set is complete, so any quantum algorithm can be decomposed into a sequence of these operations \cite{Kitaev_1997, Sol_Kit_Dawson_Nielsen}. Note that some trapped-ion platforms do not allow native \gate{ZZ} gates, but instead use \gate{XX} gates generated by bichromatic radiation fields \cite{PhysRevLett.82.1971}. Our compilation scheme is still valid for such architectures, since \gate{ZZ} gates can be generated from \gate{XX} gates using local wrapper rotations.

Furthermore, the identity \gate{I}, the Pauli gates \gate{X}, \gate{Y}, \gate{Z}, and the rotations around $X$ and $Y$, \gate{Rx} and \gate{Ry}, are special forms of the \gate{R} and \gate{Rz} gates and thus also part of the native gate set. Their relation to the gates in \linebreak
$\mathcal{M}$ is \\
\begin{subequations}
    \noindent\centering
    \begin{minipage}{0.48\columnwidth}
        \begin{align}
        \gate{I} &= \gate{R}\left(0, \phi\right), \\
        \gate{Rx}\left(\theta\right) &= \gate{R}\left(\theta, 0\right), \\
        \gate{Ry}\left(\theta\right) &= \gate{R}\left(\theta, \frac{\pi}{2}\right),
        \end{align}
    \end{minipage}
    \hfill
    \begin{minipage}{0.48\columnwidth}
        \begin{align}
        \gate{X} &= i \gate{Rx}\left(\pi\right), \\
        \gate{Y} &= i \gate{Ry}\left(\pi\right), \\
        \gate{Z} &= -\gate{Rz}\left(\pi\right).
        \end{align}
    \end{minipage}\bigskip
\end{subequations}

\gate{SWAP} gates are defined as
\begin{align}
    \gate{SWAP} = \tfrac{1}{2}\left(I \otimes I +X \otimes X + Y \otimes Y + Z \otimes Z\right).
\end{align}
These gates are required to establish full connectivity and can be realized using the logical gates contained in $\mathcal{M}$. Since our platform can store a maximum of two ions at a trap segment \cite{FaultTolerantReadoutMainz}, we remove the \gate{SWAP} gates from the circuits at compile time and reintroduce them at a later compilation stage, which we discuss in \autoref{subsec-eliminiation-swap-gates}. This is advantageous because instead of laser-driven \gate{SWAP} gates it allows us to use physical ion swapping  to reconfigure the qubit registers, which does not require the manipulation of the internal qubit states and can therefore be executed at unit fidelity \cite{KaufmannSWAP,KaufmannSWAPtheory,vanMourikSWAP}.

On our platform, laser pulses realize all gate operations in \eqref{eq-native-gates}, and the rotation angle parameters $\theta$ correspond to pulse areas, \ie integrals of intensity over time. To perform gate operations at high fidelities, these pulse areas must be carefully calibrated. We therefore restrict the set of available gates to rotation angles equal to the precalibrated pulse areas $\theta=\pi$ and $\theta=\tfrac{\pi}{2}$ for \gate{R} gates and $\tfrac{\pi}{2}$ for \gate{ZZ} gates. Note that there is no restriction on the rotation angle $\phi$ for the \gate{Rz} gates. The concatenated use of \gate{R} gates allows for all pulse area multiples of $\tfrac{\pi}{2}$. The phases $\phi$ of the gates can be chosen arbitrarily with a resolution limited only by the hardware capabilities. These considerations lead to a restriction of the gate set to:
\begin{align}
    \mathcal{N} = \left\{ 
    \gate{R}\left(\tfrac{\pi}{2}, \phi\right),
    \gate{R}\left(\pi, \phi\right), 
    \gate{Rz}\left(\phi \right), 
    \gate{ZZ}\left(\tfrac{\pi}{2}\right)
    \right\}.
\end{align}
The following sections describe the individual transformations which lead to a circuit consisting entirely of elements from $\mathcal{N}$. 

\section{Transformations}
\label{sec-transformations}
\begin{figure*}
    \centering
	\includegraphics[keepaspectratio,scale=0.9]{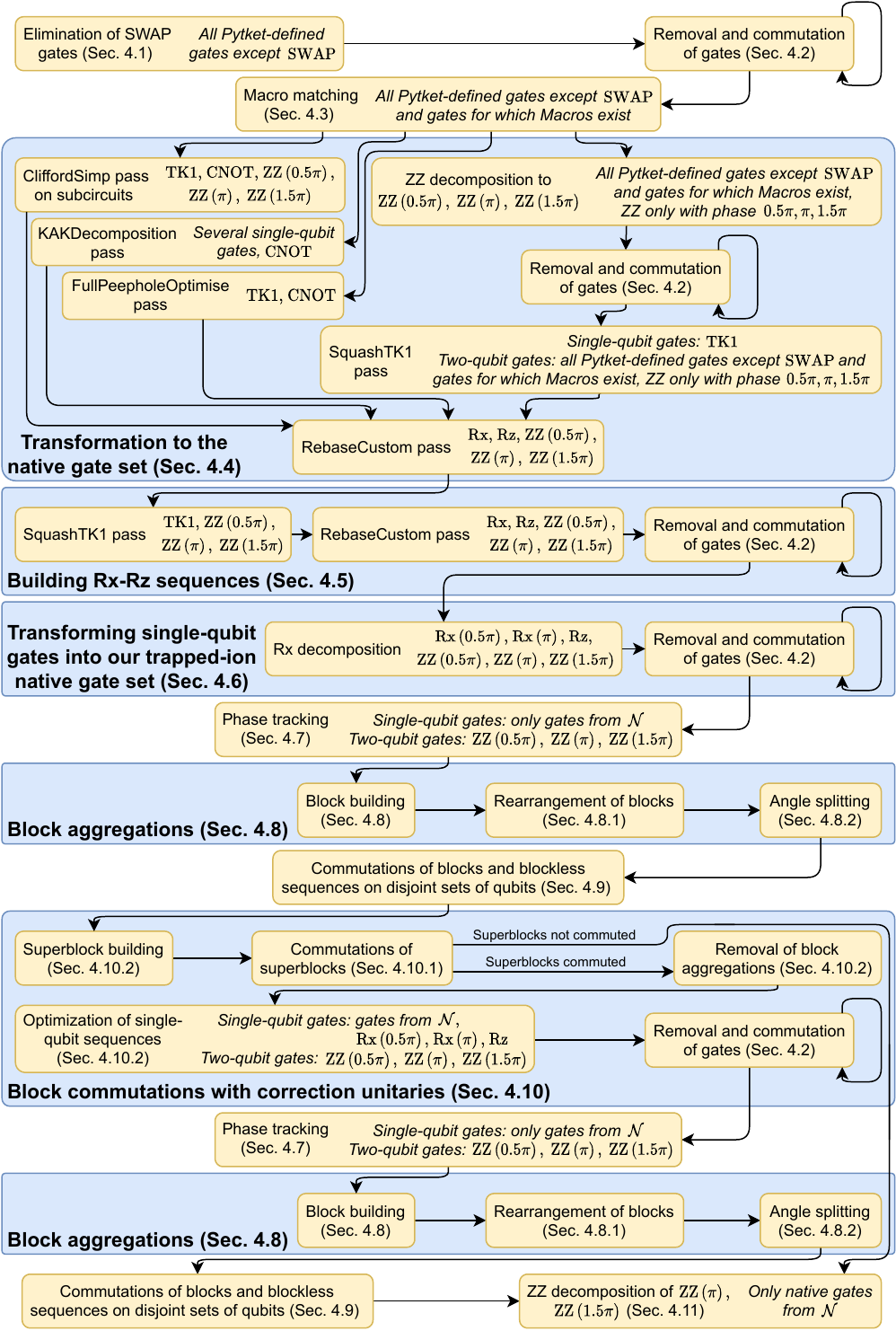}
	\caption{The compilation flow of our compiler. Each yellow box represents a transformation step of the compilation process. The boxes contain either only the name of the transformation step or the name of the transformation step on the left side and the gate set of which the circuit consists after the transformation on the right side. The blue boxes bundle several transformation steps into larger logical steps, each of which is described in a separate section.}
	\label{fig-compiler-flowchart}
\end{figure*}

\begin{figure}
    \centering
	\includegraphics[width=\columnwidth,keepaspectratio]{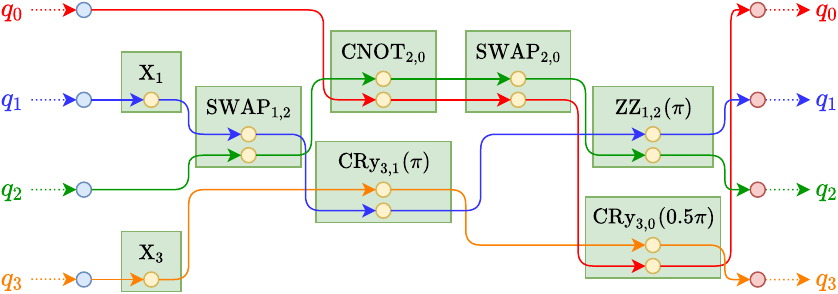}
	\caption{Example of a graph representation of a quantum circuit with four qubits. The circuit contains gates which are not part of our trapped-ion native gate set $\mathcal{N}$.}
	\label{fig-example}
\end{figure}

The overall goal of a quantum compiler is to modify and rearrange the gates in a given quantum circuit in order to obtain an equivalent circuit with a reduced total gate count after mapping to the native gate set, and more favorable operations in terms of execution resources, fidelity, and runtime. The following section describes the transformations to convert the input circuit $C$ into a circuit consisting only of gates from the allowed gate set $\mathcal{N}$, which minimizes the execution overhead on a shuttling-based platform. For some of the following transformations, we use the built-in functions of the quantum programming framework Pytket \cite{tket}. An overview of all compilation steps and the corresponding set of gates is shown in \autoref{fig-compiler-flowchart}. In the same order, each compilation step is described in a subsection. In general, transformations which affect the circuit structure on a large scale are applied earlier, while local adjustments are made later to preserve optimized structures from previous steps.
Throughout this section we assume that the input circuit $C$ consists only of quantum gates defined within Pytket \cite{pytket-gates}. This means that all high-level subcircuits, \eg a Quantum Fourier Transform \cite{nielsen_chuang_2010}, have been replaced by Pytket-defined quantum gates before the following transformations are applied. An uncompiled circuit, used as an example throughout this section, is depicted in \autoref{fig-example}. 

In addition to Pytket's built-in algorithms, we take into account the characteristics of the segmented ion trap architecture. Most importantly, gates are always executed simultaneously on all ions stored at the laser interaction zone. This allows the parallel execution of two local qubit rotations \gate{R}.

\subsection{Elimination of \gate{SWAP} gates}
\label{subsec-eliminiation-swap-gates}
Our shuttling-based ion trap quantum computer natively supports the physical swap of ions \cite{KaufmannSWAP,KaufmannSWAPtheory,vanMourikSWAP}, which is required to establish full connectivity between the qubits. Thus, in contrast to other gates, \gate{SWAP} gates are executed by physically swapping ions. In the first compilation step, our compiler replaces the function of \gate{SWAP} gates by renaming the qubits for all succeeding operations. It is the task of the Shuttling Compiler further downstream in the software stack to generate the corresponding reconfiguration operations.

\begin{figure}[t]
	\begin{subfigure}{\columnwidth}
	    \centering
    	\includegraphics[width=\columnwidth,keepaspectratio]{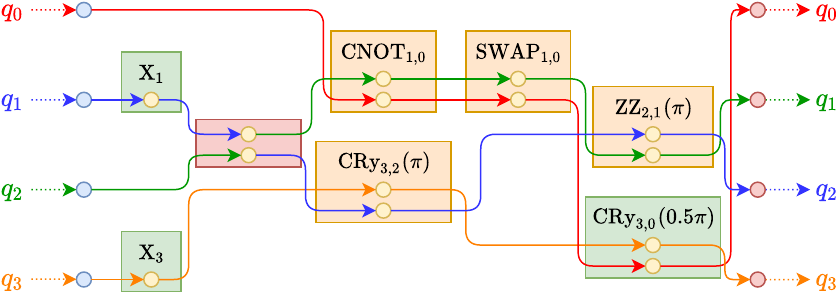}
    	\caption{The \gate{SWAP} gate elimination is applied to the first \gate{SWAP} gate (the red box). Compared to \autoref{fig-example}, the outgoing edges are exchanged. This means that after the \gate{SWAP} gate, $q_1$ follows the green path and $q_2$ the blue one. Consequently, the qubits on which the $\gate{CNOT}_{2,0}$, $\gate{CRy}_{3,1}\left(\pi\right)$, $\gate{SWAP}_{2,0}$, and $\gate{ZZ}_{1,2}\left(\pi\right)$ gates from \autoref{fig-example} act must be adapted to the new qubit. Since all gates on the path of $q_1$ and $q_2$ after the \gate{SWAP} gate are adapted to the new qubit following the path, the algorithm can eliminate the \gate{SWAP} gate from the circuit.}
    	\vspace{2mm}
    	\label{fig-example-swap-subvertices-1}
	\end{subfigure}
	\begin{subfigure}{\columnwidth}
	    \centering
    	\includegraphics[width=\columnwidth,keepaspectratio]{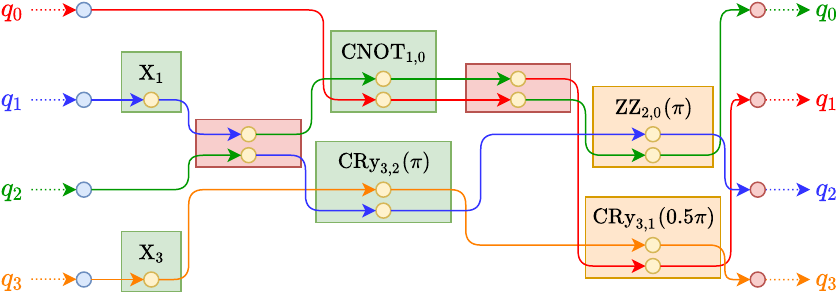}
    	\caption{The \gate{SWAP} gate elimination is also applied to the second \gate{SWAP} gate (the right red box). Consequently, the outgoing edges are exchanged compared to (a). This means that after the \gate{SWAP} gate, $q_0$ follows the green path and $q_1$ the red one.}
    	\vspace{2mm}
    	\label{fig-example-swap-subvertices-2}
	\end{subfigure}
	\begin{subfigure}{\columnwidth}
	    \centering
    	\includegraphics[height=2.9cm,keepaspectratio]{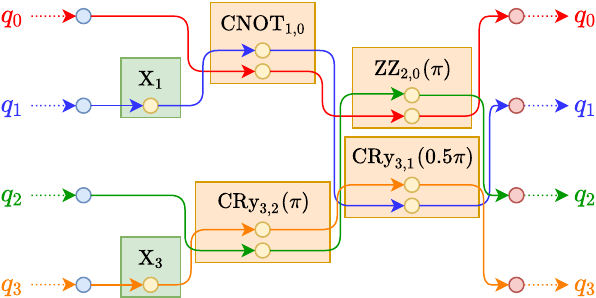}
    	\caption{The circuit after applying the \gate{SWAP} gate elimination. The orange gates now operate on different qubits than in \autoref{fig-example}.}
    	\label{fig-example-0}
    \end{subfigure}
    \caption{Applying the \gate{SWAP} gate elimination to the circuit from \autoref{fig-example}.}
    \label{fig-example-swap-subvertices}
\end{figure}

The process of \gate{SWAP} gate elimination on the circuit from \autoref{fig-example} is shown in \autoref{fig-example-swap-subvertices}. To eliminate the \gate{SWAP} gates, the elimination algorithm iterates over the gates $G \in C$ in their execution order. If $G$ is a $\gate{SWAP}_{i,j}^k$ gate acting on the qubits $q_i$ and $q_j$, $G$ has two subvertices $s_i^{k,0}$ and $s_j^{k,1}$, respectively. The algorithm exchanges the outgoing edges of these two subvertices. This means that all gates on the path from the input vertex $v_i$ to $\gate{SWAP}_{i,j}^k$ are applied to $q_i$. Since the outgoing edges of $s_i^{k,0}$ and $s_j^{k,1}$ have been exchanged, after $\gate{SWAP}_{i,j}^k$ $q_i$ follows the path originally taken by $q_j$. The same holds for $q_j$, which after $\gate{SWAP}_{i,j}^k$ follows the path originally taken by $q_i$. Consequently, all gates after $\gate{SWAP}_{i,j}^k$ which were originally executed on $q_j$ are now executed on $q_i$ and vice versa. In this way, the swap is passed through all gates succeeding $\gate{SWAP}_{i,j}^k$ and the algorithm can eliminate $\gate{SWAP}_{i,j}^k$ from the circuit.

\subsection{Repeated removal and commutation of gates}
\label{subsec-repeated-removing-and-commutation-of-gates}
Our compiler uses Pytket's \texttt{RemoveRedundancies} pass to remove redundant gates or gate sequences from the circuit. Additionally, our compiler executes Pytket's \texttt{CommuteThroughMultis} pass to commute gates. When applying commutations, single-qubit gates are commuted through two-qubit gates whenever possible. This may again introduce redundant gates, which can then be removed. The process of commuting and removing gates is repeated until the overall gate count is no longer reduced.

Our compiler executes this transformation step first after the elimination of the \gate{SWAP} gates, and will also apply it after some of the later transformation stages. The transformation preserves the property that a gate sequence is in the gate set $\mathcal{M}$ or in the gate set $\mathcal{N}$ up to multiples of $\tfrac{\pi}{2}$.

\subsection{Macro Matching}
\label{subsec-macro-matching}
Although Pytket provides a transformation for arbitrary gates into the native gate set $\mathcal{M}$, which is used in \autoref{subsec-transformation-to-extended-native-gate-set}, a custom decomposition gives better results. Therefore, in this step, our compiler applies a predefined decomposition into the native gate set -- called \emph{macro matching} -- to the quantum circuit for several gates or gate sequences. This is especially useful for large structured circuits, where known structures can be replaced by beneficial alternatives.
Our compiler performs the macro matching transformation only for gates acting on $m \geq 2$ qubits, since efficient transformations exist for local rotations with $m=1$ (see \autoref{subsec-transformation-to-extended-native-gate-set}).
Let $\mathcal{L}$ be a set of gates for which an efficient decomposition is known. All gates from $\mathcal{G}$ contained in $\mathcal{L}$ are replaced by a sequence of gates from $\mathcal{M}$. The gate sequence of the macro is defined in a way that it can be described by the same unitary matrix as the gate $G$, up to a global phase. An example of such a macro is
\begin{align}
    &\hspace{1mm}\gate{C\hspace{.2mm}Ry}_{i, j}\left(\theta\right) = \exp\left(-\tfrac{i}{4}\theta(1-Z_1)Y_2\right) \nonumber \\ \
    = &\hspace{1mm}\gate{Rx}_j\left(\tfrac{\pi}{2}\right) \gate{ZZ}_{i,j}\left(\tfrac{\theta}{2}\right) \gate{Rz}_j\left(-\tfrac{\theta}{2}\right) \gate{Rx}_j\left(\tfrac{3\pi}{2}\right). \label{eq-cry-macro}
\end{align}

Due to the angle restrictions of the gates in $\mathcal{N}$, this macro is only applied if $\theta = \ell\pi$ with $\ell \in \mathbb{Z}$ holds. Otherwise, the original \gate{C\hspace{.2mm}Ry} gate remains in the circuit and will be replaced by the transformations in \autoref{subsec-transformation-to-extended-native-gate-set}. To simplify the circuit, our compiler executes the repeated removal and commutation of gates from \autoref{subsec-repeated-removing-and-commutation-of-gates} again. The application of the macro matching to the circuit from \autoref{fig-example-0} is depicted in \autoref{fig-example-1}.

\subsection{Transformation to the native gate set}
\label{subsec-transformation-to-extended-native-gate-set}
The predefined gate decomposition into the native gate set by macro matching is followed by the conversion of all remaining gates into the native gate set. Our compiler offers four different approaches for this transformation. While they all start differently, they all end with the \texttt{RebaseCustom} pass to convert the remaining non-native gates.

\begin{figure}
    \centering
	\includegraphics[width=\columnwidth,keepaspectratio]{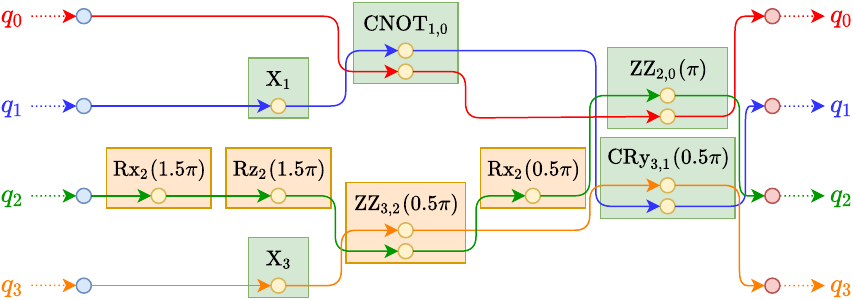}
	\caption{Example from \autoref{fig-example-0} with the macro matching applied. Since only the gate $\gate{CRy}_{3,2}\left(\pi\right)$ of the \gate{CRy} gates in \autoref{fig-example-0} satisfies the condition $\theta = \ell\pi$ with $\ell \in \mathbb{Z}$, only this gate is replaced by the orange gates.}
	\label{fig-example-1}
\end{figure}

\begin{figure*}[!t]
    \centering
    \begin{subfigure}{\textwidth}
        \centering
    	\includegraphics[height=3.7cm,keepaspectratio]{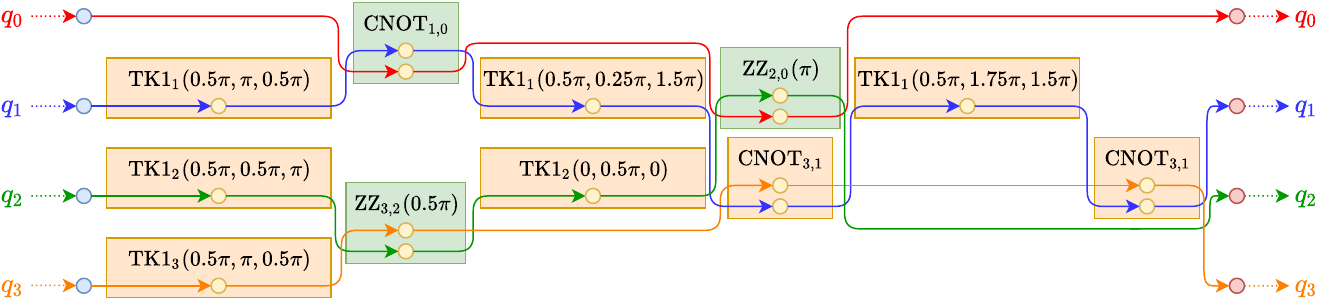}
    	\caption{Example from \autoref{fig-example-1} after applying Pytket's \texttt{CliffordSimp} pass. All sequences of single-qubit gates have been replaced by exactly one \gate{TK1} gate. Moreover, the  $\gate{CRy}_{3,1}\left(0.5\pi\right)$ gate in \autoref{fig-example-1} has been transformed into the two orange $\gate{CNOT}_{3,1}$, the $\gate{TK1}_1\left(0.5\pi,0.25\pi,1.5\pi\right)$ and the $\gate{TK1}_1\left(0.5\pi,1.75\pi,1.5\pi\right)$ gates. Since our compiler executes the \texttt{CliffordSimp} pass so that \gate{ZZ} gates with $\theta \in \{ \tfrac{\pi}{2}, \pi, \tfrac{3\pi}{2} \}$ are not converted, $\gate{ZZ}_{3,2}\left(0.5\pi\right)$ and $\gate{ZZ}_{2,0}\left(\pi\right)$ remain in the circuit.}
    	\label{fig-example-2-clifford-simp}
    	\vspace{2mm}
	\end{subfigure}
	\begin{subfigure}{\textwidth}
	    \centering
    	\includegraphics[height=3.7cm,keepaspectratio]{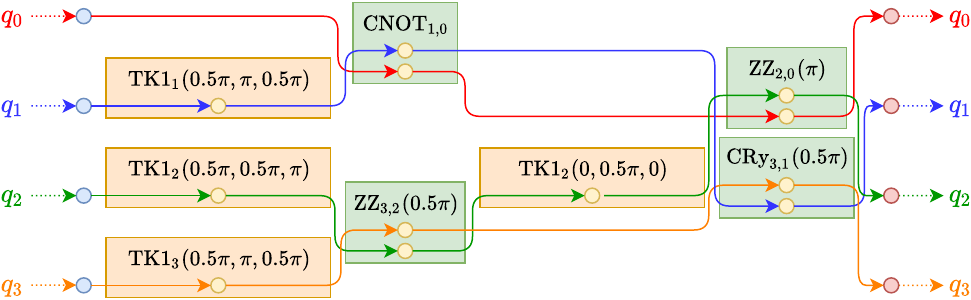}
    	\caption{Example from \autoref{fig-example-1} after applying Pytket's \texttt{SquashTK1} pass as described in \autoref{subsec-transformation-to-extended-native-gate-set}. Again, all sequences of single-qubit gates have been replaced by exactly one orange \gate{TK1} gate. In contrast to the \texttt{CliffordSimp} pass, the green two-qubit gates remain unchanged.}
    	\label{fig-example-2-squash-tk1}
    	\vspace{2mm}
	\end{subfigure}
    \begin{subfigure}{\textwidth}
        \centering
    	\includegraphics[width=\textwidth,keepaspectratio]{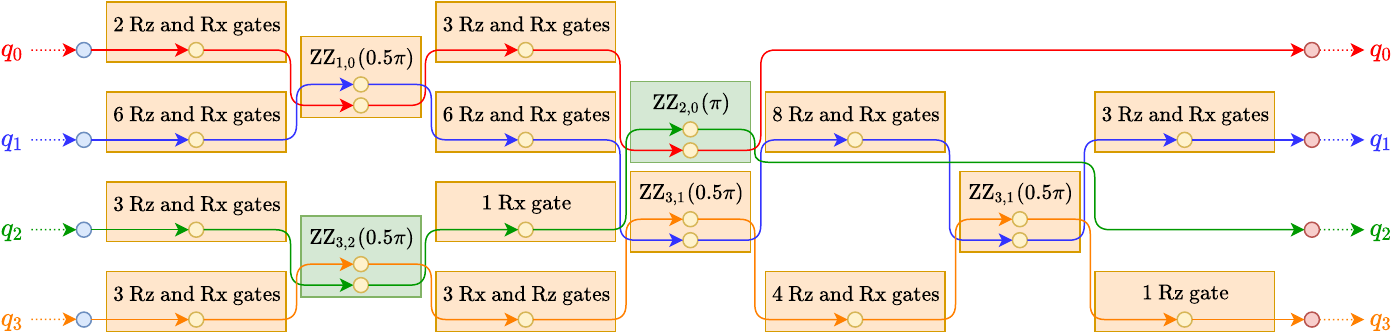}
    	\caption{Circuit after applying Pytket's \texttt{RebaseCustom} pass as described in \autoref{subsec-transformation-to-extended-native-gate-set} to the circuits from (a) and (b). In this case, both circuits result in the same transformed circuit containing only gates from the gate set $\mathcal{M}$. Each \gate{CNOT} gate has been transformed into exactly one orange \gate{ZZ} gate and eleven \gate{Rz} and \gate{Rx} gates. Additionally, each \gate{TK1} gate has been replaced by exactly three \gate{Rz} and \gate{Rx} gates. Since the green \gate{ZZ} gates are still an element of the gate set $\mathcal{M}$ and have an angle $\theta \in \{ \tfrac{\pi}{2}, \pi, \tfrac{3\pi}{2} \}$, they are the only gates which remain unchanged.}
    	\label{fig-example-3}
	\end{subfigure}
    \label{fig-example-2-3}
    \vspace{-0.36em}
    \caption{The process of transforming a circuit into the gate set $\mathcal{M}$.}
\end{figure*}

The \textbf{first approach} applies Pytket's \texttt{CliffordSimp} pass to the quantum circuit $C$. This pass contains simplifications similar to those of Duncan and Fagan \cite{Fagan_Duncan_2019}. After applying the \texttt{CliffordSimp} pass to a circuit, the resulting circuit consists only of Pytket's universal single-qubit \gate{TK1} gates and two-qubit \gate{CNOT} gates, defined as follows:
\begin{subequations}
    \begin{align}
        \gate{TK1}_i\left(\alpha, \beta, \gamma\right) &= \gate{Rz}_i\left(\alpha\right)  \gate{Rx}_i\left(\beta\right)
        \gate{Rz}_i\left(\gamma\right),
        \label{eq-tk1-decomp}
        \\
        \gate{CNOT}_{ij} &= \exp\left(i\tfrac{\pi}{4}(I-Z_i)(I-X_j)\right).
    \end{align}
\end{subequations}
Since the \texttt{CliffordSimp} pass also converts \gate{ZZ} gates with $\theta \in \{ \tfrac{\pi}{2}, \pi, \tfrac{3\pi}{2} \}$ already contained in $\mathcal{G}$ to a sequence of several \gate{TK1} and \gate{CNOT} gates, our compiler does not apply it directly to the entire circuit $C$, but executes it on subcircuits of $C$ in such a way that \gate{ZZ} gates with $\theta \in \{ \tfrac{\pi}{2}, \pi, \tfrac{3\pi}{2} \}$ are preserved. This is advantageous because the $\gate{ZZ}\left(\tfrac{\pi}{2}\right)$ gates are already part of our trapped-ion native gate set $\mathcal{N}$ and thus would not become shorter. Similarly, $\gate{ZZ}\left(\pi\right)$ as well as $\gate{ZZ}\left(\tfrac{3\pi}{2}\right)$ gates can be transformed into smaller gate sequences, see \autoref{subsec-transformation-to-native-gate-set-double}. The result of applying the \texttt{CliffordSimp} pass to the example from \autoref{fig-example-1} can be seen in \autoref{fig-example-2-clifford-simp}.

\autoref{sec-evaluation} shows that the \texttt{CliffordSimp} pass has a nonlinear runtime with the number of gates. To make even large circuits transformable in a reasonable time, a \textbf{second approach} to convert an arbitrary gate sequence to the given native gate set is to execute Pytket's \texttt{SquashTK1} pass, which simplifies single-qubit gate sequences. Since this pass does not modify the two-qubit gates, and the built-in rebasing routine used later does not allow to restrict the rotation angles to those contained in our trapped-ion native gate set $\mathcal{N}$, all \gate{ZZ} gates in $\mathcal{G}$ with $\theta \notin \{ \tfrac{\pi}{2}, \pi, \tfrac{3\pi}{2} \}$ must be replaced manually. To do this, we use the following decomposition:
\begin{align}
    \gate{ZZ}_{i,j}\left(\theta\right) = 
    &\hspace{1mm}\gate{Rz}_j\left(\tfrac{\pi}{2}\right) 
    \gate{Rx}_j\left(\tfrac{\pi}{2}\right) 
    \gate{ZZ}_{i,j}\left(\tfrac{\pi}{2}\right) \nonumber \\
    \cdot &\hspace{1mm}\gate{Rz}_j\left(\pi\right) 
    \gate{Rx}_j\left(-\theta\right)
    \gate{ZZ}_{i,j}\left(\tfrac{\pi}{2}\right)  \nonumber \\
    \cdot &\hspace{1mm}\gate{Rx}_j\left(\tfrac{3\pi}{2}\right) 
    \gate{Rz}_j\left(\tfrac{3\pi}{2}\right) 
    \gate{Rz}_i\left(\pi\right).
    \label{eq-ZZ-decomp}
\end{align}
After the substitution, our compiler executes the repeated removal and commutation of gates from \autoref{subsec-repeated-removing-and-commutation-of-gates} again to simplify the circuit. Then we apply Pytket's \texttt{SquashTK1} pass, which converts each sequence of single-qubit gates into exactly one \gate{TK1} gate. Applying these substitutions to the example from \autoref{fig-example-1} results in the circuit shown in \autoref{fig-example-2-squash-tk1}.

Besides these two compilation strategies, there are \textbf{two other approaches}. Both are passes which come with the Pytket package and generally perform well, as shown in \autoref{sec-evaluation}. The first approach uses Pytket's \texttt{KAKDecomposition} pass and performs the KAK decomposition \cite{kak} on $C$. The second approach uses Pytket's \texttt{FullPeepholeOptimise} pass, which executes Clifford simplifications, commutes single-qubit gates, and squashes subcircuits of up to three qubits \cite{pytket-passes}. Our compiler executes both approaches on the entire circuit $C$, so that the full potential of these optimizations can be exploited on deep circuits.

Regardless of the approach used, the gates in $\mathcal{G}$ are then transformed into the gates in the set $\left\{ \gate{Rx}, \gate{Rz}, \gate{ZZ} \right\}$. This process is called \emph{rebasing}. For this transformation we use Pytket's \texttt{RebaseCustom} pass. We found smaller gate counts when excluding the \gate{Ry} gate, so we excluded \gate{Ry} from the set. Since $\mathcal{G}$ may contain two-qubit gates which are not \gate{ZZ} gates when using the approach with Pytket's \texttt{SquashTK1} pass, the rebasing first replaces all two-qubit gates which are not \gate{ZZ} gates with sequences of \gate{TK1} and \gate{CNOT} gates. Then the rebasing replaces the \gate{TK1} gates with the definition in \eqref{eq-tk1-decomp} and the \gate{CNOT} gates with
\begin{align}
    \gate{CNOT}_{i, j} = &\hspace{1mm}\gate{Rz}_j\left(\tfrac{\pi}{2}\right)  \gate{Rx}_j\left(\tfrac{\pi}{2}\right) \gate{Rz}_j\left(\tfrac{\pi}{2}\right)  \gate{Rz}_i\left(\tfrac{\pi}{2}\right) \nonumber \\
    \cdot&\hspace{1mm}\gate{ZZ}_{i, j}\left(\tfrac{\pi}{2}\right) \gate{Rx}_j\left(\tfrac{\pi}{2}\right) \gate{Rz}_j\left(\tfrac{\pi}{2}\right) \nonumber \\ \cdot&\hspace{1mm}\gate{Rx}_i\left(\tfrac{\pi}{2}\right) \gate{Rz}_i\left(\pi\right) \gate{Rx}_i\left(\tfrac{\pi}{2}\right).
    \label{eq-CNOT-decomp}
\end{align}
The decomposition guarantees that the rebasing introduces only \gate{ZZ} gates with an angle $\theta = \tfrac{\pi}{2}$ into $\mathcal{G}$, so that after applying this transformation all $\gate{ZZ}$ gates have an angle $\theta \in \{ \tfrac{\pi}{2}, \pi, \tfrac{3\pi}{2} \}$. After applying the \texttt{RebaseCustom} pass to the circuits from \autoref{fig-example-2-clifford-simp} and \autoref{fig-example-2-squash-tk1}, the circuit in \autoref{fig-example-3} results, which is the same circuit for both approaches in this example. 

\begin{figure*}[!t]
    \centering
    \begin{subfigure}{\textwidth}
        \centering
    	\includegraphics[height=2.95cm,keepaspectratio]{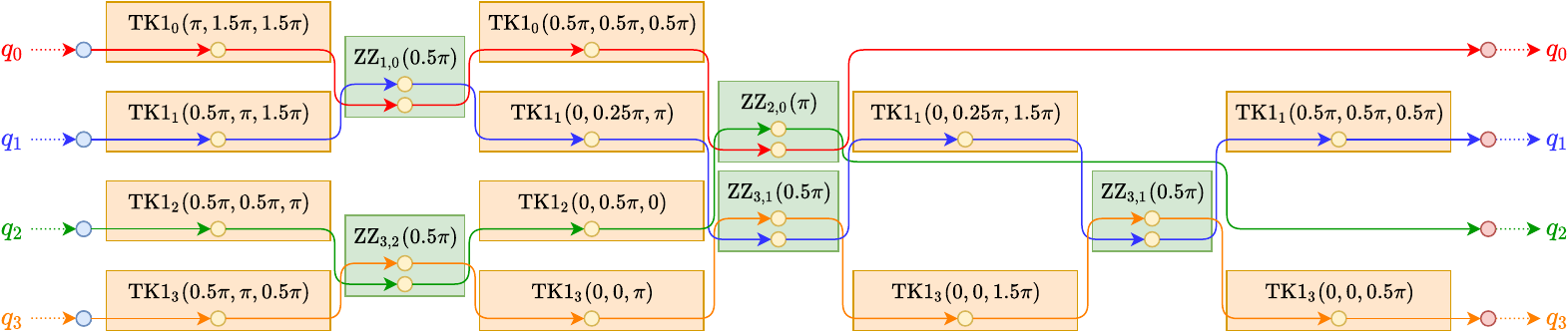}
    	\caption{Example from \autoref{fig-example-3} after applying Pytket's \texttt{SquashTK1} pass. Each single-qubit gate sequence has been transformed into exactly one orange \gate{TK1} gate.}
    	\label{fig-example-4}
    	\vspace{2mm}
	\end{subfigure}
	\begin{subfigure}{\textwidth}
	    \centering
    	\includegraphics[width=\textwidth,keepaspectratio]{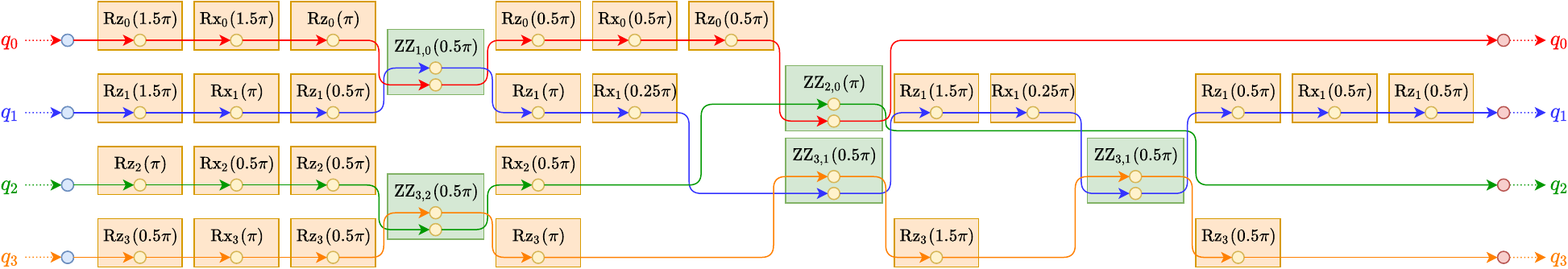}
    	\caption{Circuit after applying Pytket's \texttt{RebaseCustom} pass to the circuit from (a). Each \gate{TK1} gate has been transformed into the \gate{TK1} decomposition in \eqref{eq-tk1-decomp}, which consists of three gates. Since the angles of some gates from the decomposition are zero, and thus these gates are equal to the identity gate, some decompositions have less than three gates.}
    	\label{fig-example-5}
    	\vspace{2mm}
	\end{subfigure}
    \begin{subfigure}{\textwidth}
        \centering
    	\includegraphics[height=2.95cm,keepaspectratio]{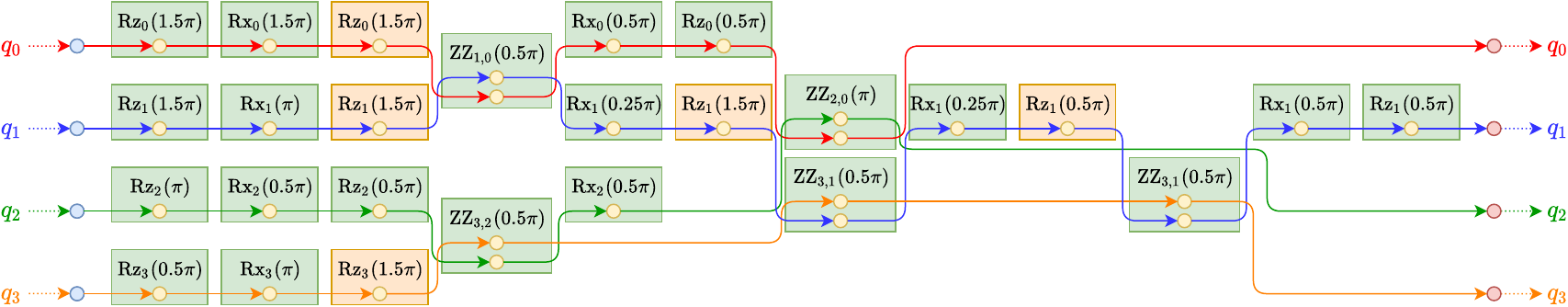}
    	\caption{The circuit from (b) after the \gate{Rz} gates have been commuted through the \gate{ZZ} gates and the resulting redundancies have been removed. Consequently, there are at most two single-qubit gates between two \gate{ZZ} gates on each qubit $q_i$ and between the last gate executed on $q_i$ and the output vertex of $q_i$. Only between the input vertex of $q_i$ and the first gate executed on $q_i$ three single-qubit gates are possible.}
    	\label{fig-example-6}
	\end{subfigure}
    \label{fig-example-4-5-6}
    \vspace{-0.29em}
    \caption{The process of building \gate{Rx}-\gate{Rz} sequences.}
\end{figure*}

\subsection{Building \gate{Rx}-\gate{Rz} sequences}
\label{subsec-building-rx-rz-sequences}
At this point, the quantum circuit contains only \gate{R} and \gate{Rz} gates with arbitrary angle parameters as single-qubit gates and \gate{ZZ} gates with an angle parameter of $\tfrac{\ell \pi}{2}$ as two-qubit gates. This makes the gate set of the quantum circuit compatible with the native gate set $\mathcal{M}$. Since exactly one \gate{TK1} gate can compactly represent any arbitrary sequence of single-qubit gates, the first step is to reduce any sequence of concatenated single-qubit gates to one \gate{TK1} gate. Such sequences start and end either at two-qubit gates or at the input or output vertices $v_i, w_i$.
The implementation of the algorithm is Pytket's \texttt{SquashTK1} pass. It is depicted in \autoref{fig-example-4} how this algorithm is applied to the circuit from \autoref{fig-example-3}. Then we use Pytket's \texttt{RebaseCustom} pass with \eqref{eq-tk1-decomp} to transform the \gate{TK1} gates into gates of set $\mathcal{M}$. The circuit unitary reduced to all gates acting along the path of the qubit $q_i$ has the following form:
\begin{align}
    U_i=\left(\prod_{\lambda=0}^{\omega_i - 1} \gate{Rz}_i  \gate{Rx}_i  \gate{Rz}_i  \gate{ZZ}_{i, j\left(\lambda\right)} \right) \gate{Rz}_i  \gate{Rx}_i  \gate{Rz}_i.
\end{align}
In this expression, $\omega_i$ is the number of \gate{ZZ} gates executed on $q_i$ and $j\left(\lambda\right)$ stands for the qubit $q_{j\left(\lambda\right)}$ on which the \gate{ZZ} gate $\lambda$ also acts. The circuit from \autoref{fig-example-4} after applying the transformation is shown in \autoref{fig-example-5}. Each path in this circuit has the above structure.

Each \gate{ZZ} gate $\gate{ZZ}_{i,j}^k \in \mathcal{G}$ is now sandwiched by two \gate{Rz} gates, which commute with the $\gate{ZZ}_{i,j}^k$ gate on both qubits $q_i$ and $q_j$. So we can commute the succeeding \gate{Rz} gates through the \gate{ZZ} gates and merge with the preceding \gate{Rz} gates:
\begin{align}
    \gate{Rz}_i(\phi_s) \gate{ZZ}_{i, j\left(\lambda\right)} \gate{Rz}_i(\phi_p) \rightarrow \gate{ZZ}_{i, j\left(\lambda\right)} \gate{Rz}_i(\phi_s+\phi_p).
\label{eq-Rz-commutation}
\end{align}
Repeating this procedure, combined with the removal of redundant gates until no further reduction in the number of gates is possible, as described in \autoref{subsec-repeated-removing-and-commutation-of-gates}, results in a circuit $C$ in which the disjoint paths of each qubit $q_i \in \mathcal{Q}$ have the following form:
\begin{figure*}[!t]
    \centering
    \begin{subfigure}{\textwidth}
        \centering
    	\includegraphics[width=\textwidth,keepaspectratio]{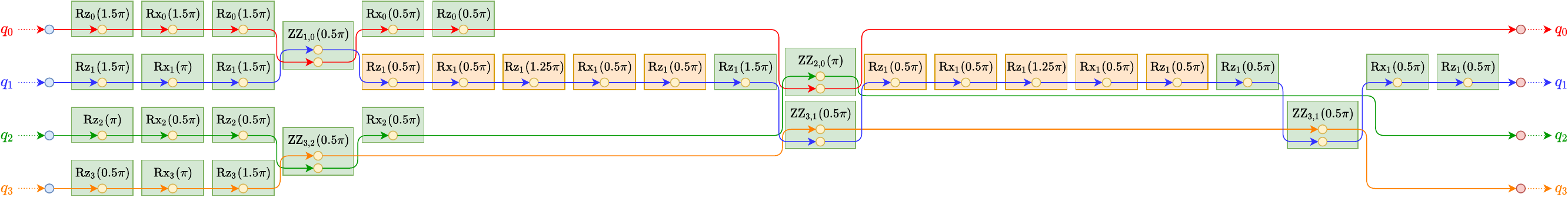}
    	\caption{Example from \autoref{fig-example-6} after applying the transformation of single-qubit gates into our trapped-ion native gate set. The two $\gate{Rx}_1\left(0.25\pi\right)$ gates in \autoref{fig-example-6} are transformed into the two orange gate sequences, and their angles are included in the phase of the $\gate{Rz}_1\left(1.25\pi\right)$ gates. All single-qubit gates of the circuit are in our trapped-ion native gate set $\mathcal{N}$.}
    	\label{fig-example-7}
    	\vspace{2mm}
	\end{subfigure}
	\begin{subfigure}{\textwidth}
	    \centering
    	\includegraphics[height=2.12cm,keepaspectratio]{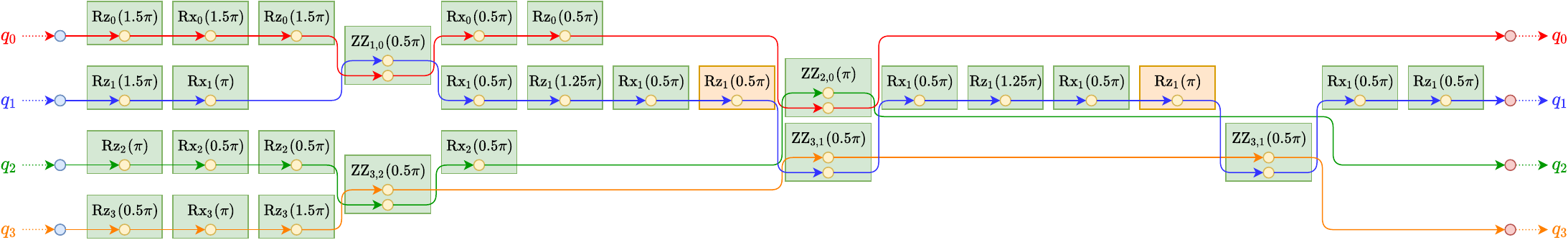}
    	\caption{Circuit from (a) after the \gate{Rz} gates have been commuted through the \gate{ZZ} gates and the resulting redundancies have been removed. Since the transformation is applied directly after the \gate{Rx}-\gate{Rz} sequences have been built, there are at most four single-qubit gates between two \gate{ZZ} gates on each qubit $q_i$ and between the last gate executed on $q_i$ and the output vertex of $q_i$. Only between the input vertex of $q_i$ and the first gate executed on $q_i$ five single-qubit gates are possible.}
    	\label{fig-example-8}
	\end{subfigure}
    \label{fig-example-7-8}
    \vspace{-0.11em}
    \caption{The process of transforming the single-qubit gates into our trapped-ion native gate set $\mathcal{N}$.}
\end{figure*}
\begin{align}
    U_i=\left(\prod_{\lambda=0}^{\omega_i - 1} \gate{Rz}_i  \gate{Rx}_i  \gate{ZZ}_{i, j\left(\lambda\right)} \right) \gate{Rz}_i \gate{Rx}_i \gate{Rz}_i.
\end{align}
Consequently, on each qubit $q_i$ at most one \gate{Rx} and one \gate{Rz} gate are executed between two \gate{ZZ} gates. Let 
\begin{align}
    \label{eq-def-omega}
    \omega = \frac{1}{2} \sum_{i=0}^{n-1} \omega_i
\end{align}
be the number of \gate{ZZ} gates in $\mathcal{G}$. Thus $\mathcal{G}$ consists of exactly $\omega$ two-qubit gates and at most $4\omega + 3n$ single-qubit gates. 
In \autoref{fig-example-6} it is shown how this transformation is applied to the circuit from \autoref{fig-example-5}.

\gate{Rx} and \gate{Rz} with arbitrary angles are the only remaining single-qubit gates in the circuit, and both are also part of $\mathcal{M}$. Since all remaining two-qubit gates are also part of this set, all gates $G_i$ of the circuit now satisfy $G_i \in \mathcal{M}$.

\subsection{Transforming single-qubit gates into our trapped-ion native gate set}
\label{subsec-transformation-to-native-gate-set}
Since all gates in the circuit are now part of $\mathcal{M}$, the next step is to make the angles conform to our trapped-ion native gate set $\mathcal{N}$, which has restrictions on the allowed rotation angles. Therefore, all \gate{Rx} gates must have $\theta \in \left\{ \tfrac{\pi}{2}, \pi \right\}$. We can use two trivial conversions for $\theta = 0$ and $\theta = \tfrac{3\pi}{2}$. In the first case, the gate is equal to the \gate{I} gate and can be eliminated. In the second case, we can replace the gate by $\gate{Rx}\left(\tfrac{\pi}{2}\right) \;\gate{Rx}\left(\pi\right)$.

Rotations \gate{Rx} with any other values of $\theta$ must be converted into a sequence of \gate{Rx} gates with allowed rotation angles and \gate{Rz} gates with freely variable rotation angles. We use the decomposition
\begin{align}
    &\hspace{1mm}\gate{Rx}\left(\theta\right) \nonumber \\
    = &\hspace{1mm}\gate{Rz}\left(\tfrac{\pi}{2}\right) \gate{Rx}\left(\tfrac{\pi}{2}\right) \gate{Rz}\left(\theta + \pi\right)  \gate{Rx}\left(\tfrac{\pi}{2}\right) \gate{Rz}\left(\tfrac{\pi}{2}\right).
\end{align}
After this substitution, the restrictions of our trapped-ion native gate set $\mathcal{N}$ are satisfied for the single-qubit gates. In \autoref{fig-example-7} it is shown how the circuit from \autoref{fig-example-6} is transformed.

Since our compiler applies this transformation directly after building the \gate{Rx}-\gate{Rz} sequences in \autoref{subsec-building-rx-rz-sequences}, the unitaries corresponding to the disjoint paths of each qubit $q_i$ have the following form:
\begin{align}
    U_i=&\left(\prod_{\lambda=0}^{\omega_i - 1} \gate{Rz}_i \gate{Rz}_i \gate{Rx}_i \gate{Rz}_i  \gate{Rx}_i  \gate{Rz}_i \gate{ZZ}_{i, j\left(\lambda\right)} \right)  \nonumber \\
    \cdot&\hspace{1mm}\gate{Rz}_i \gate{Rz}_i \gate{Rx}_i \gate{Rz}_i \gate{Rx}_i \gate{Rz}_i \gate{Rz}_i.
\end{align}
By again commuting \gate{Rz} gates through \gate{ZZ} and combining successive \gate{Rz} gates according to \eqref{eq-Rz-commutation} using the procedures from \autoref{subsec-repeated-removing-and-commutation-of-gates}, we can simplify the expression to
\begin{align}
    U_i=&\left(\prod_{\lambda=0}^{\omega_i - 1} \gate{Rz}_i  \gate{Rx}_i \gate{Rz}_i \gate{Rx}_i \gate{ZZ}_{i, j\left(\lambda\right)} \right) \nonumber \\ \cdot&\hspace{1mm}\gate{Rz}_i \gate{Rx}_i \gate{Rz}_i \gate{Rx}_i \gate{Rz}_i.
\end{align}
Consequently, on each qubit $q_i$ there are at most two \gate{Rx} gates and two \gate{Rz} gates between two \gate{ZZ} gates. Let $\omega$ be the number of \gate{ZZ} gates in $\mathcal{G}$ as defined in \eqref{eq-def-omega}. Thus $\mathcal{G}$ consists of exactly $\omega$ two-qubit gates and at most $8\omega + 5n$ single-qubit gates. It is depicted in \autoref{fig-example-8} how this transformation is applied to the circuit from \autoref{fig-example-7}.

Since all \gate{R}(\gate{x}) gates now satisfy $\theta \in \left\{ \tfrac{\pi}{2}, \pi \right\}$, all single-qubit gates are now part of our trapped-ion native gate set $\mathcal{N}$.

\begin{figure*}
    \centering
	\includegraphics[width=\textwidth,keepaspectratio]{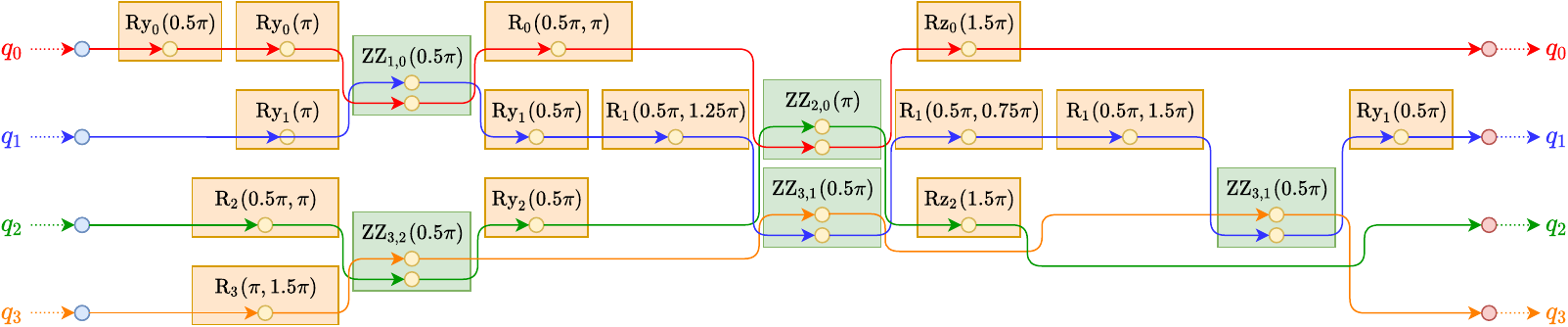}
	\caption{Example from \autoref{fig-example-8} after applying the phase tracking algorithm to the circuit. For each \gate{Rx} gate, the graph contains an \gate{R} gate depending on the angle of the original gate and the value of the tracking phase $b_i$ when the algorithm is applied to the qubit $q_i$. The \gate{R} gates can be simplified to \gate{Rx} gates if $b_i=0$ or \gate{Ry} gates if $b_i=\tfrac{\pi}{2}$. The graph contains no more \gate{Rz} gates except between the last \gate{R} gate executed on each qubit $q_i$ and the output vertex of $q_i$. Consequently, there are at most two \gate{R} gates between two \gate{ZZ} gates on each qubit $q_i$ and between the input vertex of $q_i$ and the first \gate{ZZ} gate executed on $q_i$. Only between the last gate executed on $q_i$ and the output vertex of $q_i$ two \gate{R} gates and one \gate{Rz} gate are possible. Since the values of the tracking phases $b_1$ and $b_3$ are zero after applying the algorithm to $q_1$ and $q_3$, no \gate{Rz} gate is placed before the output vertices of $q_1$ and $q_3$.}
	\label{fig-example-9}
\end{figure*}

\subsection{Phase tracking}
\label{subsec-rz-phase-tracking}
We now use the fact that an \gate{Rz} gate followed by an \gate{R} gate is equivalent to a single \gate{R} gate with a phase shifted by the rotation angle of the \gate{Rz} gate:
\begin{align}
    \gate{R}\left(\theta,\phi\right)
    \gate{Rz}\left(\theta_z\right) = \gate{Rz}\left(\theta_z\right)
    \gate{R}\left(\theta,\phi-\theta_z\right).
\end{align}
Combined with the fact that \gate{Rz} gates commute through \gate{ZZ} gates, this allows a \emph{virtual} execution of all \gate{Rz} gates using \emph{phase tracking} \cite{VirtualZGates, Pino2021}. This technique avoids any physical execution of \gate{Rz} gates and is therefore beneficial in terms of overall runtime and fidelity. All \gate{R} gates contained in $\mathcal{G}$ must be modified with respect to their phase arguments according to the procedure described in the following.

The algorithm initializes a tracking phase $b_i = 0$ for each qubit $q_i \in \mathcal{Q}$ and follows the path from the input vertex $v_i \in \mathcal{V}$ to the output vertex $w_i \in \mathcal{W}$. It applies the following rules to each gate $G_i \in \mathcal{G}$ it encounters:
\begin{itemize}
    \item If $G_i = \gate{R}_i\left(\theta,\phi\right)$, $G_i$ is replaced by $\gate{R}_i\left(\theta, \phi-b_i\right)$.
    \item If $G_i = \gate{Rz}_i\left(\phi\right)$, $G_i$ is removed from $\mathcal{G}$ and the value of $b_i$ is changed to $b_i + \phi$.
\end{itemize}
To correct the final accumulated phase, the algorithm inserts an additional gate $\gate{Rz}_i\left(b_i\right)$ into $C$ directly before $w_i$. Since the \gate{Rz} gate only changes the phase of the qubit, the measurement result in the computational basis as performed on the hardware would not be affected. However, adding the \gate{Rz} gate still has advantages when working with the unitary matrix and when using the circuit as a building block for even larger circuits. An example can be seen in \autoref{fig-example-9}.

Since our compiler executes the phase tracking algorithm immediately after the transformation in \autoref{subsec-transformation-to-native-gate-set}, the disjoint paths of each qubit $q_i \in \mathcal{Q}$ have the following form:
\begin{align}
    U_i=\gate{Rz}_i \left( \prod_{\lambda=0}^{\omega_i - 1} \gate{R}_i \gate{R}_i \gate{ZZ}_{i, j\left(\lambda\right)} \right) \gate{R}_i \gate{R}_i.
\end{align}
Note that the subsequent \gate{R} gates generally cannot be merged because they have different phase parameters. Thus, on each qubit, at most two \gate{R} gates are executed between \gate{ZZ} gates. Hence, $\mathcal{G}$ consists of exactly $\omega$ two-qubit gates and at most $4\omega + 3n$ single-qubit gates. Consequently, phase tracking can almost halve the number of single-qubit gates.

Since phase tracking only changes the phases and not pulse areas, all single-qubit gates are still in $\mathcal{N}$.

\subsection{Block aggregations}
\label{subsec-block-building}
In this section, we describe further optimization steps which are specific to our shuttling-based architecture. In our architecture, the ions are stored in a linear segmented trap, with qubit subsets stored at different trap segments. Shuttling operations can rearrange the qubit sets between gate operations \cite{FaultTolerantReadoutMainz, ShuttlingArchitecture}.
Laser beams directed to a fixed location, the laser interaction zone, perform all gate operations. There, each gate operation is executed simultaneously on all stored ions. Hence, identical single-qubit gates can be executed on multiple qubits in parallel. We use this property to reduce the number of gates. As an additional constraint, we keep shuttling operations to a minimum. Thus, we parallelize single-qubit operations only on qubits which are already in the same segment, \ie before or after a two-qubit gate is executed on them.

\begin{figure*}
    \centering
	\includegraphics[width=\textwidth,keepaspectratio]{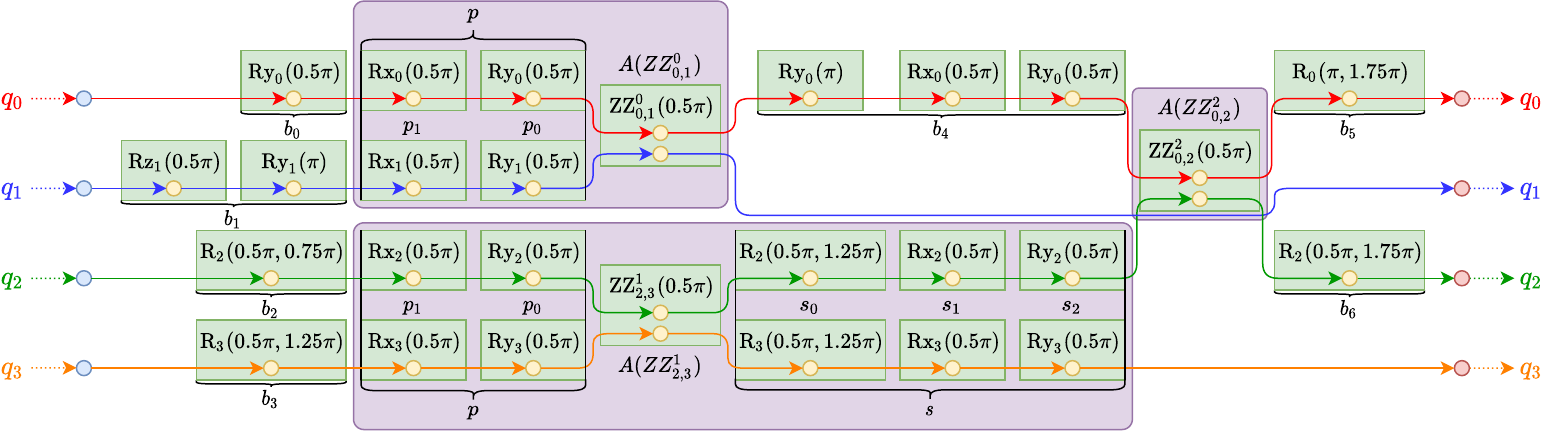}
	\caption{An example circuit with the blocks built by the algorithm from \autoref{subsec-block-building}. Since the graph contains three \gate{ZZ} gates, it also contains three blocks, represented by the purple boxes around the \gate{ZZ} gates. Each block $A\left(\gate{ZZ}_{i, j}^k\right)$ can have a predecessor sequence $p$ and/or a successor sequence $s$. The block $A\left(\gate{ZZ}_{0,2}^2\right)$ contains neither a sequence $p$ nor a sequence $s$ and consists only of the $\gate{ZZ}$ gate. Besides the three blocks, the graph consists of the seven blockless sequences $b_0$ to $b_6$.}
	\label{fig-blocks}
\end{figure*}

The optimization described in this section deals with the aggregation of single-qubit gates. It uses the property that there is a specific structure around certain \gate{ZZ} gates. Directly before and/or directly after a \gate{ZZ} gate $\gate{ZZ}_{i, j}^k$ there can be sequences of preceding single-qubit gates $p=\left(p_0, \dots, p_{\alpha-1}\right), p_0, \dots, p_{\alpha-1} \in C$ and/or succeeding single-qubit gates $s=\left(s_0,\dots, s_{\beta-1}\right), s_0, \dots, s_{\beta-1} \in C$ which are similar on both qubits $q_i, q_j \in \mathcal{Q}$ on which the gate $\gate{ZZ}_{i, j}^k$ acts. This means that in these sequences the gates as well as the angle parameters match exactly on $q_i$ and $q_j$. The advantage of these sequences is that the gates $p_0, \dots, p_{\alpha-1}, s_0, \dots, s_{\beta-1}$ can be executed simultaneously for both qubits $q_i$ and $q_j$ instead of sequentially for each qubit, which reduces the execution time and the required shuttling overhead. Each \gate{ZZ} gate, including the single-qubit gates contained in the sequences $p$ and $s$, build a block $A$. We denote the block belonging to $\gate{ZZ}_{i, j}^k$ as $A\left(\gate{ZZ}_{i, j}^k\right)$ in the following.

Iterating over all \gate{ZZ} gates in their execution order, the algorithm checks for each gate $\gate{ZZ}_{i, j}^k$ whether $q_i$ and $q_j$ undergo an identical single-qubit gate $G_p$ directly before $\gate{ZZ}_{i, j}^k$. If this is the case and $G_p$ does not already belong to another block on $q_i$ or $q_j$, the algorithm adds $G_p$ to the predecessor sequence $p$ of the block $A\left(\gate{ZZ}_{i, j}^k\right)$ and repeats the procedure with the gate directly before $G_p$. Otherwise $p$ cannot be increased. The same procedure is repeated with the gates directly after $\gate{ZZ}_{i, j}^k$ to build the sequence $s$ of $A\left(\gate{ZZ}_{i, j}^k\right)$. Since the algorithm traverses the \gate{ZZ} gates in their execution order, it is guaranteed that candidate gates for $s$ do not already belong to another block. If it is not possible to build $p$ and $s$, the block consists only of $\gate{ZZ}_{i, j}^k$.

An example of the block building process is depicted in \autoref{fig-blocks}. After building such blocks around each \gate{ZZ} gate $\gate{ZZ}^k \in \mathcal{G}$, there may exist gate sequences $b_0, \dots, b_{\ell - 1}$ which do not belong to a block. Each of these blockless sequences consists only of single-qubit gates and belongs to exactly one qubit. The goal is to minimize the number $\ell$ of blockless sequences.

To minimize the amount and length of blockless single-qubit sequences, the blocks can be rearranged and single-qubit gates can be split. These two approaches are discussed in the following two subsections.

\subsubsection{Rearrangement of blocks}
\label{subsubsec-rearranging-blocks}
\begin{figure*}[!t]
    \centering
    \begin{subfigure}{\textwidth}
        \centering
    	\includegraphics[height=3.2cm,keepaspectratio]{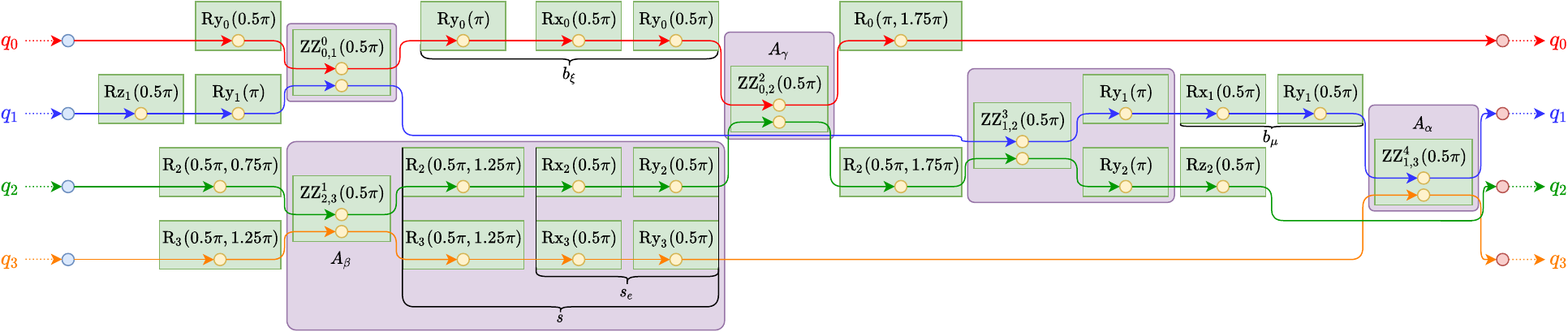}
    	\caption{The same gate sequences $b_\mu$ and $s_e$ are executed on $q_1$ and $q_3$ before the block $A_\alpha$. On $q_1$ the gate sequence $b_\mu$ is a blockless sequence, while on $q_3$ the gate sequence $s_e$ belongs to a different block $A_\beta$. To eliminate the blockless sequence $b_\mu$, $b_\mu$ and the gates on $q_3$ from $s_e$ can be added as a predecessor sequence to $A_\alpha$.}
    	\label{fig-rearranging-1}
    	\vspace{2mm}
	\end{subfigure}
	\begin{subfigure}{\textwidth}
	    \centering
    	\includegraphics[height=3.05cm,keepaspectratio]{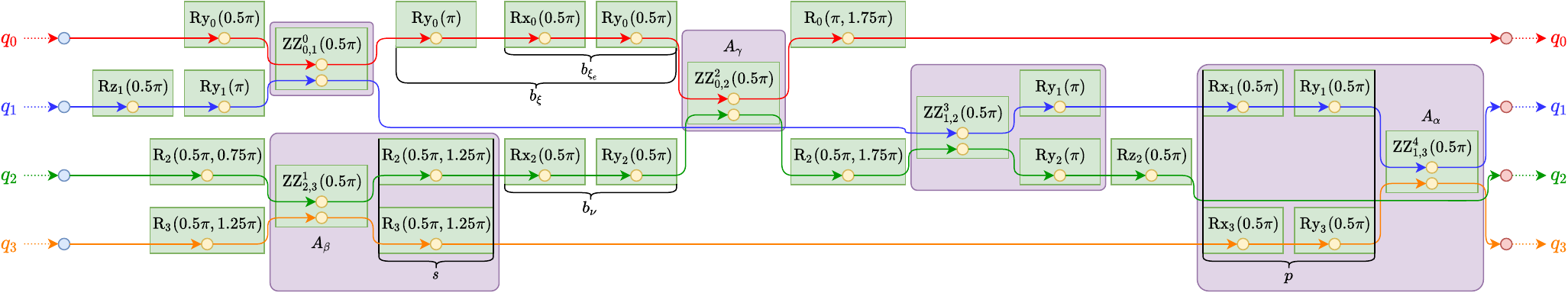}
    	\caption{$b_\mu$ and the gates on $q_3$ from $s_e$ have been added as predecessor sequence $p$ to $A_\alpha$. Consequently, the gates on $q_2$ from $s_e$ now build a new blockless sequence $b_\nu$. This blockless sequence is identical to the blockless sequence $b_{\xi_e}$ on $q_0$. Since $b_\nu$ and $b_{\xi_e}$ are both directly in front of a third block $A_\gamma$, they can be added as predecessor sequence to $A_\gamma$.}
    	\label{fig-rearranging-2}
    	\vspace{2mm}
    \end{subfigure}
	\begin{subfigure}{\textwidth}
	    \centering
    	\includegraphics[height=3.05cm,keepaspectratio]{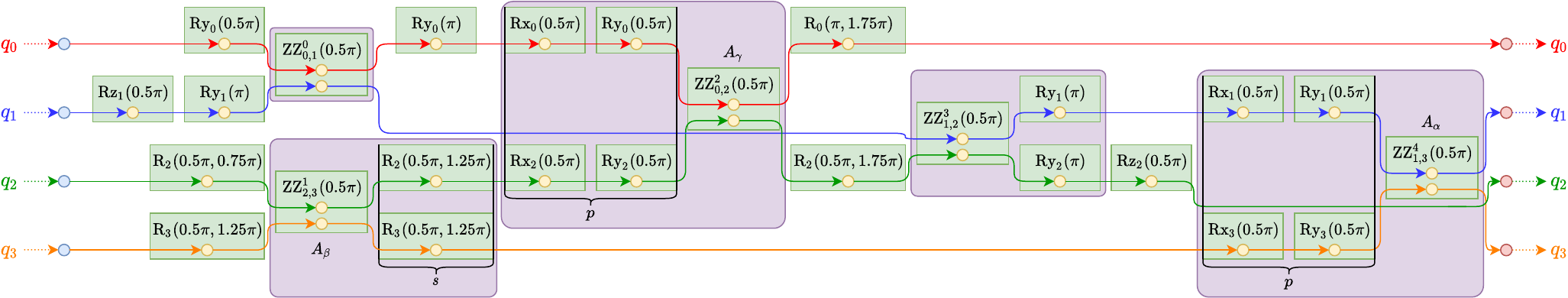}
    	\caption{$b_\mu$ and $b_{\xi_e}$ have been added as predecessor sequence $p$ to $A_\gamma$. By rearranging the blocks, the graph now has one blockless sequence less than in (a).}
    	\label{fig-rearranging-3}
    	\vspace{2mm}
	\end{subfigure}
	\begin{subfigure}{\textwidth}
	    \centering
    	\includegraphics[width=\textwidth,keepaspectratio]{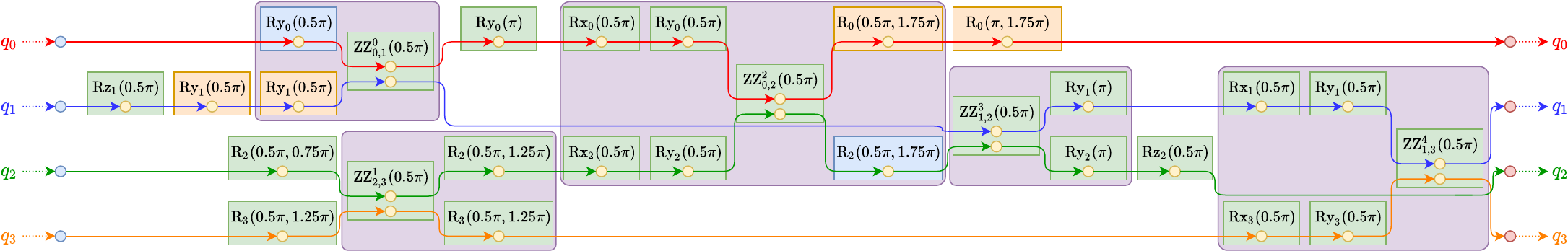}
    	\caption{In (c) the same gates are executed before the block of gate $\gate{ZZ}^0_{0,1}$ and after the block of gate $\gate{ZZ}^2_{0,2}$, but with different pulse areas $\theta$. Using the angle splitting approach, the gate with the larger angle is replaced by the two orange gates. Since the orange gate next to the \gate{ZZ} gate and the blue gate have the same angle, these two gates can be added to the block. This eliminates two blockless sequences in this example. The two \gate{R} gates before the block of gate $\gate{ZZ}^1_{2,3}$ cannot be used for angle splitting because the phase $\phi$ is different.}
    	\label{fig-rearranging-4}
	\end{subfigure}
	\caption{The process of rearranging blocks and splitting angles.}
	\label{fig-rearranging}
\end{figure*}

A block rearrangement is possible if before a sequence $p$ of a block $A_\alpha$ both qubits $q_i$ and $q_j$ undergo the same gate $G$, but for one qubit, \eg $q_j$, $G$ already belongs to a preceding block $A_\beta$ and for $q_i$ it belongs to a blockless sequence $b_\mu$. Assume $A_\beta$ operates on the qubit $q_j$ and a third qubit $q_k$. If the entire blockless sequence $b_\mu$ is equal to the end $s_e$ of the sequence $s$ of the block $A_\beta$, the gates of $b_\mu$ and $s_e$ acting on the qubit $q_j$ are appended to the front of the sequence $p$ of $A_\alpha$. Consequently, the blockless sequence $b_\mu$ is eliminated and the gates of $q_j$ in $s_e$ are removed from $s_e$, see \autoref{fig-rearranging-1} and \autoref{fig-rearranging-2}.

Since only the gates of the qubit $q_j$ in $s_e$ are removed from $s_e$, the gates of the qubit $q_k$ in $s_e$ result in a new blockless sequence $b_\nu$. To eliminate $b_\nu$, it is necessary that $b_\nu$ is adjacent to a third block $A_\gamma$ operating on the qubit $q_k$ and a qubit $q_l$. Here the case $q_i = q_l$ is possible. If before the sequence $p$ of $A_\gamma$ there is a blockless sequence $b_\xi$ on the qubit $q_l$ with the sequence $b_\nu$ at the end, the procedure appends $b_\nu$ and $b_{\xi_e}$ to the front of $p$ of $A_\gamma$ and the number of blockless sequences on the circuit is reduced by one. This can be seen in \autoref{fig-rearranging-2} and \autoref{fig-rearranging-3}. If minimizing the number of blockless sequences is not possible because a matching sequence is missing during the algorithm, the blocks are not rearranged. We repeat the rearranging procedure for all blocks on which a rearrangement is possible.

\subsubsection{Angle splitting}
\label{subsubsec-angle-splitting}
After the rearrangement, the number of blockless sequences might be further reduced by splitting the rotation angles of certain gates.
Assume that before a sequence $p$ or directly after a sequence $s$ of a block $A\left(\gate{ZZ}_{i,j}^k\right)$ both qubits $q_i$ and $q_j$ undergo rotations $\gate{R}_i(\theta_i,\phi)$ and $\gate{R}_j(\theta_j,\phi)$ with the same phase angle but different rotation angles. Then it is possible to split the gate with the larger rotation angle into two rotations, merge the resulting two identical gates into the respective set $s$ or $p$ of the block, and eventually reduce the number of blockless sequences. The phase angle is omitted from the notation in the following.

We now assume that $G_i\left(\theta_i\right)$ belongs to the blockless sequence $b_i$ and $G_j\left(\theta_j\right)$ belongs to the blockless sequence $b_j$, and that both $b_i$ and $b_j$ either directly precede or directly succeed block $A$. We also assume without loss of generality that $\theta_i < \theta_j$. The gate $G_j\left(\theta_j\right)$ can be split into two consecutive gates $G_j\left(\theta_i\right)\;G_j\left(\theta_j-\theta_i\right)$, so that a simultaneous gate $G(\theta_i)$ can be merged into either $p$ or $s$. In the former case, the procedure appends $G\left(\theta_i\right)$ to the front of $p$, while in the latter case, it appends $G\left(\theta_i\right)$ to the end of $s$. In both cases, the procedure removes $G_i(\theta_i)$ from $b_i$, and within $b_j$ it replaces $G_j(\theta_j)$ with $G_j(\theta_j-\theta_i)$. This reduces the blockless sequence $b_i$ by one gate and eliminates $b_i$ if it is now empty. If this is not the case and $b_i$ cannot be eliminated, the procedure rejects the transformation step and keeps the initial gates $G_i\left(\theta_i\right)$ and $G_j\left(\theta_j\right)$. We repeat the procedure until no more block $A$ can be increased. The execution of the angle splitting approach on the circuit from \autoref{fig-rearranging-3} is shown in \autoref{fig-rearranging-4}.

Afterwards, our compiler transforms all blockless sequences which cannot be completely eliminated according to \autoref{subsec-transformation-to-extended-native-gate-set} so that they consist of a minimal number of native gates from $\mathcal{N}$. The same transformations are applied to the sequences $p$ and $s$ of each block $A$ to minimize the number of gates used.

Note that before angle splitting the circuit consists only of gates with rotation angles $\theta \in \left\{\tfrac{\pi}{2}, \pi \right\}$, so angle splitting preserves the property that all gates are contained in the set of allowed gates $\mathcal{N}$.

\subsection{Commutations of blocks and blockless sequences on disjoint sets of qubits}
\label{subsec-simple-commutations}
\begin{figure}[!b]
    \begin{subfigure}{\columnwidth}
        \centering
    	\includegraphics[width=\columnwidth,keepaspectratio]{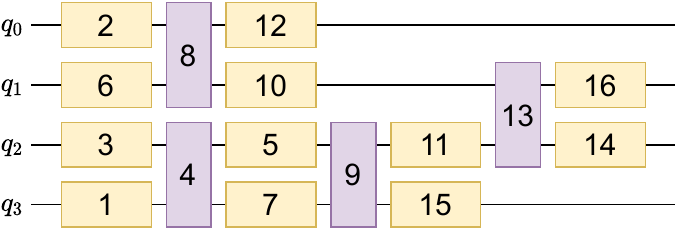}
    	\caption{The gates are executed in a semi-random order, taking care only that on each qubit $q_i$ a gate $G$ is executed after all gates placed before $G$ on $q_i$ have been executed.}
    	\vspace{2mm}
    	\label{fig-simple-commutations-example-1}
	\end{subfigure}
	\begin{subfigure}{\columnwidth}
	    \centering
    	\includegraphics[width=\columnwidth,keepaspectratio]{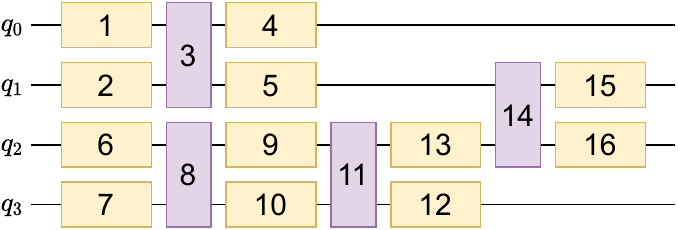}
    	\caption{The commutations have been applied. Blockless sequences of a qubit $q_i$ are always executed directly before or after a block acting on $q_i$. Moreover, since the Blocks 8 and 11 act on the same qubits, they are executed successively only with the blockless sequences 9 and 10 in between. Since the Blocks 3 and 14 have only $q_1$ in common, blockless sequence 5 is executed after blockless sequence 4. Additionally, blockless sequence 13 is executed after blockless sequence 12 because the Blocks 11 and 14 have only $q_2$ in common. The order of the blockless sequences in all other pairs executed before or after a block is arbitrary. \Eg the order of the blockless sequences 1 and 2 can be reversed.}
    	\label{fig-simple-commutations-example-2}
	\end{subfigure}
    \caption{Example of applying the commutation of blocks and blockless sequences on disjoint sets of qubits to a circuit. The purple boxes represent blocks and the yellow boxes represent blockless sequences. The numbers inside the boxes indicate the order in which the gates are executed, \ie the gate with the number 1 is executed first, then the gate with the number 2, and so on.}
    \label{fig-simple-commutations-example}
\end{figure}

The previous compilation steps ensure that the gates acting on a certain qubit are executed in a function receiving order. However, these compilation steps are not optimized for the shuttling overhead, and this generally results in an unfavorable order of blocks and blockless sequences. We now describe how the execution order of blocks and blockless sequences can be rearranged to reduce the amount of shuttling operations required. We exploit the property that the execution order of two blocks or blockless sequences can be swapped if they act on disjoint sets of qubits. The approach then tries to commute each blockless sequence so that it is executed immediately before or after a block applied to the same qubit. An example of such a rearrangement is shown in \autoref{fig-simple-commutations-example}.

Now assume that two blockless sequences $b_i$ and $b_j$ are executed after a block $A_\alpha$ and that $A_\alpha$ acts on the qubits $q_i, q_j \in \mathcal{Q}$ and $b_i$ is applied to $q_i$ and $b_j$ is applied to $q_j$. The algorithm now iterates over the initial execution order of the circuit, searching for the next block $A_\beta$ which acts on either $q_i$ or $q_j$ after $A_\alpha$. If none exists, the order of $b_i$ and $b_j$ may be arbitrary. If $A_\beta$ is applied to $q_i$ and not to $q_j$, then $b_j$ is executed before $b_i$. The analogous rule holds if $A_\beta$ acts on $q_j$ and not on $q_i$. Moreover, if $A_\beta$ acts on $q_i$ and $q_j$, the algorithm commutes $A_\beta$ so that it is executed immediately after $A_\alpha$, with only $b_i$ and $b_j$ in between. In this way, blocks which operate on exactly the same qubits are executed successively if possible.

\subsection{Block commutations with correction unitaries}
\label{subsec-advanced-commutations}
In the last subsection we have commuted blocks and blockless sequences acting on disjoint sets of qubits. In this subsection, we present an additional algorithm which commutes \emph{superblocks} with the following properties: Each superblock consists of at least one two-qubit gate and can consist of several single-qubit gates. All gates of the superblock must operate on the same two qubits and must be executed consecutively on these qubits. For the commutation, the algorithm needs two superblocks with exactly one qubit in common, so that the commutation affects gates on three qubits. Between the two superblocks, no gate may affect the common qubit. The commutation of the superblocks is performed in such a way that afterwards subsequent superblocks in the execution order should operate on the same qubits or have at least one qubit in common. If possible, these commutations can increase the number of such sequences and thus reduce the shuttling overhead.

\begin{figure}[!t]
    \begin{subfigure}{\columnwidth}
        \centering
    	\includegraphics[width=\columnwidth,keepaspectratio]{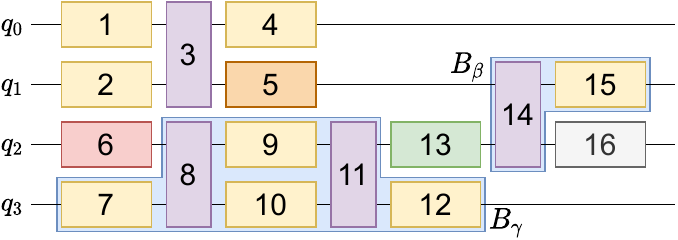}
    	\caption{Example from \autoref{fig-simple-commutations-example-2} with the superblocks $B_\beta$ and $B_\gamma$. The non-yellow blockless sequences can be added completely, partially or not to the adjacent superblocks. Since no gates are allowed on the common qubit $q_2$ between the two superblocks, blockless sequence 13 must be added completely to $B_\beta$ or $B_\gamma$ or divided between them.}
    	\vspace{2mm}
    	\label{fig-advanced-block-commutations-example-1}
	\end{subfigure}
	\begin{subfigure}{\columnwidth}
	    \centering
    	\includegraphics[width=\columnwidth,keepaspectratio]{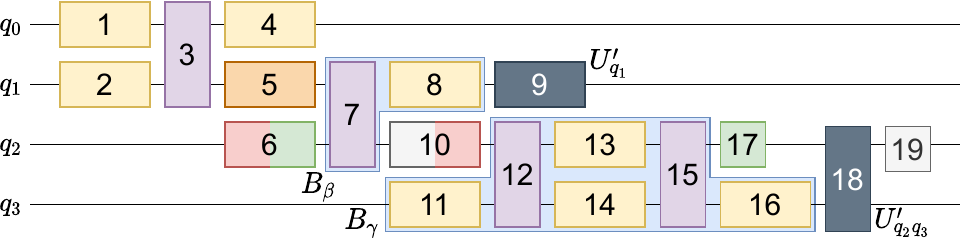}
    	\caption{The superblock $B_\beta$ has been commuted before $B_\gamma$ and the correction unitaries $U'_{q_1}$ and $U'_{q_2 q_3}$ (the dark gray boxes) has been inserted into the circuit. The brown, red, green, and light gray colors indicate to which blockless sequences the corresponding blockless sequences from (a) can be transformed, depending on whether they have been added completely, partially, or not to a superblock, and how they are ordered in the new blockless sequences. The commutation leads to a step-shaped arrangement of the gates, which further reduces the number of shuttling operations.}
    	\label{fig-advanced-block-commutations-example-2}
	\end{subfigure}
    \caption{Example of applying the block commutation with correction unitaries to a circuit. While the purple boxes represent the blocks built during block building in \autoref{subsec-block-building}, the blue boxes represent the superblocks to be swapped, and the dark gray boxes are the correction unitaries. All other boxes are blockless sequences. Analogous to \autoref{fig-simple-commutations-example}, the numbers inside the boxes indicate the order in which the gates are executed.}
    \label{fig-advanced-block-commutations-example}
\end{figure}

The order of blocks with unitaries of size $\nu \times \nu$ with $\nu \in \mathbb{N}$ cannot simply be swapped, because the corresponding matrix multiplications are non-commutative. Consequently, the commutation of the unitaries associated with the superblocks introduces an error into the circuit. To correct the error, a correction unitary is inserted into the circuit after commutation. It is placed behind the superblocks and operates on the same three qubits as the two superblocks. However, the correction unitary can insert two-qubit gates between all three qubits, breaking the structure that the gates operate only between the qubits of their superblock. To prevent this, we show that under certain conditions the correction unitary can be factorized into two correction unitaries. While one of the unitaries is applied only to one of the qubits which is not the common qubit, the other unitary is applied to the other two qubits which then act on the same superblock. This guarantees that the correction unitaries introduce at most two-qubit gates between these two qubits. If the correction unitary does not factorize, the presented algorithm rejects the commutation of the superblocks. An example is shown in \autoref{fig-advanced-block-commutations-example}.

In \autoref{subsubsec-theory-of-advanced-block-commutation} the detailed theory of the block commutation with correction unitaries is described. Afterwards, in \autoref{subsubsec-algorithm-for-executing-advanced-block-commutation} an algorithm for the execution of the block commutation with correction unitaries is presented. This is followed by a heuristic in \autoref{subsubsec-heuristic-evaluation-of-the-commutation} to measure the impact of the block commutation with correction unitaries on the shuttling operations independently of a concrete implementation of a Shuttling Compiler.

\subsubsection{Theory of the block commutation with correction unitaries}
\label{subsubsec-theory-of-advanced-block-commutation}
\begin{figure}
    \centering
	\includegraphics[width=\columnwidth,keepaspectratio]{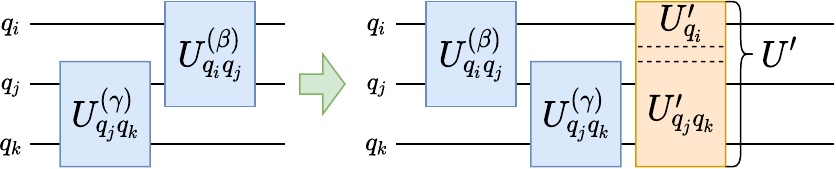}
	\caption{The two superblocks represented by the unitaries $U_{q_i q_j}^{(\beta)}$ and $U_{q_j q_k}^{(\gamma)}$ are commuted. Since in general it is not guaranteed that both unitaries are commutable, a correction unitary $U'$ is inserted, which corrects the error introduced by the commutation of the superblocks. Under certain conditions $U'$ can be factorized into two correction unitaries $U'_{q_i}$ and $U'_{q_j q_k}$.}
	\label{fig-advanced-block-commutations-theory}
\end{figure}

We consider two consecutive superblocks $B_\beta$ acting on the qubits $q_i, q_j \in \mathcal{Q}$ and $B_\gamma$ acting on $q_j, q_k \in \mathcal{Q}$, resulting in the three-qubit unitary
\begin{align}
    \label{eq-U}
    U = U_{q_i q_j}^{(\beta)} U_{q_j q_k}^{(\gamma)},
\end{align}
where $U_{q_i q_j}$ is the unitary associated with $B_\beta$ and $U_{q_j q_k}$ is the unitary associated with $B_\gamma$. \textit{Consecutive} in this context means that no gates may be executed on the common qubit $q_j$ between the two superblocks. It can be beneficial to reverse the execution order of the superblocks to reduce the shuttling overhead. In this section, we show how to determine the conditions under which the execution order can be reversed. In general, the reversal requires a correction unitary $U'$ to obtain the same total unitary:
\begin{align}
    U= U' U_{q_j q_k}^{(\gamma)} U_{q_i q_j}^{(\beta)}.
\end{align}
The three-qubit correction unitary is given by
\begin{align}
    U' &=  U_{q_i q_j}^{(\beta)} U_{q_j q_k}^{(\gamma)} U_{q_i q_j}^{(\beta)\dagger} U_{q_j q_k}^{(\gamma)\dagger}.
\end{align}
Given $U_{q_i q_j}^{(\beta)}$ and $U_{q_j q_k}^{(\gamma)}$ in matrix form, we need to check if $U'$ factorizes into a $q_i/q_jq_k$ separable form:
\begin{equation}
    U'=U_{q_i}'\otimes U_{q_j q_k}'.
\label{eq:commSeparability}
\end{equation}
We explicitly compute $U'$ and use the set of Pauli operators for a single-qubit $\mathcal{P}_{q} = \left\{I_{q}, X_{q}, Y_{q}, Z_{q}\right\}$ to decompose it into single-qubit Pauli operators acting on $q_i$ and two-qubit Pauli operators acting on $q_j$ and $q_k$:
\begin{align}
    U' = \sum_{P_{\lambda} \in \mathcal{P}_{q_i}} \sum_{P_{\mu} \in \mathcal{P}_{q_j} \otimes \mathcal{P}_{q_k} } M_{\lambda\mu} P_{\lambda} \otimes P_{\mu}.
\end{align}
Note that the $P_{\mu}$ are two-qubit unitaries from the tensor product set $\mathcal{P}_{q_j} \otimes \mathcal{P}_{q_k}$. The decomposition results in a $4 \times 16$ matrix $M$ with the entries
\begin{align}
   M_{\lambda\mu} = \frac{1}{8} \tr \left(U' \;P_{\lambda} \otimes P_{ \mu}\right),
\end{align}
which has the singular value decomposition
\begin{align}
    M = U S V^{\dagger}
\end{align}
with the $4 \times 4$ and $16 \times 16$ unitary matrices $U$ and $V$. The $4 \times 16$ matrix $S$ consists of a $4 \times 4$ diagonal matrix on the left, whose non-zero entries are the singular values of $M$, and a $4 \times 12$ zero matrix on the right. If $\rank  S = 1$, only $S_{11} \neq 0$ and $U'$ factorizes according to \eqref{eq:commSeparability}. The resulting correction unitaries are given by
\begin{subequations}
    \begin{align}
        U'_{q_i} &= \sum_{\lambda=1}^4 U_{\lambda, 1} P_{\lambda}, \\
        U'_{q_j q_k} &= \sum_{\mu = 1}^{16} V_{\mu, 1}^* P_{ \mu}.
    \end{align}
\end{subequations}
The final unitary in favorable order is
\begin{align}
    \label{eq-U-commuted}
    U=U'_{q_j q_k} U_{q_j q_k}^{(\gamma)} U'_{q_i} U_{q_i q_j}^{(\beta)},
\end{align}
which contains $U_{q_i q_j}$ and $U_{q_j q_k}$ in the commuted order compared to \eqref{eq-U}. A visualization of the commutation is given in \autoref{fig-advanced-block-commutations-theory}.

\subsubsection{Algorithm for executing the block commutation with correction unitaries}
\label{subsubsec-algorithm-for-executing-advanced-block-commutation}
In the following, we describe the algorithm for performing the block commutation with correction unitaries on an input circuit $C$. While for the theory in \autoref{subsubsec-theory-of-advanced-block-commutation} we assumed that the superblocks are already known, in practice the algorithm must first determine the superblocks on the circuit. Therefore, a block from \autoref{subsec-block-building} forms the basis of each superblock, which is enlarged in a second step. The blocks are chosen to reduce the number of shuttling operations required. The algorithm iterates over the blocks and blockless sequences in their execution order. It searches for three blocks $A_\alpha$, $A_\gamma$, and $A_\beta$, where $A_\gamma$ and $A_\beta$ are the base blocks of the superblocks $B_\gamma$ and $B_\beta$, respectively. $A_\alpha$ is the block executed directly before $B_\gamma$ in the execution order and has at least one qubit in common with $B_\beta$. Thus, commuting $B_\gamma$ and $B_\beta$ places $B_\beta$ between $A_\alpha$ and $B_\gamma$ regarding the execution order and ensures that after the commutation, $A_\alpha$ and $B_\beta$ are executed successively with at least one qubit in common. Consequently, this may lead to better locality during shuttling and reduce the shuttling overhead.

At the beginning, the algorithm searches for the block $A_\alpha$, which is the first block found in the execution order during the iteration. The block following $A_\alpha$ in the execution order is used as $A_\gamma$ acting on the qubits $q_j$ and $q_k$ and as the basis for the superblock $B_\gamma$. If $A_\alpha$ and $A_\gamma$ act on exactly the same qubits, $A_\gamma$ becomes the new $A_\alpha$ and the algorithm searches for a new $A_\gamma$. This ensures that the locality of acting on common qubits between both blocks is not broken.

If $A_\alpha$ and $A_\gamma$ act on at least one different qubit, the algorithm searches for the block $A_\beta$ as the basis for the superblock $B_\beta$. To perform the commutation as described in \autoref{subsubsec-theory-of-advanced-block-commutation}, $A_\beta$ needs exactly one common qubit, $q_j$, with $A_\gamma$. Since $B_\beta$ immediately follows $A_\alpha$ in the execution order after commutation, $A_\beta$ must also have at least $q_i$ in common with $A_\alpha$ to reduce the shuttling overhead. When searching for $A_\beta$, it need not be the direct successor block of $A_\gamma$ in the execution order. It is possible that there are other blocks between $A_\gamma$ and $A_\beta$. We use the intermediate blocks which act on $q_j$ and $q_k$ later to extend $B_\gamma$, as long as there is no other block which acts only on $q_j$ or $q_k$ in the execution order. The other intermediate blocks must not have a qubit in common with $A_\beta$ to guarantee that $A_\beta$ commutes with them. If $A_\alpha$ has exactly one qubit in common with $A_\beta$ and $A_\gamma$, a commutation of $B_\beta$ between $A_\alpha$ and $B_\gamma$ does not cause $A_\alpha$ to be followed by a block which has more qubits in common with $A_\alpha$ than $A_\gamma$. Nevertheless, the commutation is performed in this case because it may allow $B_\beta$ to be commuted with another commutation before $A_\alpha$, which may lead to better locality in shuttling. If a block $A_\beta$ cannot be found, $A_\gamma$ becomes the new $A_\alpha$ and the algorithm continues to search for a new $A_\gamma$.

After our algorithm has determined the base gates of the superblocks, there are two ways to extend them. First, a neighboring block from the blocks built during block aggregation can be merged into a superblock if it operates on the same qubits as the superblock and there is no other block between it and the superblock which operates on only one of the qubits. If another block is added to a superblock, all blockless sequences operating on the same qubits between the added block and the superblock also become part of the superblock.

The second way is to extend the superblocks by their neighboring blockless sequences or parts of them which act on a qubit of the corresponding superblock. While in general we can add blockless sequences completely or only some gates of them, there are two exceptions where the algorithm always adds blockless sequences completely to the adjacent superblock. Both exceptions have in common that the blockless sequence is executed directly before or after $B_\beta$ on $q_i$ or $B_\gamma$ on $q_k$ in the execution order. However, in the first case, the block before or after the blockless sequence in the execution order does not act on the same qubit as the blockless sequence. On the other hand, in the second case, the blockless sequence contains the first or last gates of the corresponding qubit in the circuit. Adding these blockless sequences to the superblocks leads to better locality during shuttling. All other blockless sequences which do not fall under these exceptions can be divided into two parts. When the algorithm divides a blockless sequence, only the gates closer to the superblock are added, while the other part remains outside the superblock. It is possible that one of the parts does not consist of a gate, which means that depending on the part, the entire blockless sequence may or may not be added to the superblock. An exception is the blockless sequence on the common qubit $q_j$ between the two superblocks. Since no gate is allowed at this position, this blockless sequence must be added to one of the superblocks. Therefore, it can be added completely to either $B_\beta$ or $B_\gamma$, or it can be divided so that one part is added to $B_\beta$ and the other part to $B_\gamma$.
 
 \begin{figure*}
    \centering
    \begin{subfigure}{\textwidth}
        \centering
    	\includegraphics[width=\textwidth,keepaspectratio]{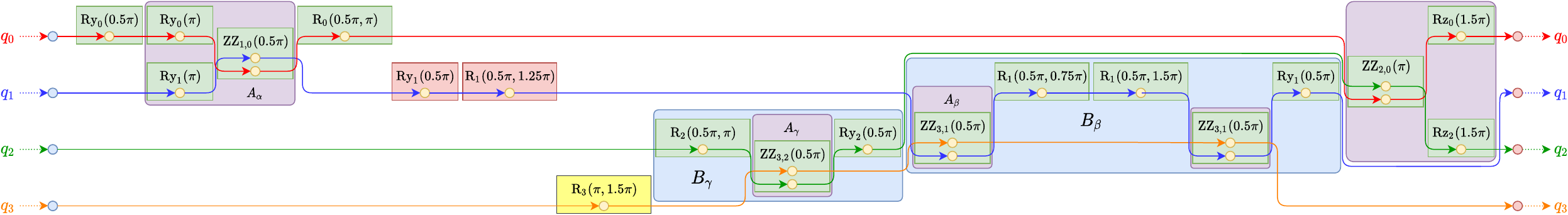}
	    \caption{Example from \autoref{fig-example-9} after applying block aggregation and the commutation of blocks and blockless sequences on disjoint sets of qubits. The purple boxes represent the blocks built by block aggregation. The left-to-right order of the gates determined by the commutation of blocks and blockless sequences on disjoint sets of qubits describes the execution order of the gates. When the algorithm in \autoref{subsubsec-algorithm-for-executing-advanced-block-commutation} is executed, it first finds the block $A_\alpha$ acting on $q_0$ and $q_1$, and $A_\gamma$ acting on $q_2$ and $q_3$. Since the third block, $A_\beta$, has $q_1$ in common with $A_\alpha$ and $q_3$ in common with $A_\gamma$, it can be used for the block commutation with correction unitaries. The blocks $A_\gamma$ and $A_\beta$ are the basis for the superblocks $B_\gamma$ and $B_\beta$, respectively. Expanding the base superblocks with other blocks and the blockless sequences, which are always added to the superblocks, results in the displayed blue superblocks $B_\gamma$ and $B_\beta$. However, it is not guaranteed that the blue blocks are commutable. It is possible to further enlarge $B_\gamma$ with the yellow gate and $B_\beta$ with the red gates. In the case of $B_\beta$, it is possible to add only the right red gate or both red gates. The value of the heuristic in \autoref{subsubsec-heuristic-evaluation-of-the-commutation} for this circuit is one.}
	\label{fig-example-10}
    	\vspace{2mm}
	\end{subfigure}
	\begin{subfigure}{\textwidth}
	    \centering
        \includegraphics[width=\textwidth,keepaspectratio]{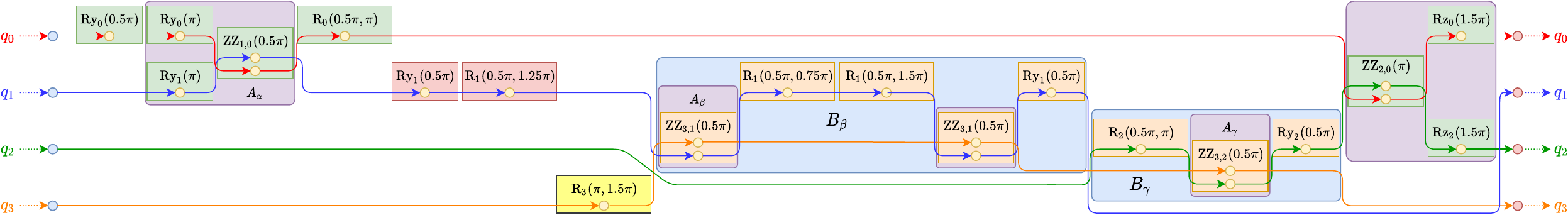}
	    \caption{Circuit from (a) after commuting $B_\beta$ before $B_\gamma$. Since $U'_{q_1}$ and $U'_{q_3 q_2}$ are identity matrices, no additional gates have been added to the circuit. The commutation has been performed without adding the red and yellow gates to their corresponding superblocks. While adding the right red gate or both red gates to $B_\beta$ would have produced the same result as without adding, adding the yellow gate to $B_\gamma$ would have made the commutation unexecutable regardless of how many red gates were added to $B_\beta$. The value of the heuristic in \autoref{subsubsec-heuristic-evaluation-of-the-commutation} for this circuit is $\tfrac{4}{3}$. In contrast to the circuit in (a), where the successive blocks $A_\alpha$ and $A_\gamma$ act on completely different qubits, after commutation each block has at least one qubit in common with its predecessor block, making the operations more local.}
	\label{fig-example-11}
    \end{subfigure}
    \caption{The process of applying the block commutation with correction unitaries.}
    \label{fig-example-10-11}
\end{figure*}

To find the best partitioning of the dividable blockless sequences, our algorithm determines all possible partitionings of these sequences. Afterwards, for both superblocks, each partitioning of each dividable blockless sequence is combined with each partitioning of the other blockless sequences. In this way, we build several candidate superblocks for $B_\beta$ and $B_\gamma$. Then we compute the unitary matrices of each candidate superblock of $B_\beta$ and $B_\gamma$, and the procedure in \autoref{subsubsec-theory-of-advanced-block-commutation} is executed for each possible combination of a candidate superblock of $B_\beta$ with a candidate superblock of $B_\gamma$. Due to the generally intertwined structure of the circuit archived by the previous compilation steps, there are only a few combinations to execute. Thus, it is not computationally infeasible to check all combinations for a typical case. In \autoref{fig-example-10} it is depicted how our algorithm builds superblocks in the example from \autoref{fig-example-9} and how they can be extended.

For all combinations where $U'$ factorizes, our algorithm transforms the unitaries $U'_{q_i}$ and $U'_{q_j q_k}$ into circuits using Pytket. We then optimize these circuits using our compilation flow up to phase tracking and using the \texttt{FullPeepholeOptimise} pass in \autoref{subsec-transformation-to-native-gate-set}.

If there are multiple combinations for which $U'$ factorizes, the algorithm determines which combination to use for the commutation. Therefore, several heuristics can be used, depending on the optimization goal, such as minimizing the overall gate count or the two-qubit gate count. We present three such heuristics in the following. The \textbf{first heuristic} prioritizes the reduction of the overall gate count over the minimization of the two-qubit gate count as the optimization goal. The heuristic is applied top-down. If multiple combinations satisfy a criterion, the next criterion is used for further selection:
\begin{enumerate}
    \item Select the combination(s) with the smallest total number of gates in the circuits of $U'_{q_i}$ and $U'_{q_j q_k}$.
    \item Select the combination(s) with the lowest number of two-qubit gates in $U'_{q_j q_k}$.
    \item Select the combination(s) with the lowest number of blockless sequences.
    \item Select one combination randomly.
\end{enumerate}

The \textbf{second heuristic} reverses the first two criteria, giving priority to reducing the two-qubit gate count over minimizing the overall gate count. A middle ground between the two heuristics is the \textbf{third heuristic}, which uses the same order of the criteria as the first heuristic. However, when counting the number of gates in $U'_{q_i}$ and $U'_{q_j q_k}$, each two-qubit gate is counted as $\xi > 1$ gates. This way the two-qubit gates have a higher weight in the gate count, prioritizing combinations with a lower number of two-qubit gates, but without completely ignoring the single-qubit gates in the first criterion as in the second heuristic. 

\begin{figure*}
    \centering
	\includegraphics[width=\textwidth,keepaspectratio]{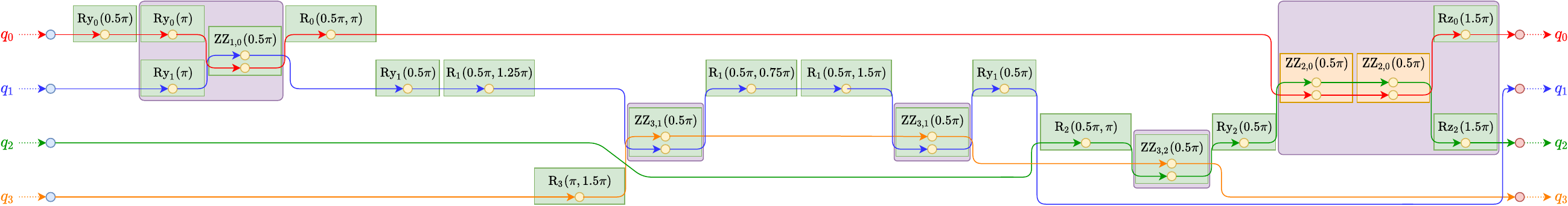}
	\caption{Example from \autoref{fig-example-11} after applying the transformation of two-qubit gates into our trapped-ion native gate set. The $\gate{ZZ}_{2,0}\left(\pi\right)$ gate in \autoref{fig-example-11} is transformed into the two orange $\gate{ZZ}_{2,0}\left(0.5\pi\right)$ gates. All gates of the circuit are in our trapped-ion native gate set $\mathcal{N}$.}
	\label{fig-example-12}
\end{figure*}

We then commute $B_\beta$ and $B_\gamma$ with respect to the selected combination and insert the gates from $U'_{q_i}$ and $U'_{q_j q_k}$ into the circuit $C$. In the following, our algorithm treats each two-qubit gate contained in the circuit of $U'_{q_j q_k}$ as a block containing only the corresponding gate, while it combines the inserted single-qubit gates into blockless sequences. As an example, \autoref{fig-example-11} shows the circuit from \autoref{fig-example-10} with commuted superblocks.

To further determine possible superblocks for commutation, the algorithm selects the new $A_\alpha$ in such a way that $A_\beta$ can serve as $A_\beta$ again and can be commuted again. If this is not possible, the algorithm uses the first block in the execution order of the circuit as the new $A_\alpha$. This allows a recursive commutation of blocks, which leads to a nonlinear runtime of the block commutation with correction unitaries. To avoid that the blocks are commuted cyclically, we check that the same two blocks are not commuted twice.

If the algorithm cannot find any more commutations and has performed at least one commutation, we optimize the circuit with the commuted blocks. Therefore, we remove the block aggregations of \autoref{subsec-block-building}. Due to the commutations, it is possible that more than two \gate{R} gates act on the same qubit between two \gate{ZZ} gates. We apply Pytket's \texttt{SquashTK1} pass and the \texttt{RebaseCustom} pass to these sequences. Then we re-run our compilation flow from the transformation pass in \autoref{subsec-transformation-to-native-gate-set}. This removes the \gate{Rx} gates introduced by the circuits of $U'_{q_i}$ and $U'_{q_j q_k}$ as well as the rebasing pass which do not have an angle of $\frac{\pi}{2}$ or $\pi$. Furthermore, phase tracking eliminates the newly introduced \gate{Rz} gates and the block aggregation builds new blocks considering the new optimizations. The new blocks and blockless sequences acting on disjoint sets of qubits are then commuted as described in \autoref{subsec-simple-commutations}.

\subsubsection{Heuristic evaluation of the commutation}
\label{subsubsec-heuristic-evaluation-of-the-commutation}
The algorithm presented above puts the circuit into a favorable structure by commuting superblocks. However, the algorithm does not measure the impact of the commutations on the shuttling operations. To allow our compiler to be used independently of a concrete implementation of a Shuttling Compiler, we present a heuristic for evaluating the impact of the commutations. We calculate this heuristic for the circuit before and after applying the block commutation with correction unitaries. The circuit with the higher heuristic value is used for the further compilation.

To compute the heuristic, the algorithm iterates over the entire circuit in the execution order, looking at each pair of consecutive blocks built in \autoref{subsec-block-building}. For each pair, the algorithm determines how many qubits the blocks have in common. These amounts are added and divided by the number of pairs. The result is a value between zero and two. The heuristic can only reach a value of zero if each block affects two different qubits compared to its immediate predecessor block in the execution order. This means that the blocks are in the most unfavorable order. In contrast, the heuristic can only reach a value of two if all blocks operate on the same two qubits. Consequently, reaching a value of two is not possible for circuits where the block commutation with correction unitaries has been applied, because it is only applicable to circuits where gates act on at least three different qubits. 

Using the block commutation with correction unitaries should increase the value of the heuristic. The higher the value, the more blocks following each other in the execution order will have qubits in common and the more local the operations will be. As a result, the number of shuttling operations will decrease. See \autoref{fig-example-10-11} for an example of using the heuristic.

\subsection{Transformation of two-qubit gates into our trapped-ion native gate set}
\label{subsec-transformation-to-native-gate-set-double}
The transformation in \autoref{subsec-transformation-to-extended-native-gate-set} has converted the \gate{ZZ} gates so that they have a phase $\theta \in \left\{ \tfrac{\pi}{2}, \pi, \tfrac{3\pi}{2} \right\}$. Since our trapped-ion native get set $\mathcal{N}$ contains only the gate $\gate{ZZ}\left(\tfrac{\pi}{2}\right)$, the \gate{ZZ} gates in $\mathcal{G}$ with a phase of $\pi$ or $\tfrac{3\pi}{2}$ must be transformed to satisfy the condition. Therefore, the following decomposition is used:
\begin{subequations}
\label{eq-ZZ-decomp-pi-3pihalf}
    \begin{align}
        \gate{ZZ}\left(\pi\right) &=
        \gate{ZZ}\left(\tfrac{\pi}{2}\right) \gate{ZZ}\left(\tfrac{\pi}{2}\right) 
        \label{eq-ZZ-decomp-pi},\\
        \gate{ZZ}\left(\tfrac{3\pi}{2}\right) &= \gate{ZZ}\left(\tfrac{\pi}{2}\right) \gate{ZZ}\left(\tfrac{\pi}{2}\right) \gate{ZZ}\left(\tfrac{\pi}{2}\right).
        \label{eq-ZZ-decomp-3pihalf}
    \end{align}
\end{subequations}
This transformation is applied to the circuit from \autoref{fig-example-11} in \autoref{fig-example-12}.

Note that this transformation could also have been applied earlier. However, not only would there be no benefit, but it would also worsen the runtimes of the block aggregation and the commutations because there are more two-qubit gates to consider.
It should also be noted that the single-qubit gates are already part of our trapped-ion native gate set $\mathcal{N}$, and since the two-qubit gates are as well, the circuit $C$ is now executable on the target quantum computing architecture.

\section{Parameterized Circuits}
\label{sec-parameterized-circuits}
Important applications of quantum processors in the NISQ era include quantum machine learning \cite{Biamonte2017} and Variational Quantum Eigensolver (VQE) algorithms \cite{Cerezo2021}. These use cases require a parameterized circuit to be executed multiple times with only small variations in some gate parameters, \ie phases and angles of the gates. A classical optimizer then compares and adjusts the results of these executions.
To make the circuits comparable, it is important that their structure is identical. However, different gate parameters can lead to different circuits after applying the compilation steps of our paper. Introducing the compilation of parameterized circuits mitigates this problem. Whenever unresolved parameters are present in the circuit to be compiled, these parameters are preserved throughout the compilation and only generally applicable optimizations are performed. Pytket natively supports such parameterized compilation and generates parameter-dependent expressions for all its transformations.

When compiling parameterized circuits, our compiler still removes all \gate{SWAP} gates (\autoref{subsec-eliminiation-swap-gates}) and the gates that cancel out (\autoref{subsec-repeated-removing-and-commutation-of-gates}), but only if the cancellation is independent of the current parameters. Additionally, macro matching is performed (\autoref{subsec-macro-matching}) only if all values of the parameters satisfy the macro condition. The same scheme is also applied to all following steps. The final circuit may contain gates with angles and phases defined by non-trivial mathematical expressions, mainly introduced when transforming the circuit into our native gate set (\autoref{subsec-transformation-to-extended-native-gate-set}).
While compiling parameterized circuits does not restrict phase tracking (\autoref{subsec-rz-phase-tracking}), block aggregation (\autoref{subsec-block-building}) and the block commutation with correction unitaries (\autoref{subsec-advanced-commutations}) lose much of their impact. The latter is not performed for parameterized circuits for performance reasons.

As an example, we used \cite{ansatz_sequence_example} to generate a parameterized circuit representing a Unitary Coupled Cluster ansatz \cite{UUC}, which is needed when executing variational algorithms such as VQE \cite{VQE}. The circuit has four qubits, 48 single-qubit and 32 two-qubit gates, and contains two parameters. When we compiled the circuit together with the parameters into our trapped-ion native gate set, we got a circuit with 69 single-qubit and 32 two-qubit gates. In this circuit, the parameters can be replaced by numeric values to get executable circuits. This allows us to execute the circuit with different values after compiling it once. If the values are known before compilation and the circuit only needs to be executed with these values, the parameters can be replaced before compilation. In this case, we got a circuit with 61 single-qubit and 32 two-qubit gates. The small increase factor of 1.13 in the number of single-qubit gates for the parameterized compilation compared to the compilation with known parameters shows that our compiler also works well for parameterized circuits.

\section{Evaluation}
\label{sec-evaluation}
This section presents the evaluation results of our compiler. We tested the compiler on a library of 153 quantum circuits \cite{circuit_repo}, which had been previously used for benchmarks \cite{8970267, 8382253, cowtan_et_al:LIPIcs:2019:10397, DBLP:conf/asplos/LiDX19}. Each circuit in the library has between 3 and 16 qubits and 5 to 207,775 gates. In addition, we benchmarked our compiler for the algorithms Quantum Approximate Optimization (QAOA) \cite{Moll_2018}, Quantum Fourier Transform (QFT) \cite{nielsen_chuang_2010}, Supremacy \cite{Boixo2018, supremacy_repo}, Sycamore \cite{Arute2019}, and Quantum Volume Estimation (QV) \cite{Cross2019} for up to 200 qubits. While we generated the circuits for the first four algorithms using the quantum circuit generator from \cite{large_circuit_repo}, we generated the circuits for QV using Qiskit \cite{qiskit}. We only evaluated QFT up to 49 qubits, because for a higher number of qubits the angles of some gates in the initial circuits are so small that they were considered as zero in the calculation. For Supremacy and Sycamore, we only used circuits where the qubits can be arranged in a square lattice with a depth of 1,000. We executed the evaluations on the Mogon II cluster of the Johannes Gutenberg University. Each evaluation was executed with one core of an Intel 2630v4 CPU and 16\,GB RAM.
We analyzed the impact of the different compilation stages on the result. Additionally, we compared our compiler with standard Pytket \cite{tket} and Qiskit \cite{qiskit} passes, using Pytket version 1.11.1 and Qiskit version 0.39.5. When measuring the runtime, we always averaged it over 10 executions of a given circuit. The resulting single-qubit and two-qubit gate counts are given in \hyperref[sec-detailed-evaluation-results]{App.~\ref*{sec-detailed-evaluation-results}}.

We executed all the circuits using the compilation flow depicted in \autoref{fig-compiler-flowchart}. For the macro matching in \autoref{subsec-macro-matching}, we used only the \gate{C\hspace{0.2mm}Ry} macro given in \eqref{eq-cry-macro}. As a heuristic for selecting the combination of blocks used for the block commutation with correction unitaries in \autoref{subsubsec-algorithm-for-executing-advanced-block-commutation}, we used the first heuristic, which prioritizes the minimization of the overall gate count as the first criterion. The reason for this is that we compared our results with those of standard Pytket and Qiskit passes, and both tools do not explicitly prioritize one type of gate over another. Since our compiler offers four alternative ways to transform the circuit into the native gate set, as discussed in \autoref{subsec-transformation-to-extended-native-gate-set}, we compared these approaches. While the first scheme performs additional optimizations using Pytket's \texttt{CliffordSimp} pass, the second scheme transforms based on the \gate{ZZ} decomposition in \eqref{eq-ZZ-decomp} and Pytket's \texttt{SquashTK1} pass without further optimizations. The third scheme performs optimizations achieved by the KAK decomposition \cite{kak}, while the fourth scheme combines various optimizations using Pytket's \texttt{FullPeepholeOptimise} pass. The four different schemes are visualized in \autoref{fig-compiler-flowchart}, where each scheme represents a different control path in the uppermost blue box. In the following, we refer to these four schemes as the \emph{CliffordSimp}, \emph{SquashTK1}, \emph{KAKDecomposition}, and \emph{FullPeepholeOptimise} approaches, respectively.

After calculating the results of our complete compilation flow, we computed the shuttling schedules for the circuits using a Shuttling Compiler \cite{umz-shuttling}. To analyze the influence of the commutations described in \autoref{subsec-simple-commutations} and  \autoref{subsec-advanced-commutations} on the shuttling, we compiled all circuits two additional times: once without both types of commutations and a second time only without the block commutation with correction unitaries. This allowed us to determine the impact of the different commutations by comparing the translation, separation/merge, and ion swaps counts of the circuits. To calculate the shuttling schedules, we assumed a linear trap with 1,401 segments, each containing a maximum of two ions, and the laser interaction zone placed in the center of the trap. This configuration ensured that there was enough space for all ions and that there was no reconfiguration overhead due to lack of space.

In the following, we analyze first the impact of the different transformations and then the impact of the different commutations on the shuttling. Finally, we compare our compiler with standard compilation passes of Pytket and Qiskit.

\subsection{Analysis of the impact of the different transformations}
\label{subsec-analysis-impact-different-transformations}
When analyzing the results of the circuit library, the FullPeepholeOptimise approach produced the circuits with the lowest overall gate count for 72\,\% of the circuits, followed by the CliffordSimp approach with 65\,\%, the KAKDecomposition approach with 40\,\%, and the SquashTK1 approach with 39\,\%. Since for some circuits multiple approaches calculated a result with the lowest gate count, the percentages add up to more than 100\,\%. When comparing the results of the different approaches pairwise, the gate counts differ by factors up to 1.15. Additionally, the FullPeepholeOptimise approach produced the lowest single-qubit gate count for 71\,\% of the circuits, followed by the CliffordSimp approach with 61\,\%, the KAKDecomposition approach with 42\,\%, and the SquashTK1 approach with 41\,\%. The single-qubit gate counts of the different approaches vary by factors up to 1.18. For 83\,\% of the circuits, the FullPeepholeOptimise approach also determined the results with the lowest two-qubit gate count, followed by the CliffordSimp approach with 59\,\%, the KAKDecomposition approach with 49\,\%, and the SquashTK1 approach with 47\,\%. The two-qubit gate counts of the different approaches differ by factors up to 1.21. The smallest differences in the overall, single-qubit, and two-qubit gate counts are between the SquashTK1 and KAKDecomposition approaches, whose counts vary by factors up to 1.04.

For QAOA, all four approaches produced the same results independently of the number of qubits. Also for QFT, all four approaches calculated the same results for the circuits up to 25 qubits, while for the larger circuits only the FullPeepholeOptimise approach determined the results with the lowest number of single- and two-qubit gates. For QV, the CliffordSimp and SquashTK1 approaches, as well as the KAKDecomposition and FullPeepholeOptimise approaches, each calculated the same results. However, the single and two-qubit gate counts of the KAKDecomposition and FullPeepholeOptimise approaches are about 1.25 times lower for the five-qubit circuit. As the number of qubits increases, these factors decrease and are only slightly greater than one for 200 qubits. The reason for the better results of these approaches is the specialization of the KAK decomposition to circuits with a structure like the QV circuits. Note that the KAK decomposition is also part of Pytket's \texttt{FullPeepholeOptimise} pass. Moreover, for Supremacy, the CliffordSimp and FullPeepholeOptimise approaches, as well as the SquashTK1 and KAKDecomposition, each determined the same gate counts. The gate counts of the CliffordSimp and FullPeepholeOptimise approaches are slightly lower than the gate counts of the other two approaches. For Sycamore, the CliffordSimp and FullPeepholeOptimise approaches calculated the same results for the circuits up to 49 qubits. The two-qubit gate counts are always the same for the CliffordSimp and FullPeepholeOptimise approaches, as well as for the SquashTK1 and KAKDecomposition approaches. While the single-qubit gate counts are up to 1.27 times lower for the SquashTK1 or KAKDecomposition approaches, the two-qubit gate counts are slightly lower for the other two approaches.

Regardless of the approach, 98\,\% of the compiled circuits in the circuit library have higher overall and single-qubit gate counts than the original circuits. The only circuits with lower overall and single-qubit gate counts are the \emph{ising\_model} circuits, which contain several \gate{Rz} gates and thus benefit greatly from phase tracking.
On the other hand, the number of circuits with a higher two-qubit gate count depends on the compilation approach used. Hence, the SquashTK1 and KAKDecomposition approaches have a higher two-qubit count for 46\,\% of the circuits, the CliffordSimp approach for 40\,\%, and the KAKDecomposition approach for only 25\,\%. The single-qubit gate counts increased because all transformed circuits must satisfy our trapped-ion native gate set $\mathcal{N}$. However, the reason for the increase in the two-qubit gate counts is the block commutation with correction unitaries, which can insert two-qubit gates through the circuit described by $U'_{q_j q_k}$. In the compilation steps before the block commutation with correction unitaries, the two-qubit gate counts can only decrease because the original circuits contain some redundant \gate{CNOT} gates, which can be removed after the single-qubit gates have been commuted through them. Then the decomposition in \eqref{eq-CNOT-decomp} replaces each remaining \gate{CNOT} gate by exactly one \gate{ZZ} gate. In the analysis below, we will see that for circuits with increasing two-qubit gate count, the block commutation with correction unitaries should be executed only if minimizing the overall gate count or the amount of shuttling operations is the optimization goal.

\begin{figure}
    \centering
	\includegraphics[width=\columnwidth,keepaspectratio]{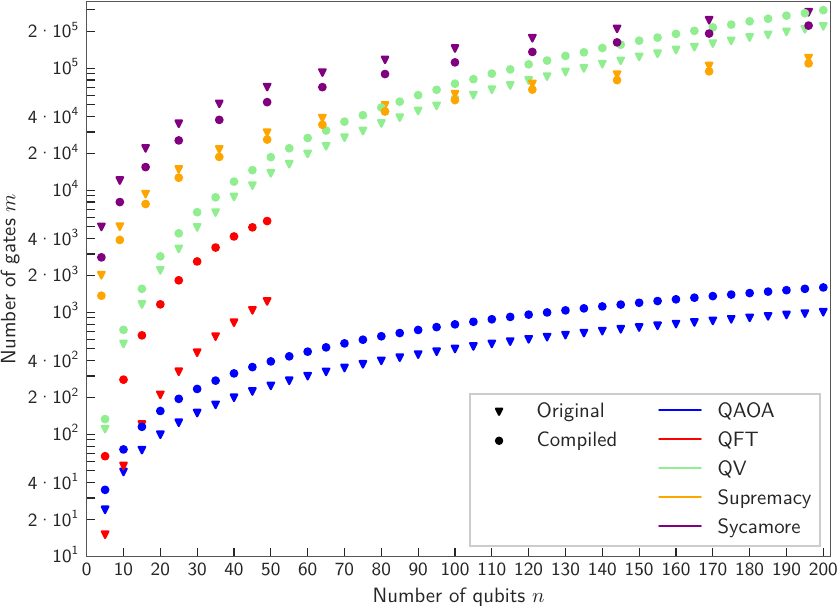}
	\caption{The overall gate counts of the five algorithms QAOA (blue), QFT (red), QV (green), Supremacy (orange), and Sycamore (purple) depending on different numbers of qubits. For each algorithm, the gate counts of the original (triangles) and compiled (circles) circuits are depicted. For the latter, the rebasing approach with the lowest gate count is used.}
	\label{fig-large-circuits-gate-counts}
\end{figure}

As depicted in \autoref{fig-large-circuits-gate-counts}, the compiled circuits of QAOA, QFT, and QV have higher overall gate counts compared to the original circuits. While for QAOA and QV the two-qubit gate counts remained the same or decreased, for QFT the two-qubit gate counts at most doubled during compilation. This is because the original circuits contain \gate{CPhase} gates as two-qubit gates, each of which was decomposed into two \gate{ZZ} gates. However, some of the \gate{ZZ} gates became redundant and could be removed after commuting the single-qubit gates through them. For Supremacy and Sycamore, the overall gate counts in the compiled circuits are lower than in the original circuits. Here, the circuits contain several \gate{T} and \gate{Z} gates, which phase tracking removed. The two-qubit \gate{CZ} gates of the original circuits were replaced by one \gate{ZZ} and two \gate{Rz} gates where phase tracking could also remove the latter. However, for 69\,\% of the circuits, the two-qubit gate counts increased slightly due to the block commutation with correction unitaries, while the single-qubit gate counts are up to 2.20 times lower. For all five algorithms, the number of qubits has no effect on the gate count relation.

\begin{figure}[!t]
	\begin{subfigure}{\columnwidth}
        \centering
    	  \includegraphics[width=\columnwidth,keepaspectratio]{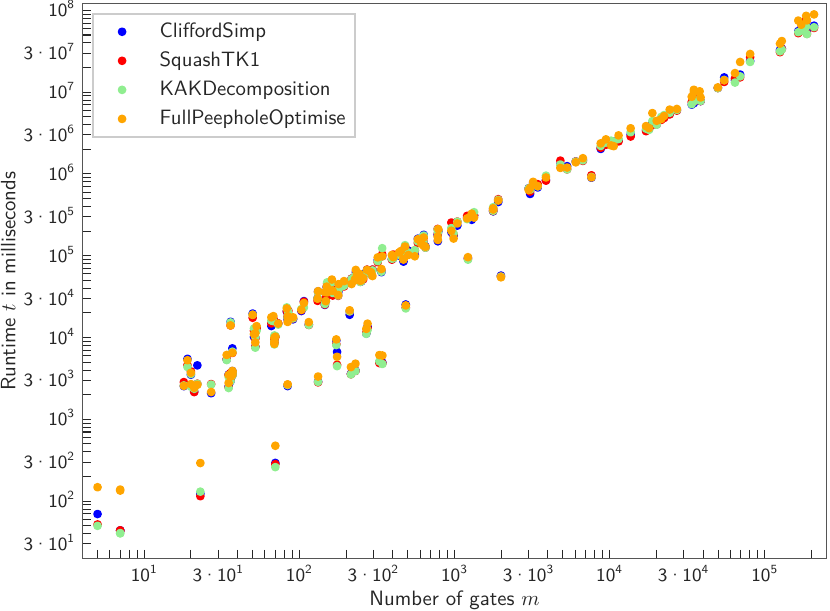}
         \caption{Runtimes of our complete compilation flow. All four compilation approaches behave nonlinearly with the number of gates.}
    	\label{fig-runtimes-with-unitary-blocks}
	\end{subfigure}
    \begin{subfigure}{\columnwidth}
        \centering
         \includegraphics[width=\columnwidth,keepaspectratio]{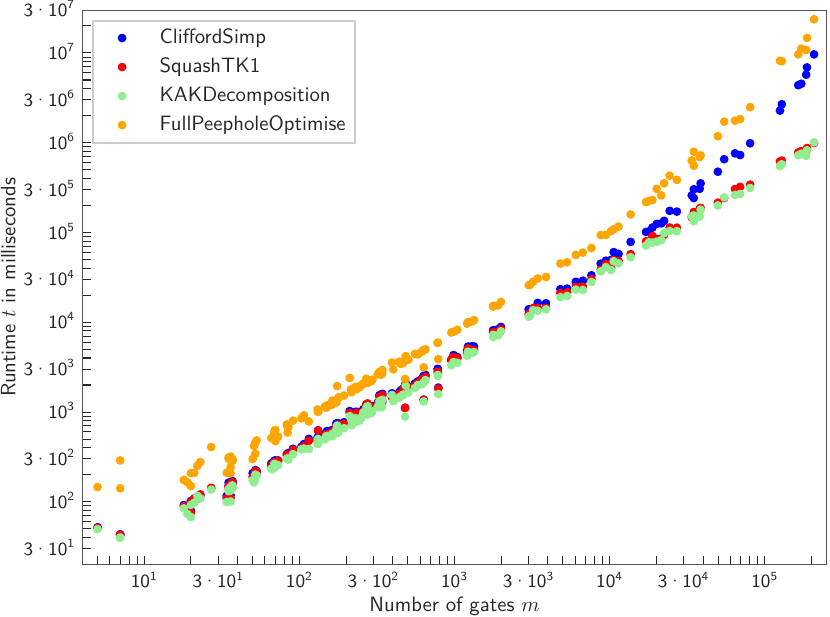}
         \caption{Runtimes of our compiler when executed without the block commutation with correction unitaries. In this case, the CliffordSimp and FullPeepholeOptimise approaches lead to a nonlinear scaling of the runtimes, while the SquashTK1 and KAKDecomposition approaches behave linearly with the number of gates.}
    	\vspace{2mm}
    	\label{fig-runtimes-without-unitary_blocks}
	\end{subfigure}
	\caption{Runtimes of our compiler for the different rebasing approaches depending on the number of gates. Each color represents one of the rebasing approaches: CliffordSimp (dark blue), SquashTK1 (red), KAKDecomposition (green), and FullPeepholeOptimise (orange). Each dot represents one of the 153 circuits in the circuit library.}
	\label{fig-runtimes}
\end{figure}

The runtimes of the four approaches versus the number of gates in the original circuits for the circuit library are shown in \autoref{fig-runtimes-with-unitary-blocks}. All four approaches had nonlinear compile times. While the compile times of the CliffordSimp, SquashTK1, and KAKDecomposition approaches were nearly the same for the different circuits, the compile time of the FullPeepholeOptimise approach was on average 1.11 times higher. The standard deviation for all four approaches was on average about 6.53\,\% of the average runtimes. The evaluation of the runtimes for the five algorithms showed the same runtime behavior without any influence of the number of qubits.

Compiling the circuits without the block commutation with correction unitaries achieved the compile times depicted in \autoref{fig-runtimes-without-unitary_blocks}. For small circuits, the SquashTK1 and KAKDecomposition approaches resulted in compile times of about 4\,ms per gate of the original circuits. In contrast, the CliffordSimp and FullPeepholeOptimise approaches showed a nonlinear growth of the compile times with the number of gates of the original circuits, with the FullPeepholeOptimise approach having on average 2.09 times higher compile times than the CliffordSimp approach. The reasons for the nonlinear growth are the additional optimizations executed by Pytket's \texttt{CliffordSimp} and \texttt{FullPeepholeOptimise} passes, which led to nonlinear runtime scaling. As mentioned in \autoref{subsubsec-algorithm-for-executing-advanced-block-commutation}, the block commutation with correction unitaries has a nonlinear runtime because it is applied to the blocks recursively. This also led to nonlinear compile times of the SquashTK1 and KAKDecomposition approaches in the cases where the block commutation with correction unitaries was executed. In contrast, the other transformations executed in the SquashTK1 and KAKDecomposition approaches only iterate a constant number of times over each gate, resulting in linear runtimes of these two approaches when the block commutation with correction unitaries was not applied.

\begin{figure}
    \centering
	\includegraphics[width=\columnwidth,keepaspectratio]{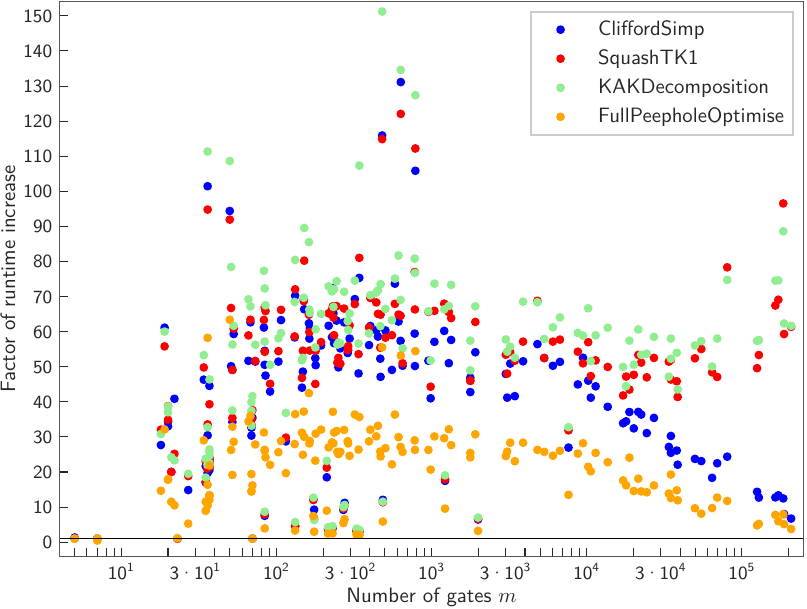}
	\caption{The runtime increases of our complete compilation flow compared to our compiler when executed without the block commutation with correction unitaries for the different rebasing approaches CliffordSimp (dark blue), SquashTK1 (red), KAKDecomposition (green), and FullPeepholeOptimise (orange) depending on the number of gates. Each dot represents one of the 153 circuits in the circuit library. For all circuits with an increase factor greater than one (the horizontal black line), our complete compilation flow has a longer runtime than our compiler when executed without the block commutation with correction unitaries.}
	\label{fig-runtime-increase}
\end{figure}

The factors of the compile time increases when executing the block commutation with correction unitaries compared to the compilation without the block commutation with correction unitaries are shown in \autoref{fig-runtime-increase}. On average, the runtimes with the block commutation with correction unitaries were 53.75 times higher for the KAKDecomposition approach, followed by the SquashTK1 approach with 48.92 times, the CliffordSimp approach with 41.49 times, and the FullPeepholeOptimse approach with 21.45 times. While the factors for the SquashTK1 and KAKDecomposition approaches are approximately constant in a certain range, the factors for the CliffordSimp and FullPeepholeOptimise approaches are approximately constant up to a circuit size of 10,000 gates, and decrease for higher number of gates.

\begin{figure}[!t]
    \begin{subfigure}{\columnwidth}
        \centering
    	\includegraphics[width=\columnwidth,keepaspectratio]{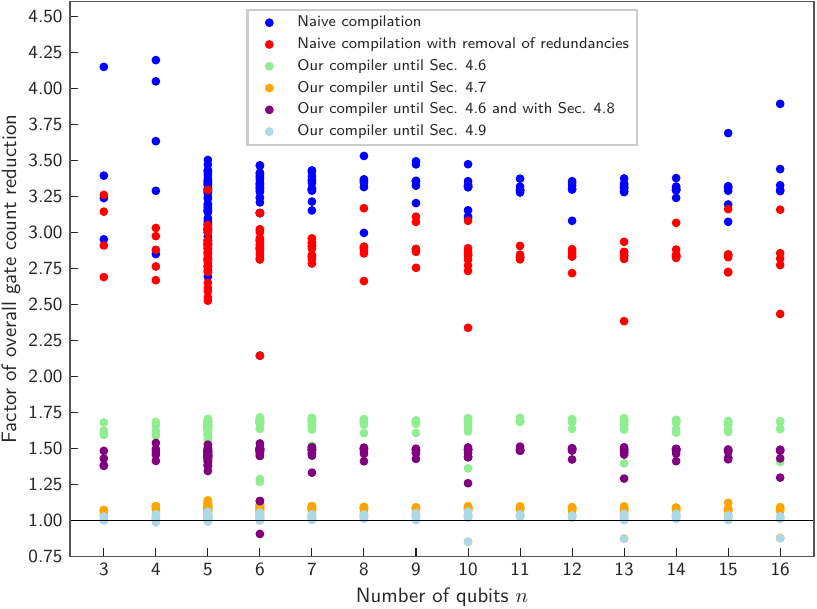}
    	\caption{The overall gate count reductions for the 153 circuits in the circuit library. The circuits are sorted by the number of qubits used.}
    	\vspace{2mm}
    	\label{fig-small-circuits-impact}
	\end{subfigure}
	\begin{subfigure}{\columnwidth}
        \centering
    	\includegraphics[width=\columnwidth,keepaspectratio]{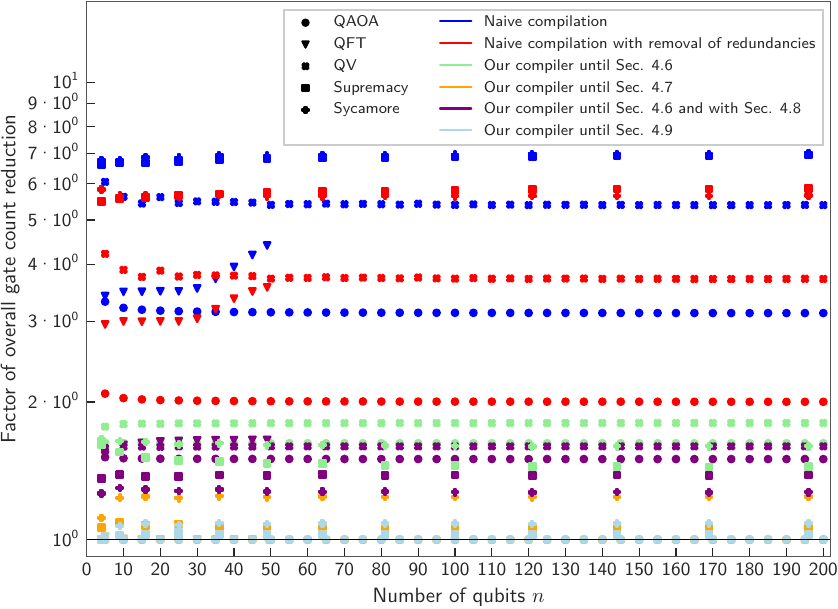}
    	\caption{The overall gate count reductions for the algorithms QAOA (circles), QFT (triangles), QV (crosses), Supremacy (squares), and Sycamore (pluses) depending on different numbers of qubits.}
    	\label{fig-large-circuits-impact}
	\end{subfigure}
	\caption{The overall gate count reductions of our complete compilation flow compared to the following less optimized compilations: (1) a naive compilation which replaces gates only with respect to our trapped-ion native gate set $\mathcal{N}$, but without any optimization (dark blue dots), (2) the naive compilation with additional removal of trivial redundancies (red dots), (3) the compilation as depicted in \autoref{fig-compiler-flowchart}, but only up to the transformation described in \autoref{subsec-transformation-to-native-gate-set} (green dots), (4) the compilation up to phase tracking (orange dots), (5) the compilation up to block aggregation but without phase tracking (purple dots), and (6) the compilation up to the commutation of blocks and blockless sequences on disjoint sets of qubits (light blue dots). For each circuit, the best compilation result of the four different rebasing approaches is used. For all circuits with a reduction factor greater than one (the horizontal black line), our complete compilation flow produces a circuit with a lower gate count than the less optimized compilations.}
	\label{fig-impact}
\end{figure}

In the following, we compare the impact of the different compilation stages on the results of the circuits. To compare the gate counts, for each circuit we used the approach (CliffordSimp, SquashTK1, KAKDecomposition, or FullPeepholeOptimise) which produced the lowest overall gate count after executing our entire compiler flow as the baseline in all evaluations. If more than one approach had the same overall gate count, we preferred the result with the lowest two-qubit gate count. The impact of the different compilation stages are depicted in \autoref{fig-small-circuits-impact} for the circuit library and in \autoref{fig-large-circuits-impact} for the five algorithms.

As a naive compilation, the gates of the original circuit were replaced using Pytket's \texttt{RebaseCustom} pass as described in \autoref{subsec-transformation-to-extended-native-gate-set} and the \gate{ZZ} decomposition given in \eqref{eq-ZZ-decomp} and \eqref{eq-ZZ-decomp-pi-3pihalf}. Afterwards, the circuit was transformed into our trapped-ion native gate set $\mathcal{N}$. Comparing this naive transformation with our compiler for the circuit library, the resulting circuits from our compiler have 2.50 to 5.38 times fewer single-qubit gates and up to 1.38 times fewer two-qubit gates. However, for 33\,\% of the circuits, our compiler produced results with up to 1.27 times more two-qubit gates. This is due to the additional gates inserted during the block commutation with correction unitaries to reduce the amount of shuttling operations. On average, our compiler produced circuits with 3.29 times fewer overall gates. For the evaluated circuits, this naive transformation produced results with about 1.10 times lower single-qubit gate counts when the \gate{Ry} was not excluded from the native gate set before executing the \texttt{RebaseCustom} pass. This is because the naive transformation need not take care of commutations, and including \gate{Ry} gates allows the gates to be replaced by shorter gate sequences.

Comparing the naive compilation with our compiler for the five algorithms shows that for QAOA and QV, the overall and single-qubit gate count reduction factors first decrease as the number of qubits increases. Then the overall gate count reduction factors converge to 3.13 for QAOA and 5.39 for QV, while the single-qubit gate count reduction factors converge to 3.43 for QAOA and 6.48 for QV. While the overall gate count reduction factors for Supremacy and Sycamore increase only slightly, the single-qubit gate count reduction factors increase from 9.85 to 11.16 for Supremacy and from 9.92 to 11.13 for Sycamore as the number of qubits increases. The reason for these large increases is that the substitution of the \gate{CZ} gates contained in the circuits used by the naive compilation requires 15 single-qubit gates, while our compiler requires only two single-qubit gates for a \gate{CZ} decomposition. Moreover, our compiler can remove the z-rotations contained in the circuit using phase tracking. While for QAOA, Supremacy, and Sycamore the gate count reduction factors for two-qubit gates are around one, meaning that the circuits calculated by our compiler have approximately as many two-qubit gates as the naive compilation, for QV the reduction factor decreases from 1.25 to a factor slightly above one. For QFT, the gate count reduction factors also increase with the number of qubits. While our compiler required 4.46 times fewer single-qubit gates and the same number of two-qubit gates compared to the naive compilation for five qubits, our compiler required 5.94 times fewer single-qubit gates and 1.27 times fewer two-qubit gates for 49 qubits. For all five algorithms, our compiler generated lower single-qubit gate counts when the \gate{Ry} was not excluded from the native gate set before the \texttt{RebaseCustom} pass was executed. In this case, the single-qubit gate counts are on average 1.04 times lower for QAOA and QV, 1.10 times lower for QFT, 1.14 times lower for Supremacy, and 1.16 times lower for Sycamore compared to a compilation without \gate{Ry} as a native gate. The higher factors for Supremacy and Sycamore are due to the fact that these circuits already contain several \gate{Ry} gates, which do not need to be replaced when the \gate{Ry} gate is added to the native gate set.

Comparing our compiler to the naive approach, but additionally removing trivial redundancies using Pytket's \texttt{RemoveRedundancies} pass, the compiled circuits in the circuit library have between 2.50 and 4.29 times fewer single-qubit gates and up to 1.13 times fewer two-qubit gates. For the same reason as for the naive approach, our compiler produced results with up to 1.27 times more two-qubit gates for 34\,\% of the circuits. The overall gate counts of our compiler are on average 2.86 times lower than the overall gate counts of the improved naive transformation. These high factors show the potential of the optimizations applied by our compiler. In the improved naive transformation, for 98\,\% of the circuits our compiler calculated results with a lower or equal gate count when the \gate{Ry} gate was excluded from the native gate set before executing the \texttt{RebaseCustom} pass. This is because excluding gates from the native gate set reduces the number of different gates. Consequently, gates of the same type are more likely to be located next to each other, allowing for better redundancy removal.

The comparison of the advanced naive compilation for QAOA, QFT, QV, and Supremacy shows the same trends as the naive compilation with lower gate count reduction factors. \Eg the overall gate count reduction factors converge to 2.00 for QAOA and 3.72 for QV as the number of qubits increases. The only exception is Sycamore, whose overall gate count reduction factor is 5.83 for four qubits and converges to 5.64 for a higher number of qubits. While the overall gate count reduction factor decreases with the number of qubits, the single-qubit gate count reduction factor increases from 8.47 to 8.89, and the two-qubit gate count reduction factors are around one as in the naive compilation. For all algorithms except QAOA, the improved naive compilation produced better results when the \gate{Ry} gate was excluded from the native gate set before executing the \texttt{RebaseCustom} pass for the reason mentioned above. For QAOA, the results of the improved naive compilation have about 1.06 times fewer single-qubit gates when including the \gate{Ry} gate. The reason for this is that the circuits benefit more from the shorter \gate{CNOT} decomposition 
\begin{align}
    \gate{CNOT}_{i, j} = &\hspace{1mm}\gate{Ry}_j\left(\tfrac{\pi}{2}\right)  \gate{Rx}_j\left(\pi\right)
    \gate{Rx}_i\left(\pi\right) \gate{ZZ}_{i, j}\left(\tfrac{\pi}{2}\right) \gate{Ry}_j\left(\tfrac{\pi}{2}\right) \nonumber \\
    \cdot&\hspace{1mm} \gate{Rx}_j\left(\tfrac{\pi}{2}\right) \gate{Ry}_i\left(\tfrac{\pi}{2}\right) \gate{Rx}_i\left(\tfrac{\pi}{2}\right) \gate{Ry}_i\left(\tfrac{\pi}{2}\right)
\end{align}
which can be used when including the \gate{Ry} gate than from the additional redundancy removal possible when excluding the \gate{Ry} gate.

For the next evaluation, our compiler was executed as depicted in \autoref{fig-compiler-flowchart}, but skipping phase tracking, block aggregation, and the block commutation with correction unitaries. Comparing our entire compiler flow with the flow skipping these three optimizations for the circuit library, our entire compiler calculated circuits with 1.38 to 2.05 times fewer single-qubit gates. For 47\,\% of the circuits, our entire compiler produced results with up to 1.27 times more two-qubit gates. Since phase tracking and block aggregation do not change the two-qubit gates, the block commutation with correction unitaries inserted the additional two-qubit gates to reduce the amount of shuttling operations. The overall gate counts of our entire compiler flow are on average 1.65 times lower than those of the reduced compiler.

For QAOA, the single-qubit reduction factors remain constant at 1.71 independently of the number of qubits. In contrast, for QFT and QV, the single-qubit gate count reduction factors increase with the number of qubits from 1.80 to 1.97 for QFT and from 1.94 to 2.00 for QV, while the factors decrease from 1.97 to 1.75 for Supremacy and from 2.05 to 2.01 for Sycamore. The overall gate count reduction factors are similar to those for single-qubit gates, but with smaller factors. For all algorithms except Sycamore, the amount of two-qubit gates for the compilation without phase tracking, block aggregation, and the block commutation with correction unitaries is as large as the amount of our entire compilation flow. Only for Sycamore, our entire compiler flow has a slightly higher two-qubit gate count than the compilation without the skipped optimizations.

In the following, phase tracking was also executed, but block aggregation and the block commutation with correction unitaries were not executed. In this case, for the circuit library, the overall gate counts of our entire compilation flow are on average only 1.07 times and the single-qubit gate counts are up to 1.20 times lower than the gate counts of the flow without block aggregation and the block commutation with correction unitaries. For the \emph{ising\_model} circuits, the single-qubit gate counts of our entire compiler flow are up to 1.15 times higher. The reason for this is that these circuits contain several \gate{Rz} gates, which phase tracking removed. However, the block commutation with correction unitaries introduced new gates to reduce shuttling operations.

Using this reduced compilation flow for QAOA and QV, the gate counts are exactly the same as for our entire compiler flow. This means that block aggregation and the block commutation with correction unitaries do not affect the gate counts of these circuits when executing the transformations after phase tracking. For QFT, the single-qubit gate count reduction factor is 1.02 for five qubits and decreases to one for 30 and more qubits. While for Supremacy the single-qubit gate count reduction factor decreases from 1.15 to 1.10 as the number of qubits increases, for Sycamore the single-qubit gate count reduction factor starts at 1.17 for the four-qubit circuit and converges to a factor of 1.41.

Compared to the compiler which did not perform phase tracking and the block commutation with correction unitaries, but performed block aggregation, the overall gate counts of our entire compiler flow for the circuit library are on average 1.45 times lower, and the single-qubit gate counts are up to 1.77 times lower. Since the reduction factors are higher than in the previous case of executing phase tracking instead of block aggregation, the impact of block aggregation is less than the impact of phase tracking. However, an exception is the circuit \mbox{\emph{graycode6\_47}}, which has a lower single-qubit gate count when not using phase tracking and the block commutation with correction unitaries. The reason for this is the construction of this circuit, which consists only of \gate{CNOT} gates. Consequently, this circuit directly benefits from the symmetric structure of the \gate{CNOT} decomposition in \eqref{eq-CNOT-decomp}, which favors block aggregations. 

For QAOA, the single-qubit gate counts of our entire compiler are about 1.57 times lower than the counts of the compilation without phase tracking and the block commutation with correction unitaries. However, these factors grow from 1.76 to 1.97 for QFT, from 1.72 to 1.75 for QV, and from 1.57 to 1.66 for Supremacy as the number of qubits increases. Only for Sycamore, they decrease slightly from 1.48 to 1.46. For QFT, the single-qubit gate counts are the same as in the compilation even without block aggregation, except for the five-qubit circuit. This shows that QFT does not benefit from block aggregation when executed without prior phase tracking. On the other hand, since the reduction factors for QAOA, QV, Supremacy, and Sycamore are lower than in the compilation even without block aggregation, these algorithms benefit from block aggregation in this case. However, because the reduction factors are higher than in the previous case of executing phase tracking instead of block aggregation, the impact of block aggregation is less pronounced than the impact of phase tracking.

\begin{figure}
    \centering
    \includegraphics[width=\columnwidth,keepaspectratio]{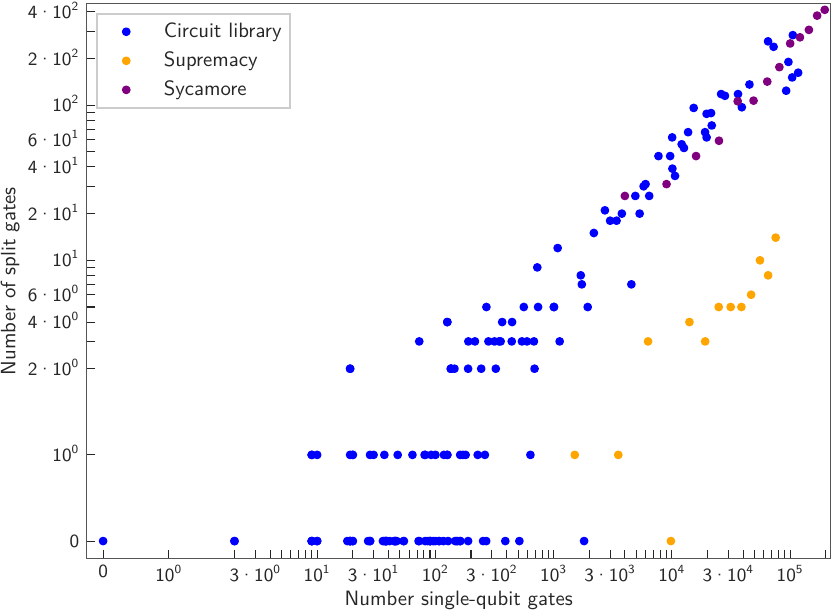}
    \caption{Number of splits performed during angle splitting in \autoref{subsubsec-angle-splitting}, depending on the number of single-qubit gates in the original circuit. For each circuit, the best compilation result of the four different rebasing approaches is used. Each dark blue dot represents at least one circuit in the circuit library, with splits applied to 97 of the 153 circuits. While angle splitting was performed on every circuit of Sycamore (purple dots), it was not applied to the 25-qubit circuit of Supremacy (orange dots). Since angle splitting was not performed on any circuit of QAOA, QFT, and QV, these circuits are not depicted.}
    \label{fig-splits}
\end{figure}

Evaluating the angle splitting in \autoref{subsubsec-angle-splitting} shows that our compiler applied this technique to 63\,\% of the circuits in the circuit library. On these circuits, our compiler performed angle splitting on 0.08\,\% to 4.65\,\% and on average on 0.72\,\% of the single-qubit gates in the original circuits. Of the five algorithms, only Supremacy and Sycamore benefit from angle splitting. The number of single-qubit gates of the original circuit which were split by angle splitting is shown in \autoref{fig-splits}. This number tends to grow with the number of single-qubit gates.

As a final case, we compare our entire compiler flow to the flow without the block commutation with correction unitaries. For the circuit library, our entire compilation flow has on average 1.02 times lower overall gate counts and up to 1.09 times lower single-qubit gate counts than the compilation without the block commutation with correction unitaries. The three \emph{ising\_model} circuits and the circuit \emph{decod24-v0\_38} have up to 1.15 times lower single-qubit gate counts when not executing the block commutation with correction unitaries. As mentioned above for 47\,\% of the circuits, performing the block commutation with correction unitaries results in up to 1.27 times higher two-qubit gate counts. The increase in the gate counts is due to the additional gates inserted during the block commutation with correction unitaries. However, the next subsection shows that the insertion of these additional gates results in a reduction of shuttling operations. Although 47\,\% of the circuits have a higher two-qubit gate count, only 3\,\% of the circuits have a higher overall gate count after executing the block commutation with correction unitaries, showing its potential to reduce single-qubit gates.

Compiling the circuits of the five algorithms without the block commutation with correction unitaries produced exactly the same results for QAOA, QFT, and QV as compiling even without block aggregation. Consequently, block aggregation has no effect on these three algorithms when executed after phase tracking. Additionally, as mentioned above, the block commutation with correction unitaries has no effect on the circuits of QAOA and QV, since the compilation without the block commutation with correction unitaries produced the same results as our entire compiler flow. However, the single-qubit gate count reduction factor for QFT decreases from 1.02 for the five-qubit circuit to one for the circuits with 30 and more qubits. This shows that for QFT, the block commutation with correction unitaries slightly reduces the single-qubit gate counts for the circuits up to 25 qubits. The single-qubit gates of our entire compiler flow are about 1.04 times less for Supremacy and 1.14 times less for Sycamore, independent of the number of qubits, than for the compilation without the block commutation with correction unitaries. Only for Sycamore, our entire compiler flow has a slightly higher number of two-qubit gates compared to the compilation without the block commutation with correction unitaries. This means that the block commutation with correction unitaries inserted some two-qubit gates to reduce the amount of shuttling operations.

\begin{figure}
    \centering
    \includegraphics[width=\columnwidth,keepaspectratio]{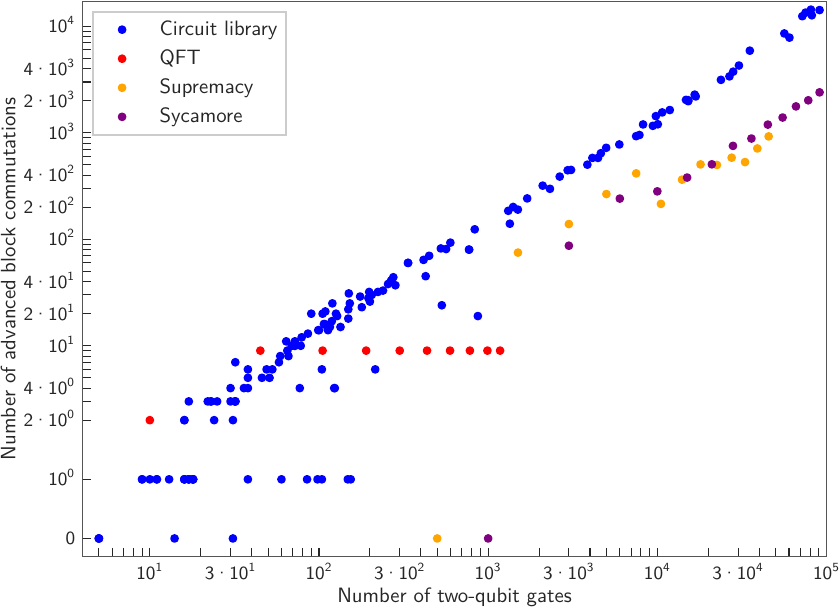}
    \caption{Number of block commutations with correction unitaries applied, depending on the number of two-qubit gates in the original circuit. For each circuit, the best compilation result of the four different rebasing approaches is used. Each dark blue dot represents at least one circuit in the circuit library where the block commutation with correction unitaries was performed on 148 of the 153 circuits. While the block commutation with correction unitaries was applied to every circuit of QFT (red dots), it was not performed on the smallest circuit of Supremacy (orange dots) and Sycamore (purple dots). Since the block commutation with correction unitaries was not applied to any circuit of QAOA and QV, these circuits are not depicted.}
    \label{fig-advanced-block-commutations}
\end{figure}

Evaluating the number of commutations executed, we found that our compiler applied commutations to 97\,\% of the circuits in the circuit library. Up to 14,345 commutations were executed on these circuits. The number of executed commutations ranges from 0.65\,\% to 21.88\,\% and averages out to 12.28\,\% of the number of two-qubit gates in the original circuits. For the five algorithms, our compiler only executed commutations on the circuits of QFT, Supremacy, and Sycamore. While the number of commutations grows with the number of two-qubit gates for the circuits in the circuit library, Supremacy, and Sycamore, for QFT our compiler executed two commutations on the five-qubit circuit and applied nine commutations constantly to the circuits with a higher number of qubits. The number of commutations depending on the number of two-qubit gates in depicted in \autoref{fig-advanced-block-commutations}.

\subsection{Impact on the shuttling}
\label{subsec-impact-on-the-shuttling}
In the following, we analyze the impact of the commutation of blocks and blockless sequences on disjoint sets of qubits and of the block commutation with correction unitaries on the shuttling by comparing the required number of translate, separation/merge, and physical ion swap operations. Therefore, we arranged as many ions in a linear chain as the circuit to be compiled has qubits. The chain was modeled as a linear graph with each vertex representing an ion. To map the logical qubits of the circuit to the ions, we used two different Pytket passes, the \texttt{GraphPlacement} pass and the \texttt{LinePlacement} pass, which use different heuristics. For each circuit, we executed both passes and obtained a mapping for each pass. Afterwards, we used Pytket's \texttt{RoutingPass} on both mappings, which inserts \gate{SWAP} gates so that two-qubit gates are only executed on ions whose vertices are adjacent.
The Shuttling Compiler executed all \gate{SWAP} gates inserted by Pytket as physical ion swaps. Due to the two mappings, we got two results, from which we selected the result with the lower number of \gate{SWAP} gates for submission to the Shuttling Compiler \cite{umz-shuttling}.

\begin{figure}[!t]
    \begin{subfigure}{\columnwidth}
        \centering
    	\includegraphics[width=\columnwidth,keepaspectratio]{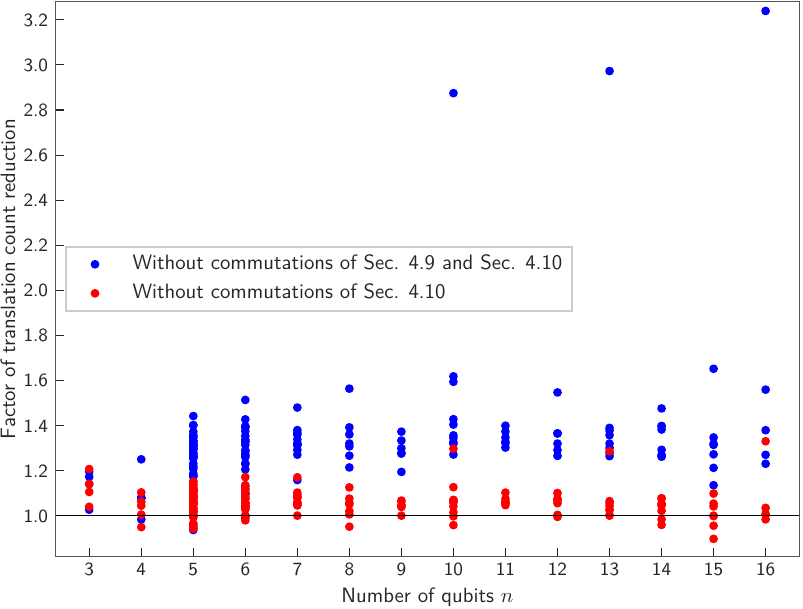}
    	\caption{The translation count reductions for the 153 circuits in the circuit library. The circuits are sorted by the number of qubits used.}
    	\vspace{2mm}
    	\label{fig-small-circuits-translations}
    \end{subfigure}
    \begin{subfigure}{\columnwidth}
        \centering
    	\includegraphics[width=\columnwidth,keepaspectratio]{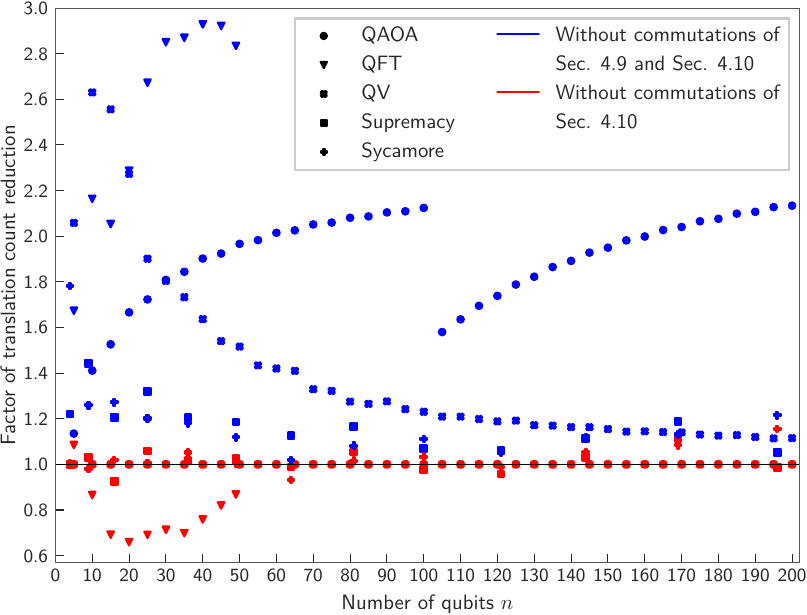}
    	\caption{The translation count reductions for the algorithms QAOA (circles), QFT (triangles), QV (crosses), Supremacy (squares), and Sycamore (pluses) depending on different numbers of qubits.}
    	\label{fig-large-circuits-translations}
    \end{subfigure}
    \caption{The translation count reductions of our complete compilation flow compared to (1) the compilation without the commutation of blocks and blockless sequences on disjoint sets of qubits as well as without the block commutation with correction unitaries (blue dots) and (2) only without the block commutation with correction unitaries (red dots). For each circuit, the translations for the best compilation result of the four different rebasing approaches were calculated. For all circuits with a reduction factor greater than one (the horizontal black line), our complete compilation flow produces a circuit with a lower translation count than the less commuted compilation flows.}
    \label{fig-translations}
\end{figure}

When both types of commutations were executed, 97\,\% of the circuits in the circuit library required fewer or the same number of translations as when both types of commutations were not executed. The commutations reduced the number of translations by factors up to 3.24, and on average by a factor of 1.33. Compared to a compiler which executed the commutation of blocks and blockless sequences on disjoint sets of qubits, but not the block commutation with correction unitaries, our compiler calculated results which lead to up to 1.33 times and on average 1.06 times fewer translations for 89\,\% of the circuits. The impact of the commutations on the amount of translations for the circuits in the circuit library is depicted in \autoref{fig-small-circuits-translations}. The differences in the factors show that the commutation of blocks and blockless sequences on disjoint sets of qubits has a higher impact on the translation counts. However, the commutation of blocks and blockless sequences on disjoint sets of qubits has no effect on the amount of separation/merge and ion swap operations. Using the block commutation with correction unitaries also reduced the amount of separation/merge operations for 94\,\% of the circuits by factors up to 1.20 and on average by 1.04. Additionally, it reduced the number of ion swaps for 97\,\% of the circuits by factors up to 1.67 and on average by 1.12.

The impact of both types of commutations on the five algorithms is shown in \autoref{fig-large-circuits-translations}. As mentioned in the last subsection, the compiler did not execute the block commutation with correction unitaries on the circuits of QAOA and QV. Moreover, as for the circuit library, the commutation of blocks and blockless sequences on disjoint sets of qubits did not affect the amount of separation/merge and ion swap operations for all five circuits. Consequently, for QAOA and QV, only the impact of the commutation of blocks and blockless sequences on disjoint sets of qubits on the translation count needs to be considered. For QV, the result of our entire compiler flow requires 2.63 times fewer translations for ten qubits compared to a compilation without the commutation of blocks and blockless sequences on disjoint sets of qubits. This translation count reduction factor decreases to 1.11 for 200 qubits. In contrast, for QAOA, the translation count reduction factor is 1.13 for five qubits and increases to 2.12 for 100 qubits. From there, it decreases rapidly to 1.58 for 105 qubits and increases again to 2.13 for 200 qubits. The reason for the rapid decrease is Pytket's placement pass, which we used to map the logical qubits to ions. The qubits in the QAOA algorithm have a nearest neighbor interaction. This means that the qubit $q_0$ interacts only with the qubit $q_1$, the qubit $q_{n-1}$ only with the qubit $q_{n-2}$, and all other qubits $q_i$ only with the qubits $q_{i-1}$ and $q_{i+1}$. This allows to map the logical qubits to the ion chain in ascending order. \Ie $q_0$ is mapped to the leftmost ion of the ion chain, $q_1$ is mapped to the ion next to $q_0$ on the right, and so on until $q_{n-1}$ is mapped to the rightmost ion of the chain. Up to the 100-qubit circuit, Pytket's placement pass could map the logical qubits exactly according to this scheme. However, when the circuits used more than 100 qubits, the placement pass mapped $q_{n-1}$ to the leftmost ion, $q_{n-2}$ to the ion next to $q_{n-1}$ on the right, and so on until the qubit $q_{100}$ was mapped. To the ion to the right of $q_{100}$ it mapped the qubit $q_0$, and from there it mapped all the qubits in ascending order to the ions until it reached the qubit $q_{99}$. Since $q_{99}$ also interacts with $q_{100}$, which the placement pass mapped 100 ions to the left, $q_{99}$ had to be swapped with 99 intermediate ions to execute the gate with $q_{100}$. The placement pass placed the ions according to this scheme, regardless of whether commutations had been used. While the additional ion swap and separation/merge operations did not affect the reduction factors of these operations, the additional translations resulted in a decreasing translation count reduction factor between the 100- and 105-qubit circuits.

For QFT, our entire compilation required between 1.67 and 2.93 times fewer translations than the compilation without both types of commutations. When executing the commutation of blocks and blockless sequences on disjoint sets of qubits, but not the block commutation with correction unitaries, the compiler required up to 1.52 times fewer translations and up to 1.22 times fewer seperation/merge operations for the circuits with 10 and more qubits compared to our entire compilation flow. Moreover, the number of ion swaps is up to 1.64 times higher for our entire compilation flow.

Furthermore, for both Supremacy and Sycamore, our entire compiler flow calculated circuits requiring fewer translations than a compilation without both types of commutations. For Supremacy, our entire compiler required between 1.05 and 1.44 times fewer translations, while for Sycamore it required between 1.02 and 1.78 times fewer. In this range, the factors alternate between the different numbers of qubits. When we compare our entire compiler flow with a compiler which used the commutation of blocks and blockless sequences on disjoint sets of qubits, but not the block commutation with correction unitaries, our compiler generated results with a lower number of translations for 54\,\% of the Supremacy circuits and 77\,\% of the Sycamore circuits. Thus, our entire compiler calculated results with up to 1.11 times fewer and up to 1.08 times more translations for Supremacy, and with up to 1.15 times fewer and up to 1.07 times more translations for Sycamore. For both algorithms, the factors alternate between the different numbers of qubits in these ranges. Regarding the amount of separation/merge and ion swap operations, our entire compiler flow calculated results with fewer separation/merge and ion swap operations for half of the circuits of Supremacy and Sycamore. While the reduction factors for separation/merge range from 0.95 to 1.08 for Supremacy and from 0.96 to 1.03 for Sycamore, the reduction factors for ion swap range from 0.93 to 1.09 for Supremacy and from 0.95 to 1.04 for Sycamore. For all circuits where our entire compiler flow produced a result with a lower number of separations/merge operations, the result also has a lower number of ion swaps. The same applies to the case where our entire compiler flow calculated a result with a higher number of separations/merge and ion swap operations.

\subsection{Comparison to other compilers}
\label{subsec-comparison-other-compilers}
We benchmarked our compiler against built-in Pytket passes, first using its \texttt{FullPeepholeOptimise} pass. Then we rebased the circuits to $\left\{ \gate{Rx}, \gate{Rz}, \gate{ZZ} \right\}$ using Pytket's \texttt{RebaseCustom} pass with the \gate{TK1} decomposition in \eqref{eq-tk1-decomp} and the \gate{CNOT} decomposition in \eqref{eq-CNOT-decomp}. There are exceptions for the circuits \emph{graycode6\_47}, \emph{ex1\_226}, and \emph{xor5\_254}, as well as all QAOA circuits and Sycamore circuits with nine and more qubits, which we rebased to $\left\{ \gate{Rx}, \gate{Ry}, \gate{Rz}, \gate{ZZ} \right\}$ because Pytket calculated significantly better results for these circuits with \gate{Ry} gate than without \gate{Ry} gate. After rebasing, we applied the repeated removal and commutation of gates based on Pytket's built-in procedures from \autoref{subsec-repeated-removing-and-commutation-of-gates}. Since Pytket does not allow the gates of the native gate set to be constrained to certain angles, it cannot directly transform the gates to our trapped-ion native gate set $\mathcal{N}$. To make the gate counts comparable, we applied the transformation from \autoref{subsec-transformation-to-native-gate-set} to the circuits.

For the Qiskit compilation, we first applied the \gate{ZZ} decomposition given in \eqref{eq-ZZ-decomp} and \eqref{eq-ZZ-decomp-pi-3pihalf} to the circuits, because Qiskit cannot transform \gate{ZZ} gates with $\theta \neq \tfrac{\pi}{2}$ already in the circuits into \gate{ZZ} gates with $\theta = \tfrac{\pi}{2}$ and a set of single-qubit gates. Afterwards, we applied Qiskit's \texttt{BasisTranslator} pass with the target basis $\left\{ \gate{Rx}, \gate{Ry}, \gate{Rz}, \gate{ZZ} \right\}$. The only exceptions are the circuits \emph{graycode6\_47}, \emph{ex1\_226}, and \emph{xor5\_254}, for which the \gate{Ry} gate was removed from the target basis because Qiskit calculated better results for these circuits without \gate{Ry} gate than with \gate{Ry} gate. After rebasing, we executed the \texttt{transpile} function of the Qiskit compiler module with optimization level three. Like Pytket, Qiskit cannot transform the gates to our trapped-ion native gate set $\mathcal{N}$, so we used \autoref{subsec-transformation-to-native-gate-set} to get comparable gate counts.

\begin{figure}[!t]
    \begin{subfigure}{\columnwidth}
        \centering
    	\includegraphics[width=\columnwidth,keepaspectratio]{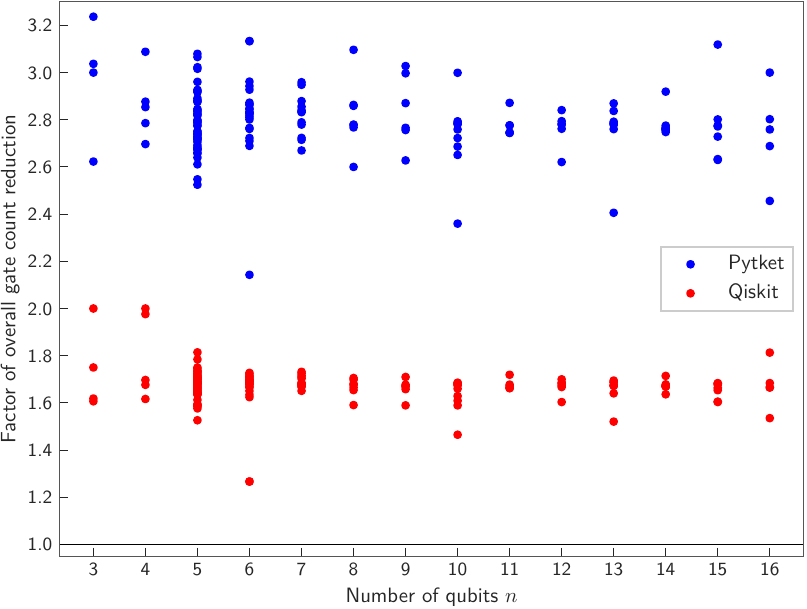}
    	\caption{The overall gate count reductions for the 153 circuits in the circuit library. The circuits are sorted by the number of qubits used.}
    	\vspace{2mm}
    	\label{fig-small-circuits-gate-reduction}
    \end{subfigure}
    \begin{subfigure}{\columnwidth}
        \centering
    	\includegraphics[width=\columnwidth,keepaspectratio]{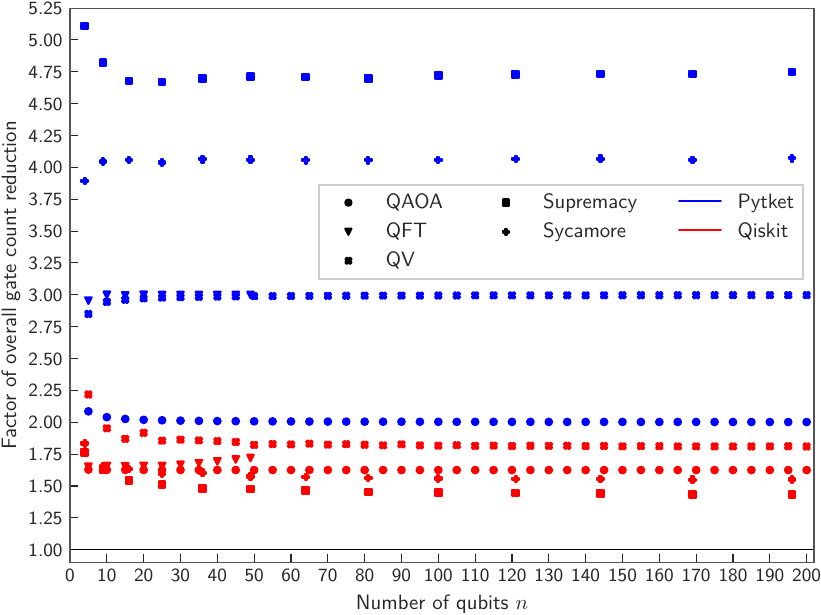}
    	\caption{The overall gate count reductions for the algorithms QAOA (circles), QFT (triangles), QV (crosses), Supremacy (squares), and Sycamore (pluses) depending on different numbers of qubits.}
    	\label{fig-large-circuits-gate-reduction}
    \end{subfigure}
    \caption{The overall gate count reductions of our compiler compared to Pytket (dark blue dots) and Qiskit (red dots) standard passes. As result for our compiler, the compilation result with the lowest gate count of the four different rebasing approaches is used for each circuit. For all circuits with a reduction factor greater than one (the horizontal black line), our compiler produces a circuit with a lower gate count than either Pytket or Qiskit.}
    \label{fig-gate-reduction}
\end{figure}

The gate count reductions achieved by our compiler compared to Pytket and Qiskit are summarized in \autoref{fig-small-circuits-gate-reduction} for the circuit library and in \autoref{fig-large-circuits-gate-reduction} for the five algorithms. 

Compared to standard Pytket, our compiler always computed results with lower gate counts. For the circuit library, there are between 2.14 and 3.24 times fewer gates. A more detailed examination shows that for all circuits our compiler produced results with 2.50 to 4.20 times fewer single-qubit gates, while the two-qubit gate counts are only up to 1.08 times less or equal for 51\,\% of the circuits. The main reason for the small fraction of less or equal two-qubit gate counts is the block commutation with correction unitaries, which added additional two-qubit gates to the circuit to reduce shuttling operations. Without the block commutation with correction unitaries, our compiler calculated results with fewer two-qubit gates for 84\,\% of the circuits.

For QAOA, our compiler required between 2.23 and 2.14 times fewer single-qubit gates than Pytket, with the factor decreasing as the number of qubits increases. In contrast, the reduction factors increase with the number of qubits from 3.80 to 3.99 for QFT and from 3.26 to 3.50 for QV. For all three algorithms, the results of our compiler have as many two-qubit gates as Pytket's results, independently of the number of qubits. The single-qubit gate count reduction factor for Supremacy increases from 7.02 for 16 qubits to 7.42 for 196 qubits. For Sycamore, our compiler calculated circuits with 5.49 to 6.20 times fewer single-qubit gates than Pytket, with the factor increasing with the number of qubits. The two-qubit gate count reduction factors for both algorithms are the same as for the naive compilation in \autoref{subsec-analysis-impact-different-transformations}, showing that Pytket calculated the same amount of two-qubit gates. Thus, for the Sycamore circuits and half of the Supremacy circuits, Pytket produced results with slightly lower two-qubit gate counts. Again, the reason for the higher two-qubit gate counts of our compiler is the block commutation with correction unitaries. Without it, the two-qubit gate counts of our compiler and Pytket are the same for all five algorithms, independently of the number of qubits.

The main reason for Pytket's higher single-qubit gate counts is that it cannot apply phase tracking. Due to this limitation, the circuits contain several \gate{Rz} gates, which in our compilation flow can only appear at the end of the circuit. Additionally, the \gate{Rz} gates are often sandwiched by \gate{Rx} gates, which prevents the \gate{Rx} gates from being merged and removed when their angle is $\theta = 0$. Since we had to execute all four approaches of our compiler to find the best result, and since one of our approaches also includes Pytket's \texttt{FullPeepholeOptimise} pass, the runtimes of our compiler were always worse than Pytket's runtimes.

The comparison with standard Qiskit shows for the circuit library that our compilation leads to 1.27 to 2.00 times fewer gates. Furthermore, the number of single-qubit gates is between 1.40 and 2.34 times less for our compiler, and the number of two-qubit gates is between 0.79 and 1.25 times less, with our compiler requiring a higher number of two-qubit gates for 33\,\% of the circuits. As in the Pytket comparison, the reason for the higher number of two-qubit gates is the block commutation with correction unitaries. Without executing it, our compiler needed fewer two-qubit gates for all circuits with reduction factors up to 1.25.

For QAOA, our compiler calculated circuits with 1.71 times fewer single-qubit gates and the same number of two-qubit gates compared to Qiskit, independently of the number of qubits. In contrast, for QFT, the single-qubit gate count reduction factor increases from 1.93 to 2.01 with the number of qubits. Regarding the amount of two-qubit gates, our compiler and Qiskit produced results with the same number of two-qubit gates for the circuits up to 25 qubits, while for 30 and more qubits the results of our compiler have between 1.01 and 1.14 times fewer two-qubit gates, with the factor increasing with the number of qubits. For the other three algorithms, the single-qubit gate count reduction factors decrease from 2.43 to 2.01 for QV, from 2.21 to 1.74 for Supremacy, and from 2.29 to 1.94 for Sycamore as the number of qubits increases. The two-qubit gate count reduction factors for all three algorithms are the same as for the naive compilation comparison in \autoref{subsec-analysis-impact-different-transformations}. Thus, Qiskit's results have the same number of two-qubit gates as the naive approach. This means that for QV, the amount of two-qubit gates of our compiler is between 1.01 and 1.25 times lower than Qiskit's, with a decreasing factor as the number of qubits increases. Moreover, for Supremacy and Sycamore, Qiskit's results have exactly the same number of two-qubit gates as Pytket. This means that for the Sycamore circuits and half of the Supremacy circuits, Qiskit produced results with slightly lower two-qubit gate counts for the same reason mentioned above.

The main reason for the single-qubit gate count reductions achieved by our compiler is the same as for Pytket. Regarding the runtime, Qiskit was always faster than our compiler. However, Qiskit's faster runtime comes at the cost of a larger number of gates.

In summary, our compiler achieved larger reductions compared to Pytket, with an average reduction factor of 2.80 for the circuit library, while the average reduction factor compared to Qiskit is only 1.67. Compared to both standard passes, our compiler always computed circuits with lower overall and single-qubit gate counts.

\section{Conclusion and outlook}
\label{sec-conclusion}
We have presented a quantum circuit compiler for a shuttling-based ion trap quantum computer that implements a number of optimization techniques such as phase tracking, block aggregation, and a block commutation with correction unitaries. These reduce the number of gates and shuttling operations when compiling a quantum circuit into the native gate set. Compared to other compilation algorithms, our compiler reduces the gate counts by factors of up to 5.11 for Pytket and up to 2.22 for Qiskit.

In the future, we plan to extend the functionality of our compiler, increase its performance using more advanced compilation techniques, and adapt it to more powerful architectures.
The next architectural step will be to adapt it to a platform which allows simultaneous addressed manipulation of larger subsets of commonly confined trapped-ion qubits, where gates can be executed in parallel.
This will allow low-resource quantum computation on sub-registers, interleaved with time-consuming register reconfiguration steps. Future multiplexed laser addressing units \cite{PRXQuantum.2.020343} will also allow the execution of multi-qubit gates, so that \eg three-qubit Toffoli gates can be realized within a single laser interaction sequence \cite{PhysRevA.101.032330} instead of being decomposed into two-qubit gates. This can massively increase the quantum computational power of a trapped-ion platform, but also requires an advanced compilation layer.

\section{Acknowledgements}
We acknowledge funding by the German Federal Ministry of Education and Research (BMBF) via the VDI within the projects HFAK and IQuAn. The research is based upon work supported by the Office of the Director of National Intelligence (ODNI), Intelligence Advanced Research Projects Activity (IARPA), via the U.S. Army Research Office grant W911NF-16-1-0070.
The views and conclusions contained herein are those of the authors and should not be interpreted as necessarily representing the official policies or endorsements, either expressed or implied, of the ODNI, IARPA, or the U.S. Government. The U.S. Government is authorized to reproduce and distribute reprints for Governmental purposes notwithstanding any copyright annotation thereon. Any opinions, findings, and conclusions or recommendations expressed in this material are those of the author(s) and do not necessarily reflect the view of the U.S. Army Research Office.

\bibliographystyle{quantum}
\bibliography{references}

\clearpage
\onecolumn
\appendix
\appendixpage
\addappheadtotoc
\section{Detailed evaluation results}
\label{sec-detailed-evaluation-results}
This appendix presents the detailed results of the evaluations described in \autoref{subsec-analysis-impact-different-transformations} and \autoref{subsec-comparison-other-compilers}. For the 153 quantum circuits in the circuit library, \autoref{table-results-1} shows the results for the circuits with three to five qubits, and \autoref{table-results-2} shows the results for the circuits with six to sixteen qubits. We took all these circuits from \cite{circuit_repo}. For QAOA, QFT, and QV, \autoref{table-results-large-circuits-1} shows the results for various numbers of qubits, while \autoref{table-results-large-circuits-2} shows the results for Supremacy and Sycamore. While we generated the circuits for QAOA, QFT, Supremacy, and Sycamore using the quantum circuit generator from \cite{large_circuit_repo}, we generated the QV circuits using Qiskit \cite{qiskit}.

\begin{table}[ht]
\tiny
\caption{Results of the evaluations for the circuits in the circuit library with three to five qubits. The first column shows the name, number of qubits (q), single-qubit gate count (1qg), and two-qubit gate count (2qg) of the original circuit. The next four columns show the single-qubit and two-qubit gate counts for our CliffordSimp, SquashTK1, KAKDecomposition, and FullPeepholeOptimise approaches. The green markers show which approaches have the lowest single-qubit and two-qubit gate counts for each circuit. If multiple approaches have the same single-qubit or two-qubit gate count, the approach with the lowest overall gate count is marked. The last two columns show the results of the standard Pytket and Qiskit passes. In each of these columns, the first two entries show the single-qubit and two-qubit gate counts. The last entry in each column shows the factors by which the best approach of our compiler reduces the overall gate count compared to Pytket or Qiskit (rg).}
\setlength{\tabcolsep}{2.31mm}
\begin{center}
\begin{tabular}{lrrr|*{4}{rr|}HHHrrr|rrr}
\toprule
\multicolumn{4}{c|}{Original Circuit}&\multicolumn{2}{c|}{\emph{CliffordSimp}}&\multicolumn{2}{c|}{\emph{SquashTK1}}&\multicolumn{2}{c|}{\emph{KAK}}&\multicolumn{2}{c|}{\emph{FullPeephole}}&\multicolumn{3}{H}{AQT}&\multicolumn{3}{c|}{Pytket}&\multicolumn{3}{c}{Qiskit} \\
Name&q&1qg&2qg&1qg&2qg&1qg&2qg&1qg&2qg&1qg&2qg&1qg&2qg&rg&1qg&2qg&rg&1qg&2qg&rg \\
\midrule
ex-1\_166&3&10&9&\cellcolor{green}20&\cellcolor{green}8&22&9&22&9&\cellcolor{green}20&\cellcolor{green}8&24&8&1.14&76&8&3.00&40&9&1.75\\
ham3\_102&3&9&11&19&\cellcolor{green}8&\cellcolor{green}18&9&\cellcolor{green}18&9&19&\cellcolor{green}8&22&7&1.07&74&8&3.04&44&10&2.00\\
3\_17\_13&3&19&17&\cellcolor{green}44&\cellcolor{green}17&\cellcolor{green}44&\cellcolor{green}17&\cellcolor{green}44&\cellcolor{green}17&\cellcolor{green}44&\cellcolor{green}17&48&17&1.07&143&17&2.62&81&17&1.61\\
miller\_11&3&27&23&54&22&\cellcolor{green}53&23&\cellcolor{green}53&23&61&\cellcolor{green}19&55&19&0.97&227&19&3.24&100&23&1.62\\
4gt11\_84&4&9&9&\cellcolor{green}24&\cellcolor{green}9&\cellcolor{green}24&\cellcolor{green}9&\cellcolor{green}24&\cellcolor{green}9&\cellcolor{green}24&\cellcolor{green}9&23&8&0.94&80&9&2.70&47&9&1.70\\
rd32-v0\_66&4&18&16&\cellcolor{green}30&\cellcolor{green}12&\cellcolor{green}30&\cellcolor{green}12&\cellcolor{green}30&\cellcolor{green}12&\cellcolor{green}30&\cellcolor{green}12&29&10&0.93&105&12&2.79&69&14&1.98\\
rd32-v1\_68&4&20&16&\cellcolor{green}29&\cellcolor{green}12&\cellcolor{green}29&\cellcolor{green}12&\cellcolor{green}29&\cellcolor{green}12&\cellcolor{green}29&\cellcolor{green}12&29&10&0.95&105&12&2.85&68&14&2.00\\
decod24-v0\_38&4&28&23&\cellcolor{green}48&\cellcolor{green}20&53&22&53&22&\cellcolor{green}48&\cellcolor{green}20&62&20&1.21&190&20&3.09&92&22&1.68\\
decod24-v2\_43&4&30&22&\cellcolor{green}51&\cellcolor{green}22&54&22&54&22&\cellcolor{green}51&\cellcolor{green}22&56&21&1.05&188&22&2.88&96&22&1.62\\
4mod5-v0\_20&5&10&10&\cellcolor{green}27&\cellcolor{green}10&\cellcolor{green}27&\cellcolor{green}10&\cellcolor{green}27&\cellcolor{green}10&\cellcolor{green}27&\cellcolor{green}10&30&9&1.05&89&10&2.68&49&10&1.59\\
4mod5-v1\_22&5&10&11&\cellcolor{green}31&\cellcolor{green}11&\cellcolor{green}31&\cellcolor{green}11&\cellcolor{green}31&\cellcolor{green}11&\cellcolor{green}31&\cellcolor{green}11&27&9&0.86&96&11&2.55&60&11&1.69\\
mod5d1\_63&5&9&13&\cellcolor{green}29&\cellcolor{green}12&31&14&31&14&31&14&33&13&1.12&107&13&2.93&55&13&1.66\\
4gt11\_83&5&9&14&\cellcolor{green}30&\cellcolor{green}14&\cellcolor{green}30&\cellcolor{green}14&\cellcolor{green}30&\cellcolor{green}14&\cellcolor{green}30&\cellcolor{green}14&24&11&0.80&119&14&3.02&63&14&1.75\\
4gt11\_82&5&9&18&\cellcolor{green}35&\cellcolor{green}16&37&18&37&18&\cellcolor{green}35&\cellcolor{green}16&24&12&0.71&135&16&2.96&73&18&1.78\\
4mod5-v0\_19&5&19&16&\cellcolor{green}40&\cellcolor{green}16&\cellcolor{green}40&\cellcolor{green}16&\cellcolor{green}40&\cellcolor{green}16&\cellcolor{green}40&\cellcolor{green}16&50&16&1.18&133&16&2.66&79&16&1.70\\
mod5mils\_65&5&19&16&\cellcolor{green}45&\cellcolor{green}16&\cellcolor{green}45&\cellcolor{green}16&\cellcolor{green}45&\cellcolor{green}16&\cellcolor{green}45&\cellcolor{green}16&51&16&1.10&138&16&2.52&84&16&1.64\\
4mod5-v1\_24&5&20&16&\cellcolor{green}41&\cellcolor{green}16&\cellcolor{green}41&\cellcolor{green}16&\cellcolor{green}41&\cellcolor{green}16&\cellcolor{green}41&\cellcolor{green}16&48&16&1.12&137&16&2.68&71&16&1.53\\
alu-v0\_27&5&19&17&\cellcolor{green}42&\cellcolor{green}17&\cellcolor{green}42&\cellcolor{green}17&\cellcolor{green}42&\cellcolor{green}17&\cellcolor{green}42&\cellcolor{green}17&51&17&1.15&148&17&2.80&84&17&1.71\\
alu-v1\_28&5&19&18&\cellcolor{green}43&\cellcolor{green}18&44&18&44&18&\cellcolor{green}43&\cellcolor{green}18&51&18&1.13&155&18&2.84&86&18&1.70\\
alu-v1\_29&5&20&17&\cellcolor{green}44&\cellcolor{green}17&\cellcolor{green}44&\cellcolor{green}17&\cellcolor{green}44&\cellcolor{green}17&\cellcolor{green}44&\cellcolor{green}17&51&17&1.11&149&17&2.72&83&17&1.64\\
alu-v2\_33&5&20&17&\cellcolor{green}43&\cellcolor{green}17&\cellcolor{green}43&\cellcolor{green}17&\cellcolor{green}43&\cellcolor{green}17&\cellcolor{green}43&\cellcolor{green}17&56&17&1.22&148&17&2.75&87&17&1.73\\
alu-v3\_35&5&19&18&43&\cellcolor{green}17&\cellcolor{green}42&18&\cellcolor{green}42&18&43&\cellcolor{green}17&50&17&1.12&146&17&2.72&83&18&1.68\\
alu-v4\_37&5&19&18&43&\cellcolor{green}17&\cellcolor{green}42&18&\cellcolor{green}42&18&43&\cellcolor{green}17&50&17&1.12&146&17&2.72&83&18&1.68\\
alu-v3\_34&5&28&24&\cellcolor{green}61&\cellcolor{green}24&\cellcolor{green}61&\cellcolor{green}24&\cellcolor{green}61&\cellcolor{green}24&\cellcolor{green}61&\cellcolor{green}24&67&24&1.07&206&24&2.71&110&24&1.58\\
mod5d2\_64&5&28&25&\cellcolor{green}65&\cellcolor{green}25&\cellcolor{green}65&\cellcolor{green}25&\cellcolor{green}65&\cellcolor{green}25&\cellcolor{green}65&\cellcolor{green}25&73&25&1.09&210&25&2.61&118&25&1.59\\
4gt13\_92&5&36&30&\cellcolor{green}75&\cellcolor{green}30&76&31&76&31&76&31&82&30&1.07&249&30&2.66&137&30&1.59\\
4gt13-v1\_93&5&38&30&\cellcolor{green}70&\cellcolor{green}29&70&30&70&30&70&30&86&29&1.16&249&30&2.82&136&30&1.68\\
4mod5-v0\_18&5&38&31&\cellcolor{green}80&\cellcolor{green}31&\cellcolor{green}80&\cellcolor{green}31&\cellcolor{green}80&\cellcolor{green}31&\cellcolor{green}80&\cellcolor{green}31&78&28&0.95&262&31&2.64&148&31&1.61\\
4mod5-v1\_23&5&37&32&\cellcolor{green}70&\cellcolor{green}32&\cellcolor{green}70&\cellcolor{green}32&\cellcolor{green}70&\cellcolor{green}32&\cellcolor{green}70&\cellcolor{green}32&87&32&1.17&266&32&2.92&140&32&1.69\\
one-two-three-v2\_100&5&37&32&\cellcolor{green}70&\cellcolor{green}31&71&32&71&32&71&32&88&31&1.18&260&32&2.89&141&32&1.71\\
one-two-three-v3\_101&5&38&32&\cellcolor{green}75&\cellcolor{green}32&\cellcolor{green}75&\cellcolor{green}32&\cellcolor{green}75&\cellcolor{green}32&\cellcolor{green}75&\cellcolor{green}32&92&32&1.16&270&32&2.82&143&32&1.64\\
4gt5\_75&5&45&38&\cellcolor{green}94&\cellcolor{green}38&\cellcolor{green}94&\cellcolor{green}38&\cellcolor{green}94&\cellcolor{green}38&\cellcolor{green}94&\cellcolor{green}38&107&37&1.09&315&38&2.67&179&38&1.64\\
alu-v0\_26&5&46&38&\cellcolor{green}92&\cellcolor{green}38&93&38&93&38&\cellcolor{green}92&\cellcolor{green}38&105&38&1.10&312&38&2.69&175&38&1.64\\
rd32\_270&5&48&36&90&37&93&37&93&37&\cellcolor{green}79&\cellcolor{green}34&95&34&1.14&314&34&3.08&169&36&1.81\\
decod24-v1\_41&5&47&38&\cellcolor{green}86&\cellcolor{green}38&92&38&92&38&88&38&107&37&1.16&337&37&3.02&171&38&1.69\\
4gt5\_76&5&45&46&\cellcolor{green}99&\cellcolor{green}46&\cellcolor{green}99&\cellcolor{green}46&\cellcolor{green}99&\cellcolor{green}46&\cellcolor{green}99&\cellcolor{green}46&106&36&0.98&377&46&2.92&207&46&1.74\\
4gt13\_91&5&54&49&110&49&110&49&\cellcolor{green}109&\cellcolor{green}47&110&47&122&45&1.07&396&47&2.84&220&49&1.72\\
4gt13\_90&5&54&53&114&51&115&53&115&51&\cellcolor{green}113&\cellcolor{green}49&122&46&1.04&412&49&2.85&228&53&1.73\\
alu-v4\_36&5&64&51&\cellcolor{green}113&49&\cellcolor{green}113&49&\cellcolor{green}113&49&116&\cellcolor{green}48&135&48&1.13&410&48&2.83&226&50&1.70\\
4gt5\_77&5&73&58&\cellcolor{green}135&\cellcolor{green}59&\cellcolor{green}135&\cellcolor{green}59&\cellcolor{green}135&\cellcolor{green}59&\cellcolor{green}135&\cellcolor{green}59&161&57&1.12&473&58&2.74&263&58&1.65\\
one-two-three-v1\_99&5&73&59&\cellcolor{green}135&\cellcolor{green}58&135&59&135&59&\cellcolor{green}135&\cellcolor{green}58&163&58&1.15&498&58&2.88&262&59&1.66\\
one-two-three-v0\_98&5&81&65&155&65&156&66&156&66&\cellcolor{green}147&\cellcolor{green}64&173&62&1.11&544&64&2.88&294&65&1.70\\
4gt10-v1\_81&5&82&66&\cellcolor{green}157&\cellcolor{green}67&\cellcolor{green}157&\cellcolor{green}67&\cellcolor{green}157&\cellcolor{green}67&\cellcolor{green}157&\cellcolor{green}67&177&65&1.08&544&66&2.72&301&66&1.64\\
decod24-v3\_45&5&86&64&\cellcolor{green}157&\cellcolor{green}65&158&65&158&65&\cellcolor{green}157&\cellcolor{green}65&179&64&1.09&530&64&2.68&288&64&1.59\\
aj-e11\_165&5&82&69&161&69&165&69&165&69&\cellcolor{green}157&\cellcolor{green}68&185&68&1.12&579&68&2.88&311&69&1.69\\
4mod7-v0\_94&5&90&72&\cellcolor{green}161&72&\cellcolor{green}161&72&163&\cellcolor{green}70&168&70&192&69&1.12&582&70&2.80&329&72&1.72\\
alu-v2\_32&5&91&72&\cellcolor{green}172&\cellcolor{green}72&\cellcolor{green}172&\cellcolor{green}72&\cellcolor{green}172&\cellcolor{green}72&\cellcolor{green}172&\cellcolor{green}72&207&72&1.14&594&72&2.73&330&72&1.65\\
4mod7-v1\_96&5&92&72&\cellcolor{green}162&\cellcolor{green}70&\cellcolor{green}162&\cellcolor{green}70&\cellcolor{green}162&\cellcolor{green}70&\cellcolor{green}162&\cellcolor{green}70&194&69&1.13&574&70&2.78&324&71&1.70\\
mod10\_176&5&100&78&\cellcolor{green}182&\cellcolor{green}79&184&79&184&79&\cellcolor{green}182&\cellcolor{green}79&209&77&1.10&640&78&2.75&357&78&1.67\\
4\_49\_16&5&118&99&217&97&218&97&218&97&\cellcolor{green}207&\cellcolor{green}96&251&95&1.14&833&96&3.07&427&98&1.73\\
hwb4\_49&5&126&107&239&\cellcolor{green}101&239&105&235&104&\cellcolor{green}233&102&266&99&1.09&833&102&2.79&468&106&1.71\\
mod10\_171&5&136&108&\cellcolor{green}250&\cellcolor{green}107&251&108&251&108&251&108&280&107&1.08&870&108&2.74&481&108&1.65\\
mini-alu\_167&5&162&126&\cellcolor{green}281&\cellcolor{green}127&\cellcolor{green}281&\cellcolor{green}127&\cellcolor{green}281&\cellcolor{green}127&\cellcolor{green}281&\cellcolor{green}127&337&126&1.13&1,011&126&2.79&558&126&1.68\\
one-two-three-v0\_97&5&162&128&295&\cellcolor{green}127&\cellcolor{green}294&128&\cellcolor{green}294&128&\cellcolor{green}294&128&348&128&1.13&1,029&128&2.74&576&128&1.67\\
alu-v2\_31&5&253&198&443&199&\cellcolor{green}441&\cellcolor{green}199&\cellcolor{green}441&\cellcolor{green}199&443&199&519&197&1.12&1,578&198&2.78&882&198&1.69\\
\bottomrule
\end{tabular}
\end{center}
\label{table-results-1}
\end{table}

\begin{table}
\tiny
\caption{Results of the evaluations for the circuits in the circuit library with six to sixteen qubits. The meaning of the columns is the same as in \autoref{table-results-1}.}
\setlength{\tabcolsep}{0.96mm}
\begin{center}
\begin{tabular}{lrrr|*{4}{rr|}HHHrrr|rrr}
\toprule
\multicolumn{4}{c|}{Original Circuit}&\multicolumn{2}{c|}{\emph{CliffordSimp}}&\multicolumn{2}{c|}{\emph{SquashTK1}}&\multicolumn{2}{c|}{\emph{KAK}}&\multicolumn{2}{c|}{\emph{FullPeephole}}&\multicolumn{3}{H}{AQT}&\multicolumn{3}{c|}{Pytket}&\multicolumn{3}{c}{Qiskit} \\
Name&q&1qg&2qg&1qg&2qg&1qg&2qg&1qg&2qg&1qg&2qg&1qg&2qg&rg&1qg&2qg&rg&1qg&2qg&rg \\
\midrule
graycode6\_47&6&0&5&\cellcolor{green}16&\cellcolor{green}5&\cellcolor{green}16&\cellcolor{green}5&\cellcolor{green}16&\cellcolor{green}5&\cellcolor{green}16&\cellcolor{green}5&20&5&1.19&40&5&2.14&30&5&1.67\\
ex1\_226&6&2&5&\cellcolor{green}10&\cellcolor{green}5&\cellcolor{green}10&\cellcolor{green}5&\cellcolor{green}10&\cellcolor{green}5&\cellcolor{green}10&\cellcolor{green}5&16&5&1.40&42&5&3.13&14&5&1.27\\
xor5\_254&6&2&5&\cellcolor{green}10&\cellcolor{green}5&\cellcolor{green}10&\cellcolor{green}5&\cellcolor{green}10&\cellcolor{green}5&\cellcolor{green}10&\cellcolor{green}5&16&5&1.40&42&5&3.13&14&5&1.27\\
decod24-bdd\_294&6&41&32&\cellcolor{green}73&\cellcolor{green}32&74&32&74&32&\cellcolor{green}73&\cellcolor{green}32&87&32&1.13&277&32&2.94&146&32&1.70\\
4gt4-v0\_80&6&100&79&183&79&\cellcolor{green}180&\cellcolor{green}79&\cellcolor{green}180&\cellcolor{green}79&183&79&225&79&1.17&655&79&2.83&360&79&1.69\\
4gt12-v0\_88&6&108&86&\cellcolor{green}202&\cellcolor{green}88&\cellcolor{green}202&\cellcolor{green}88&\cellcolor{green}202&\cellcolor{green}88&\cellcolor{green}202&\cellcolor{green}88&237&85&1.11&700&86&2.71&388&86&1.63\\
4gt12-v1\_89&6&128&100&\cellcolor{green}226&101&234&\cellcolor{green}100&234&\cellcolor{green}100&\cellcolor{green}226&101&279&100&1.16&821&100&2.82&454&100&1.69\\
4gt4-v0\_79&6&126&105&231&\cellcolor{green}102&\cellcolor{green}229&103&\cellcolor{green}229&103&231&\cellcolor{green}102&265&100&1.10&828&102&2.80&464&104&1.71\\
4gt4-v0\_78&6&126&109&\cellcolor{green}234&\cellcolor{green}107&\cellcolor{green}234&\cellcolor{green}107&\cellcolor{green}234&\cellcolor{green}107&\cellcolor{green}234&\cellcolor{green}107&270&104&1.10&870&107&2.87&481&108&1.73\\
4gt12-v0\_87&6&135&112&\cellcolor{green}249&\cellcolor{green}113&\cellcolor{green}249&\cellcolor{green}113&\cellcolor{green}249&\cellcolor{green}113&\cellcolor{green}249&\cellcolor{green}113&291&109&1.10&905&112&2.81&502&112&1.70\\
4gt12-v0\_86&6&135&116&\cellcolor{green}251&\cellcolor{green}115&254&117&254&117&\cellcolor{green}251&\cellcolor{green}115&291&110&1.10&921&114&2.83&509&116&1.71\\
4gt4-v0\_72&6&145&113&\cellcolor{green}258&\cellcolor{green}113&260&113&260&113&\cellcolor{green}258&\cellcolor{green}113&312&113&1.15&913&113&2.77&513&113&1.69\\
4gt4-v1\_74&6&154&119&\cellcolor{green}279&\cellcolor{green}120&\cellcolor{green}279&\cellcolor{green}120&\cellcolor{green}279&\cellcolor{green}120&\cellcolor{green}279&\cellcolor{green}120&332&119&1.13&954&119&2.69&529&119&1.62\\
decod24-enable\_126&6&189&149&350&\cellcolor{green}148&\cellcolor{green}349&149&\cellcolor{green}349&149&\cellcolor{green}349&149&408&149&1.12&1,207&149&2.72&673&149&1.65\\
mod8-10\_178&6&190&152&\cellcolor{green}340&\cellcolor{green}152&347&154&347&154&343&152&422&152&1.17&1,239&152&2.83&676&152&1.68\\
4gt4-v0\_73&6&216&179&\cellcolor{green}405&\cellcolor{green}181&\cellcolor{green}405&\cellcolor{green}181&\cellcolor{green}405&\cellcolor{green}181&\cellcolor{green}405&\cellcolor{green}181&468&169&1.09&1,439&179&2.76&807&179&1.68\\
ex3\_229&6&228&175&397&\cellcolor{green}174&402&178&405&178&\cellcolor{green}396&175&484&175&1.15&1,440&175&2.83&784&175&1.68\\
mod8-10\_177&6&244&196&446&\cellcolor{green}192&445&196&445&196&\cellcolor{green}433&194&526&192&1.15&1,608&193&2.87&865&196&1.69\\
alu-v2\_30&6&281&223&503&222&503&222&503&222&\cellcolor{green}478&\cellcolor{green}219&577&218&1.14&1,776&220&2.86&978&222&1.72\\
mod5adder\_127&6&316&239&\cellcolor{green}533&\cellcolor{green}241&551&241&551&241&\cellcolor{green}533&\cellcolor{green}241&644&239&1.14&1,964&239&2.85&1,051&239&1.67\\
sf\_276&6&442&336&\cellcolor{green}755&\cellcolor{green}339&778&339&778&339&\cellcolor{green}755&\cellcolor{green}339&896&336&1.13&2,778&336&2.85&1,499&336&1.68\\
sf\_274&6&445&336&\cellcolor{green}734&339&762&\cellcolor{green}338&762&\cellcolor{green}338&\cellcolor{green}734&339&891&336&1.14&2,842&336&2.96&1,489&336&1.70\\
hwb5\_53&6&738&598&1,339&595&1,343&599&1,348&599&\cellcolor{green}1,268&\cellcolor{green}593&1,539&591&1.14&4,855&593&2.93&2,591&598&1.71\\
4mod5-bdd\_287&7&39&31&79&31&80&31&80&31&\cellcolor{green}77&\cellcolor{green}29&78&29&1.01&254&29&2.67&144&31&1.65\\
alu-bdd\_288&7&46&38&\cellcolor{green}89&\cellcolor{green}38&90&38&90&38&\cellcolor{green}89&\cellcolor{green}38&108&38&1.15&315&38&2.78&174&38&1.67\\
rd53\_135&7&162&134&310&134&310&134&310&134&\cellcolor{green}291&\cellcolor{green}134&352&133&1.14&1,120&133&2.95&602&134&1.73\\
ham7\_104&7&171&149&\cellcolor{green}332&\cellcolor{green}149&\cellcolor{green}332&\cellcolor{green}149&\cellcolor{green}332&\cellcolor{green}149&\cellcolor{green}332&\cellcolor{green}149&379&145&1.09&1,214&149&2.83&658&149&1.68\\
C17\_204&7&262&205&470&206&474&206&470&206&\cellcolor{green}459&\cellcolor{green}203&534&203&1.11&1,703&203&2.88&931&205&1.72\\
rd53\_131&7&269&200&442&198&472&199&472&199&\cellcolor{green}438&\cellcolor{green}198&548&197&1.17&1,684&198&2.96&898&199&1.72\\
rd53\_133&7&324&256&591&254&591&254&591&254&\cellcolor{green}591&\cellcolor{green}251&662&249&1.08&2,035&251&2.71&1,163&254&1.68\\
majority\_239&7&345&267&\cellcolor{green}611&\cellcolor{green}267&613&267&613&267&\cellcolor{green}611&\cellcolor{green}267&710&266&1.11&2,182&267&2.79&1,205&267&1.68\\
ex2\_227&7&356&275&\cellcolor{green}612&\cellcolor{green}276&624&276&625&276&\cellcolor{green}612&\cellcolor{green}276&749&275&1.15&2,247&275&2.84&1,237&275&1.70\\
rd53\_130&7&595&448&1,012&451&1,060&\cellcolor{green}450&1,060&\cellcolor{green}450&\cellcolor{green}1,010&451&1,218&448&1.14&3,690&448&2.83&2,013&448&1.68\\
sym6\_145&7&2,187&1,701&\cellcolor{green}3,915&\cellcolor{green}1,701&\cellcolor{green}3,915&\cellcolor{green}1,701&\cellcolor{green}3,915&\cellcolor{green}1,701&\cellcolor{green}3,915&\cellcolor{green}1,701&4,544&1,701&1.11&13,593&1,701&2.72&7,673&1,701&1.67\\
hwb6\_56&7&3,771&2,952&6,654&2,959&6,651&2,963&6,657&2,963&\cellcolor{green}6,409&\cellcolor{green}2,954&7,675&2,931&1.13&23,795&2,940&2.86&13,055&2,952&1.71\\
rd53\_138&8&72&60&\cellcolor{green}155&\cellcolor{green}60&\cellcolor{green}155&\cellcolor{green}60&\cellcolor{green}155&\cellcolor{green}60&\cellcolor{green}155&\cellcolor{green}60&152&56&0.97&499&60&2.60&282&60&1.59\\
cm82a\_208&8&367&283&651&284&671&284&668&284&\cellcolor{green}638&\cellcolor{green}282&742&279&1.11&2,354&280&2.86&1,286&283&1.71\\
f2\_232&8&681&525&\cellcolor{green}1,196&526&1,208&\cellcolor{green}525&1,208&\cellcolor{green}525&\cellcolor{green}1,196&526&1,375&525&1.10&4,260&525&2.78&2,369&525&1.68\\
rd53\_251&8&727&564&\cellcolor{green}1,255&566&1,307&568&1,307&568&1,262&\cellcolor{green}565&1,505&562&1.14&4,643&563&2.86&2,528&564&1.70\\
urf2\_277&8&10,046&10,066&\cellcolor{green}19,785&9,977&20,033&10,085&19,979&10,085&19,916&\cellcolor{green}9,973&23,722&9,643&1.12&82,219&9,943&3.10&39,166&10,064&1.65\\
hwb7\_59&8&13,698&10,681&23,918&10,692&23,902&10,701&\cellcolor{green}23,895&10,701&24,294&\cellcolor{green}10,645&27,611&10,620&1.11&85,525&10,628&2.78&47,333&10,679&1.68\\
urf2\_152&8&45,270&35,210&78,442&35,187&78,289&35,334&\cellcolor{green}78,195&35,334&81,778&\cellcolor{green}35,146&90,479&35,075&1.11&279,091&35,085&2.77&153,850&35,210&1.67\\
con1\_216&9&539&415&948&416&963&415&963&415&\cellcolor{green}929&\cellcolor{green}415&1,085&411&1.11&3,445&413&2.87&1,883&415&1.71\\
urf5\_280&9&26,065&23,764&49,195&23,619&49,596&23,794&49,650&23,794&\cellcolor{green}48,820&\cellcolor{green}23,605&58,120&22,915&1.12&193,512&23,547&3.00&97,711&23,753&1.68\\
urf1\_278&9&28,074&26,692&\cellcolor{green}53,849&\cellcolor{green}26,493&54,090&26,729&54,554&26,724&54,621&26,547&63,979&25,716&1.12&216,780&26,478&3.03&106,531&26,684&1.66\\
hwb8\_113&9&39,008&30,372&\cellcolor{green}68,800&30,406&69,038&30,415&69,008&30,415&70,894&\cellcolor{green}30,308&78,674&30,209&1.10&244,129&30,272&2.77&135,277&30,370&1.67\\
urf5\_158&9&92,484&71,932&\cellcolor{green}160,665&72,011&160,994&72,096&162,607&72,096&176,112&\cellcolor{green}71,545&186,759&71,535&1.11&569,805&71,545&2.76&317,054&71,932&1.67\\
urf1\_149&9&103,986&80,878&193,449&80,771&\cellcolor{green}193,340&80,878&\cellcolor{green}193,340&80,878&196,718&\cellcolor{green}80,718&210,106&80,711&1.06&639,825&80,718&2.63&354,986&80,878&1.59\\
mini\_alu\_305&10&96&77&\cellcolor{green}186&\cellcolor{green}77&\cellcolor{green}186&\cellcolor{green}77&\cellcolor{green}186&\cellcolor{green}77&\cellcolor{green}186&\cellcolor{green}77&211&77&1.10&639&77&2.72&341&77&1.59\\
sys6-v0\_111&10&117&98&\cellcolor{green}240&\cellcolor{green}98&\cellcolor{green}240&\cellcolor{green}98&\cellcolor{green}240&\cellcolor{green}98&\cellcolor{green}240&\cellcolor{green}98&242&93&0.99&810&98&2.69&446&98&1.61\\
rd73\_140&10&126&104&\cellcolor{green}257&\cellcolor{green}104&\cellcolor{green}257&\cellcolor{green}104&\cellcolor{green}257&\cellcolor{green}104&\cellcolor{green}257&\cellcolor{green}104&264&98&1.00&853&104&2.65&484&104&1.63\\
ising\_model\_10&10&390&90&\cellcolor{green}314&\cellcolor{green}114&318&114&\cellcolor{green}314&\cellcolor{green}114&317&114&235&90&0.76&920&90&2.36&537&90&1.46\\
rd73\_252&10&3,002&2,319&\cellcolor{green}5,281&2,327&5,344&2,323&5,341&2,323&5,345&\cellcolor{green}2,319&6,165&2,314&1.11&18,844&2,314&2.78&10,442&2,319&1.68\\
sqn\_258&10&5,764&4,459&10,189&4,473&10,311&4,466&10,311&4,466&\cellcolor{green}10,083&\cellcolor{green}4,459&11,752&4,447&1.11&36,071&4,446&2.79&19,980&4,459&1.68\\
sym9\_148&10&12,096&9,408&21,836&9,431&21,835&9,431&21,831&9,431&\cellcolor{green}21,384&\cellcolor{green}9,397&24,480&9,381&1.10&76,328&9,383&2.78&42,465&9,408&1.69\\
max46\_240&10&15,282&11,844&27,216&\cellcolor{green}11,857&27,401&11,870&27,397&11,870&\cellcolor{green}26,915&11,858&31,268&11,818&1.11&95,146&11,826&2.76&53,090&11,844&1.67\\
urf3\_279&10&64,982&60,380&\cellcolor{green}123,438&\cellcolor{green}59,982&124,565&60,465&124,445&60,487&128,056&60,013&145,937&58,354&1.11&490,190&59,912&3.00&244,149&60,363&1.66\\
hwb9\_119&10&116,820&90,955&203,602&91,061&\cellcolor{green}203,271&91,100&204,174&91,100&212,025&\cellcolor{green}90,731&234,982&90,526&1.11&731,698&90,642&2.79&404,961&90,952&1.68\\
wim\_266&11&559&427&988&429&998&429&998&429&\cellcolor{green}950&\cellcolor{green}428&1,140&426&1.14&3,531&426&2.87&1,942&427&1.72\\
dc1\_220&11&1,081&833&1,939&836&\cellcolor{green}1,929&\cellcolor{green}835&\cellcolor{green}1,929&\cellcolor{green}835&1,939&836&2,182&833&1.09&6,758&833&2.75&3,767&833&1.66\\
z4\_268&11&1,730&1,343&\cellcolor{green}3,082&1,346&3,105&1,346&3,105&1,346&3,114&\cellcolor{green}1,341&3,521&1,336&1.10&10,952&1,340&2.78&6,060&1,343&1.67\\
life\_238&11&12,645&9,800&22,580&9,806&22,644&\cellcolor{green}9,804&22,639&\cellcolor{green}9,804&\cellcolor{green}22,288&9,805&25,673&9,774&1.10&79,333&9,776&2.78&44,033&9,800&1.68\\
9symml\_195&11&19,649&15,232&35,037&15,243&\cellcolor{green}35,021&\cellcolor{green}15,243&35,097&15,243&35,949&15,244&39,996&15,196&1.10&122,740&15,200&2.74&68,334&15,232&1.66\\
sym9\_193&11&19,649&15,232&\cellcolor{green}35,017&\cellcolor{green}15,243&35,031&15,243&35,101&15,243&35,947&15,244&39,996&15,196&1.10&122,740&15,200&2.74&68,334&15,232&1.66\\
sym9\_146&12&180&148&\cellcolor{green}369&\cellcolor{green}148&\cellcolor{green}369&\cellcolor{green}148&\cellcolor{green}369&\cellcolor{green}148&\cellcolor{green}369&\cellcolor{green}148&376&141&1.00&1,207&148&2.62&681&148&1.60\\
cm152a\_212&12&689&532&\cellcolor{green}1,222&532&1,238&532&1,238&532&1,227&\cellcolor{green}530&1,394&530&1.10&4,345&530&2.78&2,404&532&1.67\\
sqrt8\_260&12&1,695&1,314&2,996&1,317&3,053&1,319&3,053&1,319&\cellcolor{green}2,939&\cellcolor{green}1,315&3,441&1,310&1.12&10,774&1,311&2.84&5,917&1,314&1.70\\
cycle10\_2\_110&12&3,402&2,648&6,122&2,646&6,124&2,646&6,122&2,646&\cellcolor{green}5,952&\cellcolor{green}2,644&6,815&2,626&1.10&21,384&2,635&2.79&11,849&2,646&1.69\\
rd84\_253&12&7,698&5,960&\cellcolor{green}13,510&5,970&13,584&5,972&13,585&5,972&13,732&\cellcolor{green}5,957&15,669&5,941&1.11&48,275&5,938&2.78&26,798&5,960&1.68\\
sym10\_262&12&36,199&28,084&\cellcolor{green}64,305&28,099&64,474&28,100&64,521&28,100&65,457&\cellcolor{green}28,049&73,395&27,982&1.10&227,210&27,997&2.76&125,965&28,084&1.67\\
rd53\_311&13&151&124&285&125&283&125&285&125&\cellcolor{green}282&\cellcolor{green}124&321&123&1.09&1,029&123&2.84&542&124&1.64\\
ising\_model\_13&13&513&120&411&148&\cellcolor{green}409&\cellcolor{green}148&\cellcolor{green}409&\cellcolor{green}148&411&148&312&120&0.78&1,220&120&2.41&727&120&1.52\\
squar5\_261&13&1,124&869&2,029&869&2,102&869&2,102&869&\cellcolor{green}1,967&\cellcolor{green}866&2,267&865&1.11&7,263&866&2.87&3,932&869&1.69\\
radd\_250&13&1,808&1,405&3,264&1,414&3,278&1,415&3,274&1,415&\cellcolor{green}3,202&\cellcolor{green}1,403&3,682&1,397&1.10&11,459&1,397&2.79&6,364&1,405&1.69\\
adr4\_197&13&1,941&1,498&3,479&1,503&3,483&1,503&3,481&1,503&\cellcolor{green}3,460&\cellcolor{green}1,495&3,946&1,488&1.10&12,308&1,491&2.78&6,793&1,498&1.67\\
root\_255&13&9,666&7,493&\cellcolor{green}17,119&7,505&17,304&7,512&17,303&7,512&17,214&\cellcolor{green}7,498&19,713&7,482&1.10&60,486&7,484&2.76&33,682&7,493&1.67\\
dist\_223&13&21,422&16,624&\cellcolor{green}37,426&16,672&37,510&16,675&37,513&16,675&38,895&\cellcolor{green}16,646&43,590&16,592&1.11&133,728&16,604&2.78&74,694&16,624&1.69\\
plus63mod4096\_163&13&72,415&56,329&\cellcolor{green}128,093&56,455&128,120&56,456&128,114&56,456&130,949&\cellcolor{green}56,054&145,844&56,036&1.09&457,772&56,000&2.78&252,611&56,323&1.67\\
0410184\_169&14&107&104&\cellcolor{green}220&\cellcolor{green}102&240&102&240&102&\cellcolor{green}220&\cellcolor{green}102&249&97&1.07&838&102&2.92&449&103&1.71\\
sym6\_316&14&147&123&\cellcolor{green}283&\cellcolor{green}124&285&124&285&124&\cellcolor{green}283&\cellcolor{green}124&325&123&1.10&999&123&2.76&543&123&1.64\\
cm42a\_207&14&1,005&771&1,784&771&1,792&771&1,792&771&\cellcolor{green}1,777&\cellcolor{green}770&2,026&769&1.10&6,265&770&2.76&3,486&771&1.67\\
pm1\_249&14&1,005&771&1,784&771&1,792&771&1,792&771&\cellcolor{green}1,777&\cellcolor{green}770&2,026&769&1.10&6,265&770&2.76&3,486&771&1.67\\
cm85a\_209&14&6,428&4,986&11,606&4,990&11,666&4,989&11,666&4,989&\cellcolor{green}11,457&\cellcolor{green}4,976&12,806&4,950&1.08&40,340&4,959&2.76&22,486&4,986&1.67\\
clip\_206&14&19,055&14,772&\cellcolor{green}33,578&14,794&33,931&14,801&33,935&14,801&34,850&\cellcolor{green}14,772&38,539&14,726&1.10&119,515&14,734&2.78&66,417&14,772&1.68\\
sao2\_257&14&21,713&16,864&38,709&\cellcolor{green}16,869&38,885&16,870&38,887&16,870&\cellcolor{green}38,596&16,894&44,436&16,816&1.10&135,670&16,816&2.75&75,744&16,864&1.67\\
plus63mod8192\_164&14&105,247&81,865&186,494&82,002&\cellcolor{green}186,457&82,002&186,509&82,001&193,114&\cellcolor{green}81,461&211,594&81,437&1.09&661,503&81,407&2.77&366,546&81,859&1.67\\
rd84\_142&15&189&154&\cellcolor{green}386&\cellcolor{green}154&\cellcolor{green}386&\cellcolor{green}154&\cellcolor{green}386&\cellcolor{green}154&\cellcolor{green}386&\cellcolor{green}154&392&147&1.00&1,266&154&2.63&712&154&1.60\\
misex1\_241&15&2,713&2,100&4,789&2,103&4,887&\cellcolor{green}2,101&4,887&\cellcolor{green}2,101&\cellcolor{green}4,760&2,112&5,424&2,095&1.09&17,156&2,096&2.80&9,470&2,100&1.68\\
square\_root\_7&15&4,541&3,089&\cellcolor{green}6,193&3,110&6,396&3,110&6,396&3,110&6,219&\cellcolor{green}3,109&6,654&2,550&0.99&25,927&3,088&3.12&12,294&3,089&1.65\\
ham15\_107&15&4,905&3,858&8,863&3,860&8,864&3,863&8,864&3,863&\cellcolor{green}8,857&\cellcolor{green}3,858&10,062&3,835&1.09&30,841&3,854&2.73&17,294&3,857&1.66\\
dc2\_222&15&5,331&4,131&9,483&4,146&9,519&4,142&9,527&4,142&\cellcolor{green}9,424&\cellcolor{green}4,134&10,706&4,122&1.09&33,454&4,124&2.77&18,601&4,131&1.68\\
co14\_215&15&10,096&7,840&18,211&7,853&18,232&7,851&18,281&7,851&\cellcolor{green}17,914&\cellcolor{green}7,802&20,539&7,785&1.10&63,601&7,785&2.78&35,412&7,840&1.68\\
urf6\_160&15&96,660&75,180&181,389&75,143&181,341&75,180&181,341&75,180&\cellcolor{green}180,729&\cellcolor{green}74,951&195,771&74,908&1.06&598,342&74,951&2.63&335,227&75,180&1.61\\
cnt3-5\_179&16&90&85&160&\cellcolor{green}70&\cellcolor{green}156&75&\cellcolor{green}156&75&\cellcolor{green}156&75&192&75&1.16&615&75&3.00&337&80&1.81\\
cnt3-5\_180&16&270&215&\cellcolor{green}500&\cellcolor{green}206&\cellcolor{green}500&\cellcolor{green}206&\cellcolor{green}500&\cellcolor{green}206&\cellcolor{green}500&\cellcolor{green}206&563&205&1.09&1,693&205&2.69&966&210&1.67\\
ising\_model\_16&16&636&150&519&188&515&188&\cellcolor{green}496&\cellcolor{green}184&519&188&388&150&0.79&1,520&150&2.46&894&150&1.54\\
inc\_237&16&5,983&4,636&10,532&\cellcolor{green}4,642&10,808&4,643&10,787&4,643&\cellcolor{green}10,488&4,652&12,015&4,632&1.10&37,795&4,634&2.80&20,859&4,636&1.68\\
mlp4\_245&16&10,620&8,232&\cellcolor{green}18,973&8,241&19,133&8,236&19,133&8,236&19,334&\cellcolor{green}8,223&21,338&8,208&1.09&66,874&8,207&2.76&37,082&8,232&1.67\\
\bottomrule
\end{tabular}
\end{center}
\label{table-results-2}
\end{table}

\begin{table}
\tiny
\caption{Results of the evaluations for the algorithms QAOA, QFT, and QV for different numbers of qubits. The meaning of the columns is the same as in \autoref{table-results-1}.}
\setlength{\tabcolsep}{1.33mm}
\begin{center}
\begin{tabular}{lrrr|*{4}{rr|}HHHrrr|rrr}
\toprule
\multicolumn{4}{c|}{Original Circuit}&\multicolumn{2}{c|}{\emph{CliffordSimp}}&\multicolumn{2}{c|}{\emph{SquashTK1}}&\multicolumn{2}{c|}{\emph{KAK}}&\multicolumn{2}{c|}{\emph{FullPeephole}}&\multicolumn{3}{H}{AQT}&\multicolumn{3}{c|}{Pytket}&\multicolumn{3}{c}{Qiskit} \\
Name&q&1qg&2qg&1qg&2qg&1qg&2qg&1qg&2qg&1qg&2qg&1qg&2qg&rg&1qg&2qg&rg&1qg&2qg&rg \\
\midrule
QAOA&5&20&4&\cellcolor{green}31&\cellcolor{green}4&\cellcolor{green}31&\cellcolor{green}4&\cellcolor{green}31&\cellcolor{green}4&\cellcolor{green}31&\cellcolor{green}4&30&4&0.97&69&4&2.09&53&4&1.63\\
QAOA&10&40&9&\cellcolor{green}66&\cellcolor{green}9&\cellcolor{green}66&\cellcolor{green}9&\cellcolor{green}66&\cellcolor{green}9&\cellcolor{green}66&\cellcolor{green}9&65&9&0.99&144&9&2.04&113&9&1.63\\
QAOA&15&60&14&\cellcolor{green}101&\cellcolor{green}14&\cellcolor{green}101&\cellcolor{green}14&\cellcolor{green}101&\cellcolor{green}14&\cellcolor{green}101&\cellcolor{green}14&100&14&0.99&219&14&2.03&173&14&1.63\\
QAOA&20&80&19&\cellcolor{green}136&\cellcolor{green}19&\cellcolor{green}136&\cellcolor{green}19&\cellcolor{green}136&\cellcolor{green}19&\cellcolor{green}136&\cellcolor{green}19&135&19&0.99&294&19&2.02&233&19&1.63\\
QAOA&25&100&24&\cellcolor{green}171&\cellcolor{green}24&\cellcolor{green}171&\cellcolor{green}24&\cellcolor{green}171&\cellcolor{green}24&\cellcolor{green}171&\cellcolor{green}24&170&24&0.99&369&24&2.02&293&24&1.63\\
QAOA&30&120&29&\cellcolor{green}206&\cellcolor{green}29&\cellcolor{green}206&\cellcolor{green}29&\cellcolor{green}206&\cellcolor{green}29&\cellcolor{green}206&\cellcolor{green}29&205&29&1.00&444&29&2.01&353&29&1.63\\
QAOA&35&140&34&\cellcolor{green}241&\cellcolor{green}34&\cellcolor{green}241&\cellcolor{green}34&\cellcolor{green}241&\cellcolor{green}34&\cellcolor{green}241&\cellcolor{green}34&240&34&1.00&519&34&2.01&413&34&1.63\\
QAOA&40&160&39&\cellcolor{green}276&\cellcolor{green}39&\cellcolor{green}276&\cellcolor{green}39&\cellcolor{green}276&\cellcolor{green}39&\cellcolor{green}276&\cellcolor{green}39&275&39&1.00&594&39&2.01&473&39&1.63\\
QAOA&45&180&44&\cellcolor{green}311&\cellcolor{green}44&\cellcolor{green}311&\cellcolor{green}44&\cellcolor{green}311&\cellcolor{green}44&\cellcolor{green}311&\cellcolor{green}44&310&44&1.00&669&44&2.01&533&44&1.63\\
QAOA&50&200&49&\cellcolor{green}346&\cellcolor{green}49&\cellcolor{green}346&\cellcolor{green}49&\cellcolor{green}346&\cellcolor{green}49&\cellcolor{green}346&\cellcolor{green}49&345&49&1.00&744&49&2.01&593&49&1.63\\
QAOA&55&220&54&\cellcolor{green}381&\cellcolor{green}54&\cellcolor{green}381&\cellcolor{green}54&\cellcolor{green}381&\cellcolor{green}54&\cellcolor{green}381&\cellcolor{green}54&380&54&1.00&819&54&2.01&653&54&1.63\\
QAOA&60&240&59&\cellcolor{green}416&\cellcolor{green}59&\cellcolor{green}416&\cellcolor{green}59&\cellcolor{green}416&\cellcolor{green}59&\cellcolor{green}416&\cellcolor{green}59&415&59&1.00&894&59&2.01&713&59&1.63\\
QAOA&65&260&64&\cellcolor{green}451&\cellcolor{green}64&\cellcolor{green}451&\cellcolor{green}64&\cellcolor{green}451&\cellcolor{green}64&\cellcolor{green}451&\cellcolor{green}64&450&64&1.00&969&64&2.01&773&64&1.63\\
QAOA&70&280&69&\cellcolor{green}486&\cellcolor{green}69&\cellcolor{green}486&\cellcolor{green}69&\cellcolor{green}486&\cellcolor{green}69&\cellcolor{green}486&\cellcolor{green}69&485&69&1.00&1,044&69&2.01&833&69&1.63\\
QAOA&75&300&74&\cellcolor{green}521&\cellcolor{green}74&\cellcolor{green}521&\cellcolor{green}74&\cellcolor{green}521&\cellcolor{green}74&\cellcolor{green}521&\cellcolor{green}74&520&74&1.00&1,119&74&2.01&893&74&1.63\\
QAOA&80&320&79&\cellcolor{green}556&\cellcolor{green}79&\cellcolor{green}556&\cellcolor{green}79&\cellcolor{green}556&\cellcolor{green}79&\cellcolor{green}556&\cellcolor{green}79&555&79&1.00&1,194&79&2.00&953&79&1.63\\
QAOA&85&340&84&\cellcolor{green}591&\cellcolor{green}84&\cellcolor{green}591&\cellcolor{green}84&\cellcolor{green}591&\cellcolor{green}84&\cellcolor{green}591&\cellcolor{green}84&590&84&1.00&1,269&84&2.00&1,013&84&1.63\\
QAOA&90&360&89&\cellcolor{green}626&\cellcolor{green}89&\cellcolor{green}626&\cellcolor{green}89&\cellcolor{green}626&\cellcolor{green}89&\cellcolor{green}626&\cellcolor{green}89&625&89&1.00&1,344&89&2.00&1,073&89&1.63\\
QAOA&95&380&94&\cellcolor{green}661&\cellcolor{green}94&\cellcolor{green}661&\cellcolor{green}94&\cellcolor{green}661&\cellcolor{green}94&\cellcolor{green}661&\cellcolor{green}94&660&94&1.00&1,419&94&2.00&1,133&94&1.63\\
QAOA&100&400&99&\cellcolor{green}696&\cellcolor{green}99&\cellcolor{green}696&\cellcolor{green}99&\cellcolor{green}696&\cellcolor{green}99&\cellcolor{green}696&\cellcolor{green}99&695&99&1.00&1,494&99&2.00&1,193&99&1.63\\
QAOA&105&420&104&\cellcolor{green}731&\cellcolor{green}104&\cellcolor{green}731&\cellcolor{green}104&\cellcolor{green}731&\cellcolor{green}104&\cellcolor{green}731&\cellcolor{green}104&730&104&1.00&1,569&104&2.00&1,253&104&1.63\\
QAOA&110&440&109&\cellcolor{green}766&\cellcolor{green}109&\cellcolor{green}766&\cellcolor{green}109&\cellcolor{green}766&\cellcolor{green}109&\cellcolor{green}766&\cellcolor{green}109&765&109&1.00&1,644&109&2.00&1,313&109&1.63\\
QAOA&115&460&114&\cellcolor{green}801&\cellcolor{green}114&\cellcolor{green}801&\cellcolor{green}114&\cellcolor{green}801&\cellcolor{green}114&\cellcolor{green}801&\cellcolor{green}114&800&114&1.00&1,719&114&2.00&1,373&114&1.63\\
QAOA&120&480&119&\cellcolor{green}836&\cellcolor{green}119&\cellcolor{green}836&\cellcolor{green}119&\cellcolor{green}836&\cellcolor{green}119&\cellcolor{green}836&\cellcolor{green}119&835&119&1.00&1,794&119&2.00&1,433&119&1.63\\
QAOA&125&500&124&\cellcolor{green}871&\cellcolor{green}124&\cellcolor{green}871&\cellcolor{green}124&\cellcolor{green}871&\cellcolor{green}124&\cellcolor{green}871&\cellcolor{green}124&870&124&1.00&1,869&124&2.00&1,493&124&1.63\\
QAOA&130&520&129&\cellcolor{green}906&\cellcolor{green}129&\cellcolor{green}906&\cellcolor{green}129&\cellcolor{green}906&\cellcolor{green}129&\cellcolor{green}906&\cellcolor{green}129&905&129&1.00&1,944&129&2.00&1,553&129&1.63\\
QAOA&135&540&134&\cellcolor{green}941&\cellcolor{green}134&\cellcolor{green}941&\cellcolor{green}134&\cellcolor{green}941&\cellcolor{green}134&\cellcolor{green}941&\cellcolor{green}134&940&134&1.00&2,019&134&2.00&1,613&134&1.63\\
QAOA&140&560&139&\cellcolor{green}976&\cellcolor{green}139&\cellcolor{green}976&\cellcolor{green}139&\cellcolor{green}976&\cellcolor{green}139&\cellcolor{green}976&\cellcolor{green}139&975&139&1.00&2,094&139&2.00&1,673&139&1.63\\
QAOA&145&580&144&\cellcolor{green}1,011&\cellcolor{green}144&\cellcolor{green}1,011&\cellcolor{green}144&\cellcolor{green}1,011&\cellcolor{green}144&\cellcolor{green}1,011&\cellcolor{green}144&1,010&144&1.00&2,169&144&2.00&1,733&144&1.63\\
QAOA&150&600&149&\cellcolor{green}1,046&\cellcolor{green}149&\cellcolor{green}1,046&\cellcolor{green}149&\cellcolor{green}1,046&\cellcolor{green}149&\cellcolor{green}1,046&\cellcolor{green}149&1,045&149&1.00&2,244&149&2.00&1,793&149&1.63\\
QAOA&155&620&154&\cellcolor{green}1,081&\cellcolor{green}154&\cellcolor{green}1,081&\cellcolor{green}154&\cellcolor{green}1,081&\cellcolor{green}154&\cellcolor{green}1,081&\cellcolor{green}154&1,080&154&1.00&2,319&154&2.00&1,853&154&1.63\\
QAOA&160&640&159&\cellcolor{green}1,116&\cellcolor{green}159&\cellcolor{green}1,116&\cellcolor{green}159&\cellcolor{green}1,116&\cellcolor{green}159&\cellcolor{green}1,116&\cellcolor{green}159&1,115&159&1.00&2,394&159&2.00&1,913&159&1.63\\
QAOA&165&660&164&\cellcolor{green}1,151&\cellcolor{green}164&\cellcolor{green}1,151&\cellcolor{green}164&\cellcolor{green}1,151&\cellcolor{green}164&\cellcolor{green}1,151&\cellcolor{green}164&1,150&164&1.00&2,469&164&2.00&1,973&164&1.63\\
QAOA&170&680&169&\cellcolor{green}1,186&\cellcolor{green}169&\cellcolor{green}1,186&\cellcolor{green}169&\cellcolor{green}1,186&\cellcolor{green}169&\cellcolor{green}1,186&\cellcolor{green}169&1,185&169&1.00&2,544&169&2.00&2,033&169&1.63\\
QAOA&175&700&174&\cellcolor{green}1,221&\cellcolor{green}174&\cellcolor{green}1,221&\cellcolor{green}174&\cellcolor{green}1,221&\cellcolor{green}174&\cellcolor{green}1,221&\cellcolor{green}174&1,220&174&1.00&2,619&174&2.00&2,093&174&1.63\\
QAOA&180&720&179&\cellcolor{green}1,256&\cellcolor{green}179&\cellcolor{green}1,256&\cellcolor{green}179&\cellcolor{green}1,256&\cellcolor{green}179&\cellcolor{green}1,256&\cellcolor{green}179&1,255&179&1.00&2,694&179&2.00&2,153&179&1.63\\
QAOA&185&740&184&\cellcolor{green}1,291&\cellcolor{green}184&\cellcolor{green}1,291&\cellcolor{green}184&\cellcolor{green}1,291&\cellcolor{green}184&\cellcolor{green}1,291&\cellcolor{green}184&1,290&184&1.00&2,769&184&2.00&2,213&184&1.63\\
QAOA&190&760&189&\cellcolor{green}1,326&\cellcolor{green}189&\cellcolor{green}1,326&\cellcolor{green}189&\cellcolor{green}1,326&\cellcolor{green}189&\cellcolor{green}1,326&\cellcolor{green}189&1,325&189&1.00&2,844&189&2.00&2,273&189&1.63\\
QAOA&195&780&194&\cellcolor{green}1,361&\cellcolor{green}194&\cellcolor{green}1,361&\cellcolor{green}194&\cellcolor{green}1,361&\cellcolor{green}194&\cellcolor{green}1,361&\cellcolor{green}194&1,360&194&1.00&2,919&194&2.00&2,333&194&1.63\\
QAOA&200&800&199&\cellcolor{green}1,396&\cellcolor{green}199&\cellcolor{green}1,396&\cellcolor{green}199&\cellcolor{green}1,396&\cellcolor{green}199&\cellcolor{green}1,396&\cellcolor{green}199&1,395&199&1.00&2,994&199&2.00&2,393&199&1.63\\
QFT&5&5&10&\cellcolor{green}46&\cellcolor{green}20&\cellcolor{green}46&\cellcolor{green}20&\cellcolor{green}46&\cellcolor{green}20&\cellcolor{green}46&\cellcolor{green}20&50&20&1.06&175&20&2.95&89&20&1.65\\
QFT&10&10&45&\cellcolor{green}190&\cellcolor{green}90&\cellcolor{green}190&\cellcolor{green}90&\cellcolor{green}190&\cellcolor{green}90&\cellcolor{green}190&\cellcolor{green}90&200&90&1.04&750&90&3.00&374&90&1.66\\
QFT&15&15&105&\cellcolor{green}436&\cellcolor{green}210&\cellcolor{green}436&\cellcolor{green}210&\cellcolor{green}436&\cellcolor{green}210&\cellcolor{green}436&\cellcolor{green}210&450&210&1.02&1,725&210&3.00&859&210&1.65\\
QFT&20&20&190&\cellcolor{green}780&\cellcolor{green}380&\cellcolor{green}780&\cellcolor{green}380&\cellcolor{green}780&\cellcolor{green}380&\cellcolor{green}780&\cellcolor{green}380&800&380&1.02&3,100&380&3.00&1,543&380&1.66\\
QFT&25&25&300&\cellcolor{green}1,226&\cellcolor{green}600&\cellcolor{green}1,226&\cellcolor{green}600&\cellcolor{green}1,226&\cellcolor{green}600&\cellcolor{green}1,226&\cellcolor{green}600&1,250&600&1.01&4,875&600&3.00&2,426&600&1.66\\
QFT&30&30&435&1,770&870&1,770&870&1,770&870&\cellcolor{green}1,747&\cellcolor{green}858&1,776&858&1.01&6,954&858&3.00&3,475&865&1.67\\
QFT&35&35&595&2,416&1,190&2,416&1,190&2,416&1,190&\cellcolor{green}2,272&\cellcolor{green}1,118&2,306&1,118&1.01&9,049&1,118&3.00&4,538&1,155&1.68\\
QFT&40&40&780&3,107&1,548&3,107&1,548&3,107&1,548&\cellcolor{green}2,797&\cellcolor{green}1,378&2,836&1,378&1.01&11,144&1,378&3.00&5,600&1,470&1.69\\
QFT&45&45&990&3,797&1,908&3,797&1,908&3,797&1,908&\cellcolor{green}3,322&\cellcolor{green}1,638&3,366&1,638&1.01&13,239&1,638&3.00&6,663&1,810&1.71\\
QFT&49&49&1,176&4,349&2,196&4,349&2,196&4,349&2,196&\cellcolor{green}3,742&\cellcolor{green}1,846&3,790&1,846&1.01&14,915&1,846&3.00&7,513&2,100&1.72\\
QV&5&80&30&135&30&135&30&\cellcolor{green}109&\cellcolor{green}24&\cellcolor{green}109&\cellcolor{green}24&111&24&1.02&355&24&2.85&265&30&2.22\\
QV&10&400&150&630&150&630&150&\cellcolor{green}579&\cellcolor{green}138&\cellcolor{green}579&\cellcolor{green}138&582&138&1.00&1,973&138&2.94&1,250&150&1.95\\
QV&15&840&315&1,305&315&1,305&315&\cellcolor{green}1,253&\cellcolor{green}303&\cellcolor{green}1,253&\cellcolor{green}303&1,257&303&1.00&4,305&303&2.96&2,595&315&1.87\\
QV&20&1,600&600&2,460&600&2,460&600&\cellcolor{green}2,304&\cellcolor{green}564&\cellcolor{green}2,304&\cellcolor{green}564&2,316&564&1.00&7,960&564&2.97&4,900&600&1.92\\
QV&25&2,400&900&3,675&900&3,675&900&\cellcolor{green}3,560&\cellcolor{green}873&\cellcolor{green}3,560&\cellcolor{green}873&3,567&873&1.00&12,326&873&2.98&7,325&900&1.86\\
QV&30&3,600&1,350&5,490&1,350&5,490&1,350&\cellcolor{green}5,295&\cellcolor{green}1,305&\cellcolor{green}5,295&\cellcolor{green}1,305&5,310&1,305&1.00&18,375&1,305&2.98&10,950&1,350&1.86\\
QV&35&4,760&1,785&7,245&1,785&7,245&1,785&\cellcolor{green}7,011&\cellcolor{green}1,731&\cellcolor{green}7,011&\cellcolor{green}1,731&7,029&1,731&1.00&24,355&1,731&2.98&14,455&1,785&1.86\\
QV&40&6,400&2,400&9,720&2,400&9,720&2,400&\cellcolor{green}9,435&\cellcolor{green}2,334&\cellcolor{green}9,435&\cellcolor{green}2,334&9,456&2,334&1.00&32,813&2,334&2.99&19,400&2,400&1.85\\
QV&45&7,920&2,970&12,015&2,970&12,015&2,970&\cellcolor{green}11,703&\cellcolor{green}2,898&\cellcolor{green}11,703&\cellcolor{green}2,898&11,727&2,898&1.00&40,725&2,898&2.99&23,985&2,970&1.85\\
QV&50&10,000&3,750&15,150&3,750&15,150&3,750&\cellcolor{green}14,955&\cellcolor{green}3,705&\cellcolor{green}14,955&\cellcolor{green}3,705&14,970&3,705&1.00&52,075&3,705&2.99&30,250&3,750&1.82\\
QV&55&11,880&4,455&17,985&4,455&17,985&4,455&\cellcolor{green}17,686&\cellcolor{green}4,386&\cellcolor{green}17,686&\cellcolor{green}4,386&17,709&4,386&1.00&61,610&4,386&2.99&35,915&4,455&1.83\\
QV&60&14,400&5,400&21,780&5,400&21,780&5,400&\cellcolor{green}21,457&\cellcolor{green}5,325&\cellcolor{green}21,457&\cellcolor{green}5,325&21,480&5,325&1.00&74,781&5,325&2.99&43,500&5,400&1.83\\
QV&65&16,640&6,240&25,155&6,240&25,155&6,240&\cellcolor{green}24,701&\cellcolor{green}6,135&\cellcolor{green}24,701&\cellcolor{green}6,135&24,735&6,135&1.00&86,113&6,135&2.99&50,245&6,240&1.83\\
QV&70&19,600&7,350&29,610&7,350&29,610&7,350&\cellcolor{green}29,194&\cellcolor{green}7,254&\cellcolor{green}29,194&\cellcolor{green}7,254&29,226&7,254&1.00&101,810&7,254&2.99&59,150&7,350&1.82\\
QV&75&22,200&8,325&33,525&8,325&33,525&8,325&\cellcolor{green}32,992&\cellcolor{green}8,202&\cellcolor{green}32,992&\cellcolor{green}8,202&33,033&8,202&1.00&115,080&8,202&2.99&66,975&8,325&1.83\\
QV&80&25,600&9,600&38,640&9,600&38,640&9,600&\cellcolor{green}38,120&\cellcolor{green}9,480&\cellcolor{green}38,120&\cellcolor{green}9,480&38,160&9,480&1.00&133,000&9,480&2.99&77,200&9,600&1.82\\
QV&85&28,560&10,710&43,095&10,710&43,095&10,710&\cellcolor{green}42,615&\cellcolor{green}10,599&\cellcolor{green}42,615&\cellcolor{green}10,599&42,651&10,599&1.00&148,703&10,599&2.99&86,105&10,710&1.82\\
QV&90&32,400&12,150&48,870&12,150&48,870&12,150&\cellcolor{green}48,130&\cellcolor{green}11,979&\cellcolor{green}48,130&\cellcolor{green}11,979&48,186&11,979&1.00&167,988&11,979&2.99&97,650&12,150&1.83\\
QV&95&35,720&13,395&53,865&13,395&53,865&13,395&\cellcolor{green}53,293&\cellcolor{green}13,263&\cellcolor{green}53,293&\cellcolor{green}13,263&53,337&13,263&1.00&186,025&13,263&2.99&107,635&13,395&1.82\\
QV&100&40,000&15,000&60,300&15,000&60,300&15,000&\cellcolor{green}59,741&\cellcolor{green}14,871&\cellcolor{green}59,741&\cellcolor{green}14,871&59,784&14,871&1.00&208,565&14,871&2.99&120,500&15,000&1.82\\
QV&105&43,680&16,380&65,835&16,380&65,835&16,380&\cellcolor{green}65,082&\cellcolor{green}16,206&\cellcolor{green}65,082&\cellcolor{green}16,206&65,139&16,206&1.00&227,238&16,206&2.99&131,565&16,380&1.82\\
QV&110&48,400&18,150&72,930&18,150&72,930&18,150&\cellcolor{green}72,332&\cellcolor{green}18,012&\cellcolor{green}72,332&\cellcolor{green}18,012&72,378&18,012&1.00&252,580&18,012&3.00&145,750&18,150&1.81\\
QV&115&52,440&19,665&79,005&19,665&79,005&19,665&\cellcolor{green}78,226&\cellcolor{green}19,485&\cellcolor{green}78,226&\cellcolor{green}19,485&78,285&19,485&1.00&273,188&19,485&3.00&157,895&19,665&1.82\\
QV&120&57,600&21,600&86,760&21,600&86,760&21,600&\cellcolor{green}86,084&\cellcolor{green}21,444&\cellcolor{green}86,084&\cellcolor{green}21,444&86,136&21,444&1.00&300,660&21,444&3.00&173,400&21,600&1.81\\
QV&125&62,000&23,250&93,375&23,250&93,375&23,250&\cellcolor{green}92,530&\cellcolor{green}23,055&\cellcolor{green}92,530&\cellcolor{green}23,055&92,595&23,055&1.00&323,200&23,055&3.00&186,625&23,250&1.82\\
QV&130&67,600&25,350&101,790&25,350&101,790&25,350&\cellcolor{green}100,906&\cellcolor{green}25,146&\cellcolor{green}100,906&\cellcolor{green}25,146&100,974&25,146&1.00&352,490&25,146&3.00&203,450&25,350&1.82\\
QV&135&72,360&27,135&108,945&27,135&108,945&27,135&\cellcolor{green}107,971&\cellcolor{green}26,910&\cellcolor{green}107,971&\cellcolor{green}26,910&108,045&26,910&1.00&377,193&26,910&3.00&217,755&27,135&1.82\\
QV&140&78,400&29,400&118,020&29,400&118,020&29,400&\cellcolor{green}117,163&\cellcolor{green}29,202&\cellcolor{green}117,163&\cellcolor{green}29,202&117,228&29,202&1.00&409,333&29,202&3.00&235,900&29,400&1.81\\
QV&145&83,520&31,320&125,715&31,320&125,715&31,320&\cellcolor{green}124,714&\cellcolor{green}31,089&\cellcolor{green}124,714&\cellcolor{green}31,089&124,791&31,089&1.00&435,740&31,089&3.00&251,285&31,320&1.81\\
QV&150&90,000&33,750&135,450&33,750&135,450&33,750&\cellcolor{green}134,566&\cellcolor{green}33,546&\cellcolor{green}134,566&\cellcolor{green}33,546&134,634&33,546&1.00&470,190&33,546&3.00&270,750&33,750&1.81\\
QV&155&95,480&35,805&143,685&35,805&143,685&35,805&\cellcolor{green}142,632&\cellcolor{green}35,562&\cellcolor{green}142,632&\cellcolor{green}35,562&142,713&35,562&1.00&498,400&35,562&3.00&287,215&35,805&1.81\\
QV&160&102,400&38,400&154,080&38,400&154,080&38,400&\cellcolor{green}152,988&\cellcolor{green}38,148&\cellcolor{green}152,988&\cellcolor{green}38,148&153,072&38,148&1.00&534,620&38,148&3.00&308,000&38,400&1.81\\
QV&165&108,240&40,590&162,855&40,590&162,855&40,590&\cellcolor{green}161,893&\cellcolor{green}40,368&\cellcolor{green}161,893&\cellcolor{green}40,368&161,967&40,368&1.00&565,755&40,368&3.00&325,545&40,590&1.81\\
QV&170&115,600&43,350&173,910&43,350&173,910&43,350&\cellcolor{green}172,858&\cellcolor{green}43,107&\cellcolor{green}172,858&\cellcolor{green}43,107&172,938&43,107&1.00&604,108&43,107&3.00&347,650&43,350&1.81\\
QV&175&121,800&45,675&183,225&45,675&183,225&45,675&\cellcolor{green}182,107&\cellcolor{green}45,417&\cellcolor{green}182,107&\cellcolor{green}45,417&182,193&45,417&1.00&636,455&45,417&3.00&366,275&45,675&1.81\\
QV&180&129,600&48,600&194,940&48,600&194,940&48,600&\cellcolor{green}193,796&\cellcolor{green}48,336&\cellcolor{green}193,796&\cellcolor{green}48,336&193,884&48,336&1.00&677,340&48,336&3.00&389,700&48,600&1.81\\
QV&185&136,160&51,060&204,795&51,060&204,795&51,060&\cellcolor{green}203,599&\cellcolor{green}50,784&\cellcolor{green}203,599&\cellcolor{green}50,784&203,691&50,784&1.00&711,625&50,784&3.00&409,405&51,060&1.81\\
QV&190&144,400&54,150&217,170&54,150&217,170&54,150&\cellcolor{green}215,806&\cellcolor{green}53,835&\cellcolor{green}215,806&\cellcolor{green}53,835&215,910&53,835&1.00&754,328&53,835&3.00&434,150&54,150&1.81\\
QV&195&151,320&56,745&227,565&56,745&227,565&56,745&\cellcolor{green}225,993&\cellcolor{green}56,382&\cellcolor{green}225,993&\cellcolor{green}56,382&226,113&56,382&1.00&789,963&56,382&3.00&454,935&56,745&1.81\\
QV&200&160,000&60,000&240,600&60,000&240,600&60,000&\cellcolor{green}239,209&\cellcolor{green}59,679&\cellcolor{green}239,209&\cellcolor{green}59,679&239,316&59,679&1.00&836,185&59,679&3.00&481,000&60,000&1.81\\
\bottomrule
\end{tabular}
\end{center}
\label{table-results-large-circuits-1}
\end{table}

\begin{table}
\tiny
\caption{Results of the evaluations for the algorithms Supremacy and Sycamore for different numbers of qubits. The meaning of the columns is the same as in \autoref{table-results-1}.}
\setlength{\tabcolsep}{1.2mm}
\begin{center}
\begin{tabular}{lrrr|*{4}{rr|}HHHrrr|rrr}
\toprule
\multicolumn{4}{c|}{Original Circuit}&\multicolumn{2}{c|}{\emph{CliffordSimp}}&\multicolumn{2}{c|}{\emph{SquashTK1}}&\multicolumn{2}{c|}{\emph{KAK}}&\multicolumn{2}{c|}{\emph{FullPeephole}}&\multicolumn{3}{H}{AQT}&\multicolumn{3}{c|}{Pytket}&\multicolumn{3}{c}{Qiskit} \\
Name&q&1qg&2qg&1qg&2qg&1qg&2qg&1qg&2qg&1qg&2qg&1qg&2qg&rg&1qg&2qg&rg&1qg&2qg&rg \\
\midrule
Supremacy&4&1,508&500&\cellcolor{green}865&\cellcolor{green}500&866&500&866&500&\cellcolor{green}865&\cellcolor{green}500&1,502&500&1.47&6,474&500&5.11&1,909&500&1.76\\
Supremacy&9&3,520&1,500&2,430&\cellcolor{green}1,508&2,404&1,510&\cellcolor{green}2,400&1,510&2,428&\cellcolor{green}1,508&3,757&1,500&1.34&17,348&1,500&4.82&4,851&1,500&1.62\\
Supremacy&16&6,282&3,000&\cellcolor{green}4,704&\cellcolor{green}3,000&4,730&3,000&4,730&3,000&\cellcolor{green}4,704&\cellcolor{green}3,000&6,872&3,000&1.28&33,033&3,000&4.68&8,874&3,000&1.54\\
Supremacy&25&9,800&5,000&7,663&5,000&\cellcolor{green}7,653&\cellcolor{green}5,000&\cellcolor{green}7,653&\cellcolor{green}5,000&7,663&5,000&10,872&5,000&1.25&54,084&5,000&4.67&14,099&5,000&1.51\\
Supremacy&36&14,072&7,500&\cellcolor{green}11,229&7,505&11,382&\cellcolor{green}7,503&11,382&\cellcolor{green}7,503&\cellcolor{green}11,229&7,505&15,779&7,500&1.24&80,507&7,500&4.70&20,252&7,500&1.48\\
Supremacy&49&19,100&10,500&\cellcolor{green}15,489&\cellcolor{green}10,500&15,768&10,504&15,768&10,504&\cellcolor{green}15,489&\cellcolor{green}10,500&21,546&10,500&1.23&111,929&10,500&4.71&27,855&10,500&1.48\\
Supremacy&64&24,878&14,000&\cellcolor{green}20,403&\cellcolor{green}14,000&20,906&14,000&20,906&14,000&\cellcolor{green}20,403&\cellcolor{green}14,000&28,231&14,000&1.23&148,003&14,000&4.71&36,402&14,000&1.47\\
Supremacy&81&31,412&18,000&\cellcolor{green}26,156&\cellcolor{green}18,000&26,616&18,000&26,616&18,000&\cellcolor{green}26,156&\cellcolor{green}18,000&35,577&18,000&1.21&189,398&18,000&4.70&46,115&18,000&1.45\\
Supremacy&100&38,700&22,500&\cellcolor{green}32,302&22,504&33,108&\cellcolor{green}22,502&33,108&\cellcolor{green}22,502&\cellcolor{green}32,302&22,504&44,055&22,500&1.21&236,359&22,500&4.72&56,842&22,500&1.45\\
Supremacy&121&46,744&27,500&\cellcolor{green}39,247&\cellcolor{green}27,501&40,087&27,503&40,087&27,503&\cellcolor{green}39,247&\cellcolor{green}27,501&53,188&27,500&1.21&288,154&27,500&4.73&68,903&27,500&1.44\\
Supremacy&144&55,538&33,000&\cellcolor{green}46,825&\cellcolor{green}33,000&48,139&33,007&48,139&33,007&\cellcolor{green}46,825&\cellcolor{green}33,000&63,500&33,000&1.21&344,787&33,000&4.73&82,011&33,000&1.44\\
Supremacy&169&65,088&39,000&\cellcolor{green}55,233&\cellcolor{green}39,001&56,564&39,001&56,564&39,001&\cellcolor{green}55,233&\cellcolor{green}39,001&74,454&39,000&1.20&406,870&39,000&4.73&96,178&39,000&1.43\\
Supremacy&196&75,392&45,500&\cellcolor{green}63,913&\cellcolor{green}45,502&65,797&45,504&65,797&45,504&\cellcolor{green}63,913&\cellcolor{green}45,502&86,304&45,500&1.20&474,101&45,500&4.75&111,513&45,500&1.44\\
Sycamore&4&4,000&1,000&2,312&1,000&2,102&1,000&\cellcolor{green}1,815&\cellcolor{green}1,000&2,312&1,000&940&383&0.47&9,956&1,000&3.89&4,164&1,000&1.83\\
Sycamore&9&9,000&3,000&5,100&3,016&\cellcolor{green}4,982&\cellcolor{green}3,016&4,984&3,016&5,100&3,016&5,096&2,133&0.90&29,356&3,000&4.05&10,212&3,000&1.65\\
Sycamore&16&16,000&6,000&9,599&\cellcolor{green}6,034&\cellcolor{green}9,422&6,045&9,429&6,045&9,599&\cellcolor{green}6,034&10,658&4,441&0.98&56,747&6,000&4.06&19,257&6,000&1.63\\
Sycamore&25&25,000&10,000&15,684&\cellcolor{green}10,055&\cellcolor{green}15,514&10,064&\cellcolor{green}15,514&10,064&15,684&\cellcolor{green}10,055&18,194&7,590&1.01&93,285&10,000&4.04&30,815&10,000&1.60\\
Sycamore&36&36,000&15,000&23,096&15,079&\cellcolor{green}22,641&\cellcolor{green}15,074&\cellcolor{green}22,641&\cellcolor{green}15,074&23,096&15,079&27,645&11,587&1.04&138,234&15,000&4.06&45,510&15,000&1.60\\
Sycamore&49&49,000&21,000&31,870&\cellcolor{green}21,095&\cellcolor{green}31,593&21,096&31,595&21,096&31,869&\cellcolor{green}21,095&39,065&16,258&1.05&192,924&21,000&4.06&62,003&21,000&1.58\\
Sycamore&64&64,000&28,000&42,035&\cellcolor{green}28,107&\cellcolor{green}41,792&28,147&41,812&28,147&42,035&\cellcolor{green}28,107&52,145&21,910&1.06&255,701&28,000&4.06&81,862&28,000&1.57\\
Sycamore&81&81,000&36,000&54,104&\cellcolor{green}36,144&\cellcolor{green}53,413&36,173&53,414&36,173&54,117&\cellcolor{green}36,144&67,749&28,267&1.07&327,387&36,000&4.06&103,916&36,000&1.56\\
Sycamore&100&100,000&45,000&67,062&\cellcolor{green}45,191&66,413&45,208&\cellcolor{green}66,405&45,208&67,079&\cellcolor{green}45,191&84,985&35,583&1.08&407,913&45,000&4.06&129,082&45,000&1.56\\
Sycamore&121&121,000&55,000&81,513&\cellcolor{green}55,261&\cellcolor{green}80,620&55,285&80,631&55,285&81,494&\cellcolor{green}55,261&104,200&43,518&1.09&497,481&55,000&4.07&156,210&55,000&1.55\\
Sycamore&144&144,000&66,000&97,142&\cellcolor{green}66,329&96,398&66,342&\cellcolor{green}96,383&66,342&97,185&\cellcolor{green}66,329&124,986&52,345&1.09&595,933&66,000&4.07&187,088&66,000&1.56\\
Sycamore&169&169,000&78,000&114,469&\cellcolor{green}78,329&113,984&78,385&\cellcolor{green}113,981&78,385&114,470&\cellcolor{green}78,329&147,763&61,938&1.09&702,782&78,000&4.06&220,132&78,000&1.55\\
Sycamore&196&196,000&91,000&133,317&\cellcolor{green}91,413&\cellcolor{green}132,046&91,445&132,047&91,445&133,311&\cellcolor{green}91,413&172,753&72,314&1.10&818,792&91,000&4.07&255,704&91,000&1.55\\
\bottomrule
\end{tabular}
\end{center}
\label{table-results-large-circuits-2}
\end{table}

\end{document}